\documentclass[longauth]{aa}

\usepackage{graphicx}
\usepackage{txfonts}
\usepackage{xcolor}
\usepackage{url}
\usepackage[yyyymmdd]{datetime}
\usepackage{hyperref}
\usepackage{rotating}
\usepackage{amsmath,amssymb}
\usepackage{multirow}

\usepackage{listings}
\definecolor{dkgreen}{rgb}{0,0.6,0}
\definecolor{gray}{rgb}{0.5,0.5,0.5}
\definecolor{mauve}{rgb}{0.58,0,0.82}
\definecolor{Magenta}{rgb}{0.75,0,0.75}
\definecolor{Blue}{rgb}{0,0,0.55}
\hypersetup{
    colorlinks=true,
    linkcolor=Blue,
    urlcolor=Magenta,
}

\providecommand{\kms}{\ensuremath{\rm \,km\,s^{-1}}\xspace}

\providecommand{\afe}{\ensuremath{[\alpha/\mathrm{Fe}]}\xspace}
\providecommand{\teff}{\ensuremath{{{T_{\rm eff}}}}\xspace}

\providecommand{\logg}{\ensuremath{{\log g}}\xspace}
\providecommand{\lum}{\ensuremath{{L}}\xspace}

\providecommand{\loggrav}{\ensuremath{\log\,g}\xspace}
\providecommand{\mass}{\ensuremath{{M}}\xspace}
\providecommand{\radius}{\ensuremath{{R}}\xspace}
\providecommand{\mh}{\ensuremath{{[\text{M}/\text{H}]}}\xspace}

\providecommand{\vsini}{\ensuremath{v\sin i}\xspace}
\providecommand{\gravshift}{\ensuremath{{rv_{\rm GR}}}\xspace}
\providecommand{\age}{\ensuremath{{\tau}}\xspace}
\providecommand{\evolstage}{\ensuremath{{\epsilon}}\xspace}

\providecommand{\azero}{\ensuremath{A_0}\xspace}

\providecommand{\ag}{\ensuremath{A_G}\xspace}
\providecommand{\abp}{\ensuremath{A_\mathrm{BP}}\xspace}
\providecommand{\arp}{\ensuremath{A_\mathrm{RP}}\xspace}
\providecommand{\ebpminrp}{\ensuremath{E(G_{\rm BP} - G_{\rm RP})}\xspace}

\providecommand{\ebpminrp}{\ensuremath{E(G_{\rm BP} - G_{\rm RP})}\xspace}
\providecommand{\gmag}{\ensuremath{G}}
\providecommand{\bpmag}{\ensuremath{G_\mathrm{BP}}}
\providecommand{\rpmag}{\ensuremath{G_\mathrm{RP}}}
\providecommand{\mg}{$M_\gmag$}
\providecommand{\ag}{\ensuremath{A_G}\xspace}

\providecommand{\nm}{\ensuremath{\,\mathrm{nm}}\xspace}

\providecommand{\pc}{\ensuremath{\,\rm pc}\xspace}
\providecommand{\kpc}{\ensuremath{\,\rm kpc}\xspace}

\providecommand{\Lsun}{\ensuremath{\,{\lum}_{\odot}}\xspace}
\providecommand{\Msun}{\ensuremath{\,{\mass}_{\odot}}\xspace}

\providecommand{\kms}{\ensuremath{\textrm{km\,s}^{-1}}}

\providecommand{\msun}{\mass_\odot}

\providecommand{\modulename}[1]{#1\xspace}
\providecommand{\apsis}{\modulename{Apsis}}

\providecommand{\dsc}{\modulename{DSC}}
\providecommand{\gspphot}{\modulename{GSP-Phot}}

\providecommand{\gspspec}{\modulename{GSP-Spec}}
\providecommand{\msc}{\modulename{MSC}}
\providecommand{\flame}{\modulename{FLAME}}
\providecommand{\espels}{\modulename{ESP-ELS}}
\providecommand{\esphs}{\modulename{ESP-HS}}
\providecommand{\espcs}{\modulename{ESP-CS}}
\providecommand{\espucd}{\modulename{ESP-UCD}}

\providecommand{\oa}{\modulename{OA}}

\providecommand{\tge}{\modulename{TGE}}

\providecommand{\gaia}{\textit{Gaia}}
\providecommand{\gdr}[1]{Gaia~DR{#1}}
\providecommand{\gedr}[1]{Gaia~eDR{#1}}

\providecommand{\linktoparam}[2]{\href{\linktodoc/Gaia_archive/chap_datamodel/sec_dm_main_tables/ssec_dm_#1.html\##1-#2}{\fieldName{#2}\xspace}}
\providecommand{\linktoparampath}[2]{\href{\linktodoc/Gaia_archive/chap_datamodel/sec_dm_main_tables/ssec_dm_#1.html\##1-#2}{\fieldName{#1.#2}}}

\providecommand{\linktotable}[1]{\href{\linktodoc/Gaia_archive/chap_datamodel/sec_dm_main_tables/ssec_dm_#1.html}{\fieldName{#1}\xspace}}

\providecommand{\linktodoc}{https://gea.esac.esa.int/archive/documentation/GDR3}

\providecommand{\linksec}[2]{\href{\linktodoc/Data_analysis/chap_cu8par/#1}{#2\xspace}}
\providecommand{\linkfig}[1]{\href{\linktodoc/Data_analysis/chap_cu8par/#1}{see table\xspace}}

\makeatletter
\DeclareRobustCommand*{\fieldName}[1]{%
  \begingroup\@fieldName\scantokens{\texttt{\small {#1}}\noexpand}\endgroup}
\begingroup\lccode`\~=`\_\relax
   \lowercase{\endgroup\def\@fieldName{\catcode`\_=\active \let~\_}}
\makeatother

\begin{document}

\title{Gaia Data Release 3: Apsis II - Stellar Parameters}

\providecommand{\orcit}[1]{\protect\href{https://orcid.org/#1}{\protect\includegraphics[width=8pt]{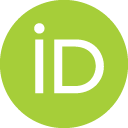}}}
\providecommand{\instref}[1]{\inst{\ref{#1}}}

\authorrunning{M. Fouesneau et al.}
\author{
M.~ Fouesneau\orcit{0000-0001-9256-5516}\inst{\ref{inst:0001}}
     \thanks{Corresponding author: Morgan Fouesneau, fouesneau@mpia.de}
\and
Y.~ Fr\'{e}mat\orcit{0000-0002-4645-6017}\inst{\ref{inst:0002}} \and
R.~ Andrae\orcit{0000-0001-8006-6365}\inst{\ref{inst:0001}} \and
A.J.~ Korn\orcit{0000-0002-3881-6756}\inst{\ref{inst:0004}} \and
C.~ Soubiran\orcit{0000-0003-3304-8134}\inst{\ref{inst:0005}} \and
G.~ Kordopatis\orcit{0000-0002-9035-3920}\inst{\ref{inst:0006}} \and
A.~ Vallenari\orcit{0000-0003-0014-519X}\inst{\ref{inst:0007}} \and
U.~ Heiter\orcit{0000-0001-6825-1066}\inst{\ref{inst:0004}} \and
O.L.~ Creevey\orcit{0000-0003-1853-6631}\inst{\ref{inst:0006}} \and
L.M.~ Sarro\orcit{0000-0002-5622-5191}\inst{\ref{inst:0010}} \and
P.~ de Laverny\orcit{0000-0002-2817-4104}\inst{\ref{inst:0006}} \and
A.C.~ Lanzafame\orcit{0000-0002-2697-3607}\inst{\ref{inst:0012},\ref{inst:0013}} \and
A.~ Lobel\orcit{0000-0001-5030-019X}\inst{\ref{inst:0002}} \and
R.~ Sordo\orcit{0000-0003-4979-0659}\inst{\ref{inst:0007}} \and
J.~ Rybizki\orcit{0000-0002-0993-6089}\inst{\ref{inst:0001}} \and
I.~ Slezak\inst{\ref{inst:0006}} \and
M.A.~ \'{A}lvarez\orcit{0000-0002-6786-2620}\inst{\ref{inst:0018}} \and
R.~ Drimmel\orcit{0000-0002-1777-5502}\inst{\ref{inst:0019}} \and
D.~ Garabato\orcit{0000-0002-7133-6623}\inst{\ref{inst:0018}} \and
L.~ Delchambre\orcit{0000-0003-2559-408X}\inst{\ref{inst:0021}} \and
C.A.L.~ Bailer-Jones\inst{\ref{inst:0001}} \and
D.~ Hatzidimitriou\orcit{0000-0002-5415-0464}\inst{\ref{inst:0023},\ref{inst:0024}} \and
A.~ Lorca\inst{\ref{inst:0025}} \and
Y.~ Le Fustec\inst{\ref{inst:0026}} \and
F.~ Pailler\orcit{0000-0002-4834-481X}\inst{\ref{inst:0027}} \and
N.~ Mary\inst{\ref{inst:0028}} \and
C.~ Robin\inst{\ref{inst:0028}} \and
E.~ Utrilla\inst{\ref{inst:0025}} \and
A.~ Abreu Aramburu\inst{\ref{inst:0031}} \and
J.~Bakker\inst{\ref{inst:0030}} \and
I.~ Bellas-Velidis\inst{\ref{inst:0024}} \and
A.~ Bijaoui\inst{\ref{inst:0006}} \and
R.~Blomme\instref{inst:0002} \and
J.-C.~Bouret\inst{\ref{inst:0166}} \and
N.~ Brouillet\orcit{0000-0002-3274-7024}\inst{\ref{inst:0005}} \and
E.~ Brugaletta\orcit{0000-0003-2598-6737}\inst{\ref{inst:0012}} \and
A.~ Burlacu\inst{\ref{inst:0026}} \and
R.~ Carballo\orcit{0000-0001-7412-2498}\inst{\ref{inst:0037}} \and
L.~ Casamiquela\orcit{0000-0001-5238-8674}\inst{\ref{inst:0005},\ref{inst:0039}} \and
L.~ Chaoul\inst{\ref{inst:0027}} \and
A.~ Chiavassa\orcit{0000-0003-3891-7554}\inst{\ref{inst:0006}} \and
G.~ Contursi\orcit{0000-0001-5370-1511}\inst{\ref{inst:0006}} \and
W.J.~ Cooper\orcit{0000-0003-3501-8967}\inst{\ref{inst:0043},\ref{inst:0019}} \and
C.~ Dafonte\orcit{0000-0003-4693-7555}\inst{\ref{inst:0018}} \and
C.~ Demouchy\inst{\ref{inst:0047}} \and
T.E.~ Dharmawardena\orcit{0000-0002-9583-5216}\inst{\ref{inst:0001}} \and
P.~ Garc\'{i}a-Lario\orcit{0000-0003-4039-8212}\inst{\ref{inst:0050}} \and
M.~ Garc\'{i}a-Torres\orcit{0000-0002-6867-7080}\inst{\ref{inst:0051}} \and
A.~ Gomez\orcit{0000-0002-3796-3690}\inst{\ref{inst:0018}} \and
I.~ Gonz\'{a}lez-Santamar\'{i}a\orcit{0000-0002-8537-9384}\inst{\ref{inst:0018}} \and
A.~ Jean-Antoine Piccolo\orcit{0000-0001-8622-212X}\inst{\ref{inst:0027}} \and
M.~ Kontizas\orcit{0000-0001-7177-0158}\inst{\ref{inst:0023}} \and
Y.~ Lebreton\orcit{0000-0002-4834-2144}\inst{\ref{inst:0057},\ref{inst:0058}} \and
E.L.~ Licata\orcit{0000-0002-5203-0135}\inst{\ref{inst:0019}} \and
H.E.P.~ Lindstr{\o}m\inst{\ref{inst:0019},\ref{inst:0061},\ref{inst:0062}} \and
E.~ Livanou\orcit{0000-0003-0628-2347}\inst{\ref{inst:0023}} \and
A.~ Magdaleno Romeo\inst{\ref{inst:0026}} \and
M.~ Manteiga\orcit{0000-0002-7711-5581}\inst{\ref{inst:0065}} \and
F.~ Marocco\orcit{0000-0001-7519-1700}\inst{\ref{inst:0066}} \and
C. Martayan \inst{\ref{inst:0165}}\and
D.J.~ Marshall\orcit{0000-0003-3956-3524}\inst{\ref{inst:0067}} \and
C.~ Nicolas\inst{\ref{inst:0027}} \and
C.~ Ordenovic\inst{\ref{inst:0006}} \and
P.A.~ Palicio\orcit{0000-0002-7432-8709}\inst{\ref{inst:0006}} \and
L.~ Pallas-Quintela\orcit{0000-0001-9296-3100}\inst{\ref{inst:0018}} \and
B.~ Pichon\orcit{0000 0000 0062 1449}\inst{\ref{inst:0006}} \and
E.~ Poggio\orcit{0000-0003-3793-8505}\inst{\ref{inst:0006},\ref{inst:0019}} \and
A.~ Recio-Blanco\orcit{0000-0002-6550-7377}\inst{\ref{inst:0006}} \and
F.~ Riclet\inst{\ref{inst:0027}} \and
R.~ Santove\~{n}a\orcit{0000-0002-9257-2131}\inst{\ref{inst:0018}} \and
M.S.~ Schultheis\orcit{0000-0002-6590-1657}\inst{\ref{inst:0006}} \and
M.~ Segol\inst{\ref{inst:0047}} \and
A.~ Silvelo\orcit{0000-0002-5126-6365}\inst{\ref{inst:0018}} \and
R.L.~ Smart\orcit{0000-0002-4424-4766}\inst{\ref{inst:0019}} \and
M.~ S\"{ u}veges\orcit{0000-0003-3017-5322}\inst{\ref{inst:0083}} \and
F.~ Th\'{e}venin\inst{\ref{inst:0006}} \and
G.~ Torralba Elipe\orcit{0000-0001-8738-194X}\inst{\ref{inst:0018}} \and
A.~ Ulla\orcit{0000-0001-6424-5005}\inst{\ref{inst:0086}} \and
E.~ van Dillen\inst{\ref{inst:0047}} \and
H.~ Zhao\orcit{0000-0003-2645-6869}\inst{\ref{inst:0006}} \and
J.~ Zorec\inst{\ref{inst:0089}}
}
\institute{
Max Planck Institute for Astronomy, K\"{ o}nigstuhl 17, 69117 Heidelberg, Germany\relax \label{inst:0001}
\and Royal Observatory of Belgium, Ringlaan 3, 1180 Brussels, Belgium\relax \label{inst:0002}\vfill
\and Observational Astrophysics, Division of Astronomy and Space Physics, Department of Physics and Astronomy, Uppsala University, Box 516, 751 20 Uppsala, Sweden\relax \label{inst:0004}\vfill
\and Laboratoire d'astrophysique de Bordeaux, Univ. Bordeaux, CNRS, B18N, all{\'e}e Geoffroy Saint-Hilaire, 33615 Pessac, France\relax \label{inst:0005}\vfill
\and Universit\'{e} C\^{o}te d'Azur, Observatoire de la C\^{o}te d'Azur, CNRS, Laboratoire Lagrange, Bd de l'Observatoire, CS 34229, 06304 Nice Cedex 4, France\relax \label{inst:0006}\vfill
\and INAF - Osservatorio astronomico di Padova, Vicolo Osservatorio 5, 35122 Padova, Italy\relax \label{inst:0007}\vfill
\and Dpto. de Inteligencia Artificial, UNED, c/ Juan del Rosal 16, 28040 Madrid, Spain\relax \label{inst:0010}\vfill
\and INAF - Osservatorio Astrofisico di Catania, via S. Sofia 78, 95123 Catania, Italy\relax \label{inst:0012}\vfill
\and Dipartimento di Fisica e Astronomia ""Ettore Majorana"", Universit\`{a} di Catania, Via S. Sofia 64, 95123 Catania, Italy\relax \label{inst:0013}\vfill
\and CIGUS CITIC - Department of Computer Science and Information Technologies, University of A Coru\~{n}a, Campus de Elvi\~{n}a s/n, A Coru\~{n}a, 15071, Spain\relax \label{inst:0018}\vfill
\and INAF - Osservatorio Astrofisico di Torino, via Osservatorio 20, 10025 Pino Torinese (TO), Italy\relax \label{inst:0019}\vfill
\and Institut d'Astrophysique et de G\'{e}ophysique, Universit\'{e} de Li\`{e}ge, 19c, All\'{e}e du 6 Ao\^{u}t, B-4000 Li\`{e}ge, Belgium\relax \label{inst:0021}\vfill
\and Department of Astrophysics, Astronomy and Mechanics, National and Kapodistrian University of Athens, Panepistimiopolis, Zografos, 15783 Athens, Greece\relax \label{inst:0023}\vfill
\and National Observatory of Athens, I. Metaxa and Vas. Pavlou, Palaia Penteli, 15236 Athens, Greece\relax \label{inst:0024}\vfill
\and Aurora Technology for European Space Agency (ESA), Camino bajo del Castillo, s/n, Urbanizacion Villafranca del Castillo, Villanueva de la Ca\~{n}ada, 28692 Madrid, Spain\relax \label{inst:0025}\vfill
\and Telespazio for CNES Centre Spatial de Toulouse, 18 avenue Edouard Belin, 31401 Toulouse Cedex 9, France\relax \label{inst:0026}\vfill
\and CNES Centre Spatial de Toulouse, 18 avenue Edouard Belin, 31401 Toulouse Cedex 9, France\relax \label{inst:0027}\vfill
\and Thales Services for CNES Centre Spatial de Toulouse, 18 avenue Edouard Belin, 31401 Toulouse Cedex 9, France\relax \label{inst:0028}\vfill
\and European Space Agency (ESA), European Space Astronomy Centre (ESAC), Camino bajo del Castillo, s/n, Urbanizacion Villafranca del Castillo, Villanueva de la Ca\~{n}ada, 28692 Madrid, Spain\relax 
 \label{inst:0030}\vfill
\and ATG Europe for European Space Agency (ESA), Camino bajo del Castillo, s/n, Urbanizacion Villafranca del Castillo, Villanueva de la Ca\~{n}ada, 28692 Madrid, Spain\relax \label{inst:0031}\vfill
\and Aix Marseille Univ, CNRS, CNES, LAM, Marseille, France\relax \label{inst:0166}\vfill
\and Dpto. de Matem\'{a}tica Aplicada y Ciencias de la Computaci\'{o}n, Univ. de Cantabria, ETS Ingenieros de Caminos, Canales y Puertos, Avda. de los Castros s/n, 39005 Santander, Spain\relax \label{inst:0037}\vfill
\and GEPI, Observatoire de Paris, Universit\'{e} PSL, CNRS, 5 Place Jules Janssen, 92190 Meudon, France\relax \label{inst:0039}\vfill
\and Centre for Astrophysics Research, University of Hertfordshire, College Lane, AL10 9AB, Hatfield, United Kingdom\relax \label{inst:0043}\vfill
\and APAVE SUDEUROPE SAS for CNES Centre Spatial de Toulouse, 18 avenue Edouard Belin, 31401 Toulouse Cedex 9, France\relax \label{inst:0047}\vfill
\and Theoretical Astrophysics, Division of Astronomy and Space Physics, Department of Physics and Astronomy, Uppsala University, Box 516, 751 20 Uppsala, Sweden\relax \label{inst:0049}\vfill
\and European Space Agency (ESA), European Space Astronomy Centre (ESAC), Camino bajo del Castillo, s/n, Urbanizacion Villafranca del Castillo, Villanueva de la Ca\~{n}ada, 28692 Madrid, Spain\relax \label{inst:0050}\vfill
\and Data Science and Big Data Lab, Pablo de Olavide University, 41013, Seville, Spain\relax \label{inst:0051}\vfill
\and LESIA, Observatoire de Paris, Universit\'{e} PSL, CNRS, Sorbonne Universit\'{e}, Universit\'{e} de Paris, 5 Place Jules Janssen, 92190 Meudon, France\relax \label{inst:0057}\vfill
\and Universit\'{e} Rennes, CNRS, IPR (Institut de Physique de Rennes) - UMR 6251, 35000 Rennes, France\relax \label{inst:0058}\vfill
\and Niels Bohr Institute, University of Copenhagen, Juliane Maries Vej 30, 2100 Copenhagen {\O}, Denmark\relax \label{inst:0061}\vfill
\and DXC Technology, Retortvej 8, 2500 Valby, Denmark\relax \label{inst:0062}\vfill
\and CIGUS CITIC, Department of Nautical Sciences and Marine Engineering, University of A Coru\~{n}a, Paseo de Ronda 51, 15071, A Coru\~{n}a, Spain\relax \label{inst:0065}\vfill
\and IPAC, Mail Code 100-22, California Institute of Technology, 1200 E. California Blvd., Pasadena, CA 91125, USA\relax \label{inst:0066}\vfill
\and European Organisation for Astronomical Research in the Southern Hemisphere, Alonso de C\'{o}rdova 3107, Vitacura, 19001 Casilla, Santiago de Chile, Chile\relax \label{inst:0165}\vfill
\and IRAP, Universit\'{e} de Toulouse, CNRS, UPS, CNES, 9 Av. colonel Roche, BP 44346, 31028 Toulouse Cedex 4, France\relax \label{inst:0067}\vfill
\and Institute of Global Health, University of Geneva\relax \label{inst:0083}\vfill
\and Applied Physics Department, Universidade de Vigo, 36310 Vigo, Spain\relax \label{inst:0086}\vfill
\and Sorbonne Universit\'{e}, CNRS, UMR7095, Institut d'Astrophysique de Paris, 98bis bd. Arago, 75014 Paris, France\relax \label{inst:0089}\vfill
}

\date{Version: \today~\currenttime}
\abstract
%
{The third \gaia\ data release (\gdr{3}) contains, beyond the astrometry and photometry, dispersed light for hundreds of millions of sources from the Gaia prism spectra (BP and RP) and the spectrograph (RVS). This data release opens a new window on the chemo-dynamical properties of stars in our Galaxy, essential knowledge for understanding the structure, formation, and evolution of the Milky Way.}
{To provide insight into the physical properties of Milky Way stars, we used these data to produce a uniformly-derived, all-sky catalog of stellar astrophysical parameters (APs): atmospheric properties (\teff, \logg, \mh, \afe, activity index, emission lines, rotation), $13$ chemical abundance estimates, evolution characteristics (radius, age, mass, bolometric luminosity), distance, and dust extinction.}
{We developed the astrophysical parameters inference system (\apsis) pipeline to infer APs of \gaia\ objects by analyzing their astrometry, photometry, BP/RP, and RVS spectra. We validate our results against other works in the literature, including benchmark stars, interferometry, and asteroseismology. Here we assessed the stellar analysis performance from \apsis\ statistically.}
{We describe the quantities we obtained, including the underlying assumptions and the limitations of our results.
We provide guidance and identify regimes in which our parameters should and should not be used.}
{Despite some limitations, this is the most extensive catalog of uniformly-inferred stellar parameters to date.
These comprise \teff, \logg, and \mh (470 million using BP/RP, 6 million using RVS), radius (470 million), mass (140 million), age (120 million),
chemical abundances (5 million), diffuse interstellar band analysis (1/2 million), activity indices (2 million), H$\alpha$ equivalent widths (200 million), and further classification of spectral types (220 million) and emission-line stars (50 thousand).
More precise and detailed astrophysical parameters based on epoch BP, RP, and RVS spectrophotometry are planned for the next Gaia data release.
\\
\textit{Our catalog is available from the Gaia Archive and partner data centers}.
}


\maketitle


\section{Introduction}\label{sec:introduction}

Studying the present-day structure and substructures of the Milky Way is one of the most direct ways of understanding the true nature of the Galaxy formation mechanism and evolutionary history. Gaia is an ambitious space mission of the European Space Agency (ESA) to primarily provide a three-dimensional map of the Milky Way with an unprecedented volume and precision \citep{GaiaMission}. It represents a revolution in galactic archaeology and a leap forward to reveal how galaxies take shape and investigate our own's exciting complexities.
Although it observes only one percent of our Galaxy's stellar population, Gaia still characterizes $\sim1.8$ billion stars across the Milky Way, measuring their positions, parallaxes, and proper motions.
It provides us with not only their three-dimensional positions but also their two- or three-dimensional velocities through the proper-motion for $\sim 1.4$ billion stars and radial velocity measurements for $\sim 33$ million bright stars, respectively \citep[\gedr{3}]{EDR3-DPACP-130}.

\gdr{3} \citep{DR3-DPACP-185} improves upon the previous releases by both improving the quality of the previously released data and providing entirely new data products:
(i) the dispersed light spectra from the spectro-photometry (BP, blue photometer $[330-680]\nm$; RP, red photometer $[640-1050\nm$) for $\sim 100$\,million stars, in addition to their integrated photometry (\bpmag, \rpmag) and the white light \gmag-band published in \gedr{3} \citep{DR3-DPACP-118};
(ii) the medium resolution spectroscopy (RVS, radial velocity spectrometer $[845-872]\nm$, $\lambda/\Delta\lambda\sim ~ 11500$) for $\sim 1$\, million stars \citep{DR3-DPACP-154}.

In the previously released data, \citet{GDR2Andrae2018} published the first set of stellar parameters from the analysis of the integrated photometry and parallaxes available in \gdr{2} \citep{2018A&A...616A...1G}.
In contrast, \gdr{3} provides a complex set of astrophysical parameters obtained from the analysis of the Gaia's astrometry measurements and the BP, RP, and RVS spectra.
This wide variety of information enables us to conduct a hyper-dimensional analysis of the Milky Way populations that have never been possible before the Gaia era.

\medskip

The present work is one of a series of three papers on the \gdr{3} astrophysical parameters. \citet{DR3-DPACP-157} presents an overview of the astrophysical parameters inference system (\apsis) and its overall contributions to \gdr{3}. This paper focuses on the stellar content description and quality assessments. The non-stellar content is presented in \citet{DR3-DPACP-158}.
For more technical details on the \apsis\ modules, we refer readers to the online documentation\footnote{\gdr{3} online documentation: \url{\linktodoc}} \citep{onlinedocdr3} and specific publications describing some of the modules (%
\gspphot\ in \citealt{DR3-DPACP-156};
\gspspec\ in \citealt{DR3-DPACP-186}; and
\espcs\ in \citealt{DR3-DPACP-175}%
). We listed the relevant module acronyms in Table~\ref{tab:modules}.

\begin{figure*}
    \centering
    \includegraphics[width=2\columnwidth, clip, trim=0 1.3cm 0 0]{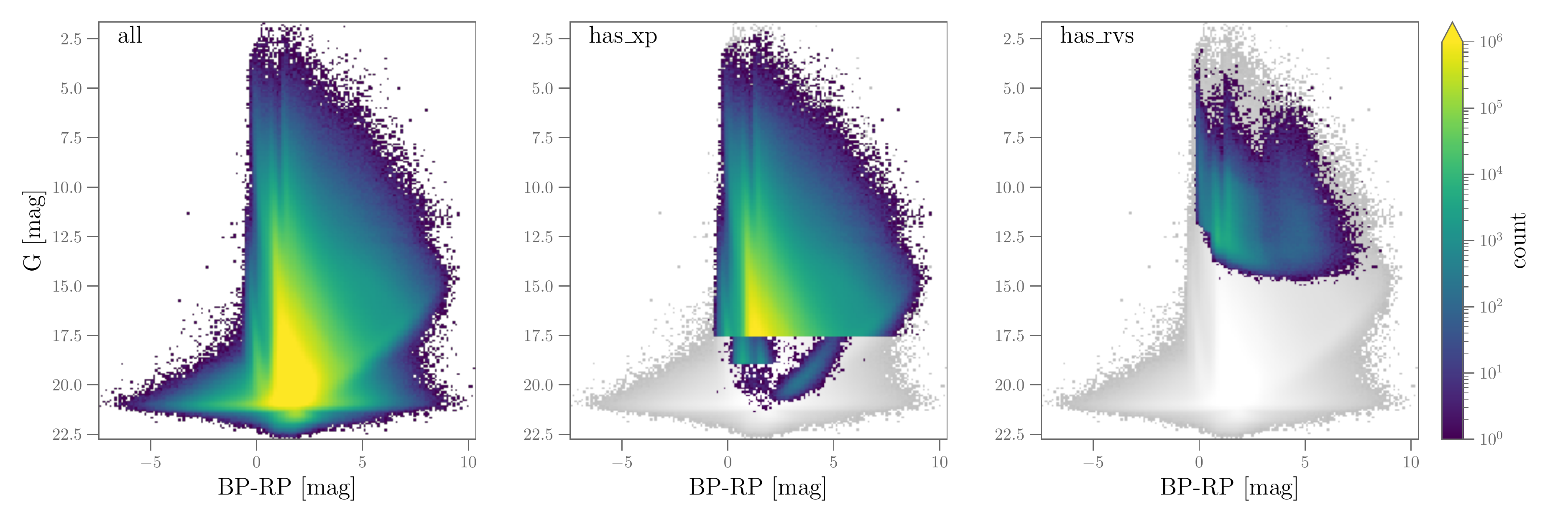}
    \includegraphics[width=2\columnwidth]{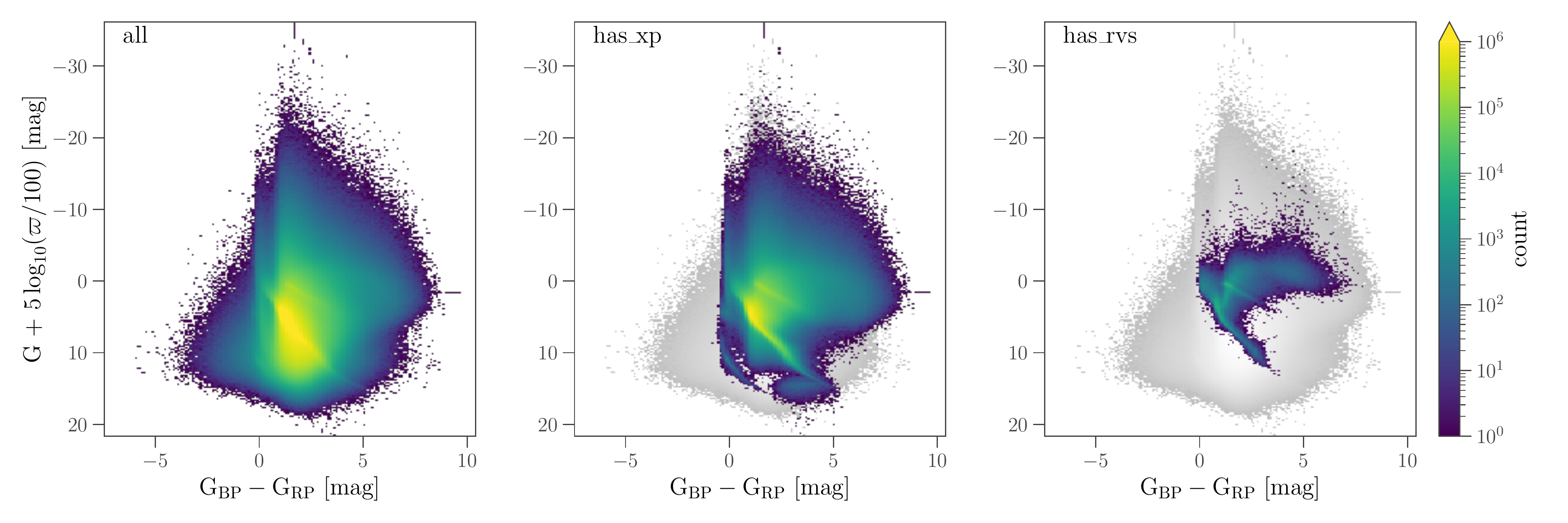}
    \caption{Distribution of the sources in color--magnitude space processed by \apsis\ according to the available measurements. The top panels show the observed color--magnitude diagram. In contrast, the bottom panels show their absolute magnitude computed using the inverse parallax as the distance and assuming zero extinction for sources with positive parallax measurements. From left to right, the sources with G, BP, and RP photometry ("all"), those with published BP/RP spectra (\linktoparampath{gaia_source}{has_xp}) and with RVS spectra (\linktoparampath{gaia_source}{has_rvs}), respectively. The gray density in the middle and right panels indicates the whole sample's distribution for reference. We note the peculiar distribution of BP/RP fainter than $\gmag=17.65$\,mag in the top middle panel, corresponding to selected UCDs (red sources) and extragalactic sources (blue sources). The inverse parallax used in the bottom panels includes low-quality parallaxes responsible for the non-physical high brightness of many sources.
    }\label{fig:cmd_per_data}
\end{figure*}

We only process stellar sources down to $\gmag=19$\,mag for which Gaia provides us with a BP/RP or RVS spectrum, except for ultra-cool dwarfs (UCDs) where we selectively processed {78\,739} sources fainter than this limit (see Fig.~\ref{fig:cmd_per_data}). This limiting magnitude choice was driven primarily by the limited processing time of the BP/RP spectra. The astrophysical parameter dataset contains stellar spectroscopic and evolutionary parameters for 470 million sources.  These comprise \teff, \logg, and \mh (470 million using BP/RP, 6 million using RVS), radius (470 million), mass (140 million), age (120 million),
chemical abundances (up to 5 million), diffuse interstellar band analysis (0.5 million), activity indices (2 million), H$\alpha$ equivalent widths (200 million), and further classification of spectral types (220 million) and emission-line stars (50 thousand).

The work described here was carried out under the umbrella of the Gaia Data Processing and Analysis Consortium (DPAC) within Coordination Unit 8 (CU8; see \citealt{GaiaMission} for an overview of the DPAC).
We realize that one can create more precise, and possibly more accurate, estimates of the stellar parameters by cross-matching Gaia with other survey data, such as GALEX \citep{Morrissey2007}, Pan-STARRS \citep{panstarrs1}, or catWISE \citep{Eisenhardt2020} and spectroscopic surveys such as LAMOST \citep{Luo2019}, GALAH \citep{Buder2021}, or APOGEE \citep{Jonsson2020}. For example, \citet{2022arXiv220103252F}, \citet{2022A&A...658A..91A}, and \citet{2022ApJ...925..164H} combined Gaia data with other photometry and spectroscopic surveys to derive APs for millions of stars.\footnote{Survey acronyms:
GALEX: the Galaxy Evolution Explorer;
Pan-STARRS: the Panoramic Survey Telescope and Rapid Response System;
APOGEE: the Apache Point Observatory Galactic Evolution Experiment;
catWISE: the catalog from the Wide-field Infrared Survey Explorer;
LAMOST: the Large Sky Area Multi-Object Fibre Spectroscopic Telescope;
and
GALAH: the Galactic archaeology with HERMES.
}
However, the remit of the Gaia-DPAC is to process the Gaia data. Further exploitation, for instance, including data from other catalogs, is left to the community at large. Yet, these ``Gaia-only'' stellar parameters will assist the exploitation of \gdr{3} and the validation of such extended analyses.

We continue this article in Sect. \ref{sec:overview} with a brief overview of our assumptions and key processing aspects. In Sect.~\ref{sec:APcontent}, we describe the \gdr{3} AP content and the validation of our results, their internal consistency, and we compare them against other published results (e.g., benchmark stars, interferometry, and asteroseismology). Finally, we highlight a few applications of our catalog in Sect.~\ref{sec:deeper} and its limitations in Sect.~\ref{sec:limitations} before we summarize in Sect. \ref{sec:summary}.

\section{Overview of Stellar APs in GDR3}\label{sec:overview}

\begin{figure}
    \centering
    \includegraphics[width=\columnwidth]{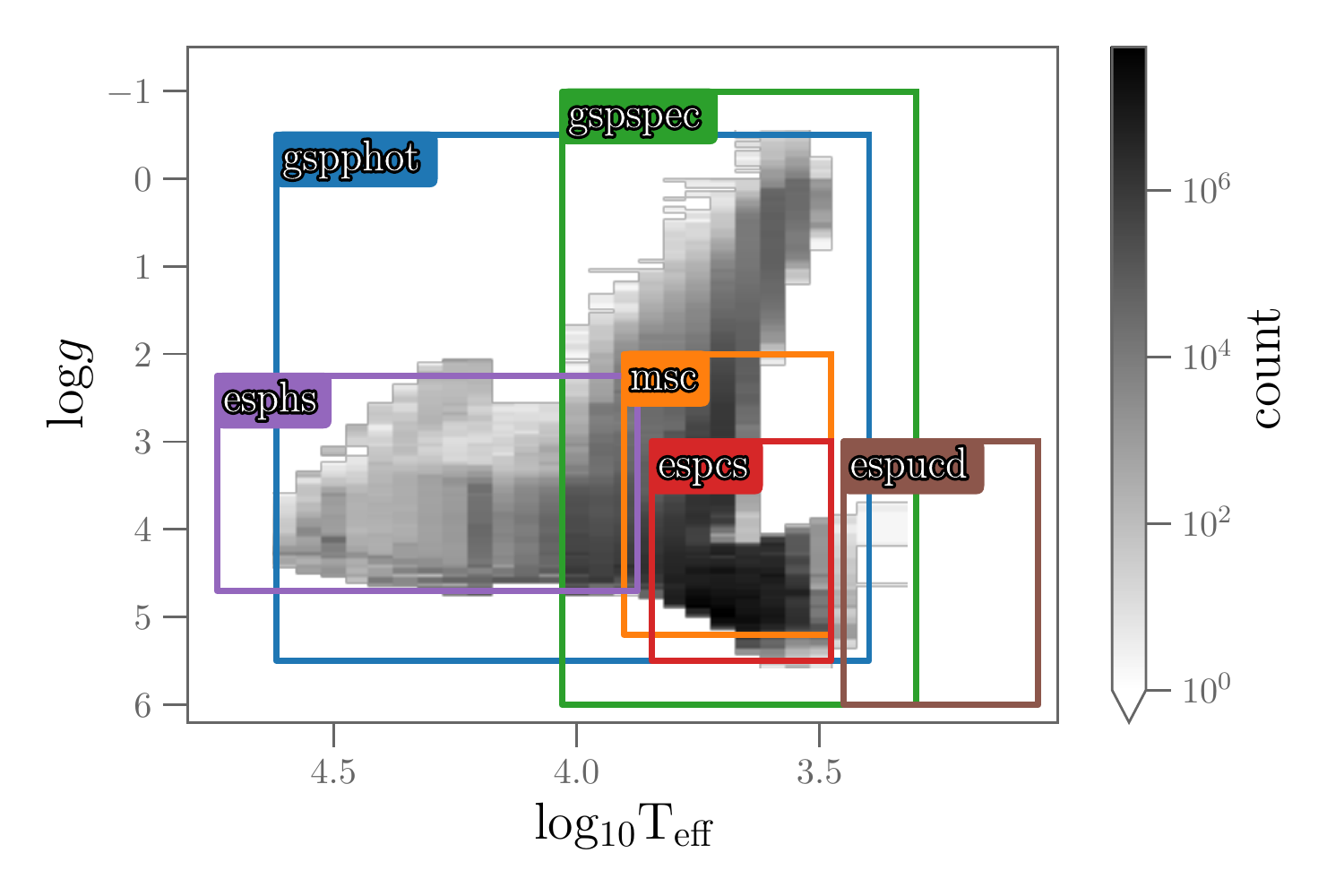}\\
    \caption{Stellar parameter space spanned by \apsis modules in the Kiel diagram. Boxes indicate the modules producing estimates for either \teff, \loggrav or both.
    The density distribution represents the content from \linktotable{gaiadr3.gaia_source}. }
    \label{fig:apsis_kiel_modules}
\end{figure}

The goal of \apsis\ is to classify and estimate astrophysical parameters for the Gaia sources using (only) the Gaia data \citep{BailerJones2013, DR3-DPACP-157}. In addition to assisting the exploitation of \gdr{3}, the DPAC data processing itself uses these APs internally, for example, to help the template-based radial velocity extraction from the RVS spectra, the identification of quasars used to fix the astrometric reference frame or the optimization of the BP/RP calibration.

We designed the \apsis software to provide estimates for a broad class of objects covering a significant fraction of the Gaia catalog, rather than treating specific types of objects. \apsis consists of several modules with different functions and source selections. \citet{DR3-DPACP-157} presents the architecture and the modules of \apsis separately. We provide in Fig. \ref{fig:apsis_kiel_modules} a schematic overview of the source selection per \apsis module in the Kiel diagram. Some modules do not appear on this diagram as they have a more complex role (e.g., emission lines, classification).

\subsection{Source processing selection function}\label{sec:stellar_selection_function}

\begin{table}
    \caption{\apsis\ module acronyms we mention in this manuscript.}
    \label{tab:modules}
    {\centering
    \begin{tabular}{rl}
        \hline
        acronym         & description                                        \\
        \hline\hline
        \textit{\apsis} & \textit{Astrophysical Parameters Inference System} \\
        \dsc            & Discrete Source Classifier                         \\
        \textit{GSP}    & \textit{Generalized Stellar Parametrizer}          \\
        \gspphot        & -- from Photometry (photometry \& BP, RP spectra)  \\
        \gspspec        & -- from Spectroscopy  (RVS spectra)                \\
        \textit{ESP}    & \textit{Extended Stellar Parametrizer}             \\
        \espcs          & -- for Cool Stars                                  \\
        \espels         & -- for Emission Line Stars                         \\
        \esphs          & -- for Hot Stars                                   \\
        \espucd         & -- for Ultra Cool Dwarfs                           \\
        \flame          & Final Luminosity Age Mass Estimator                \\
        \oa             & Outlier Analysis                                   \\
        \msc            & Multiple Star Classifier                           \\
        \tge            & Total Galactic Extinction                          \\
        \hline
    \end{tabular}}
    Refer to Sect.~\ref{sec:APcontent} for module descriptions.
\end{table}

This section details the source selection and assumptions we applied during the processing of stellar objects.

First, we processed only sources for which one of the BP, RP, or RVS spectra was available with at least 10 focal plane transits (repeated observations). Which sources are processed by which modules depends on (1) the availability of the necessary data; (2) the signal-to-noise ratio (SNR) of the data, brightness to first order; and (3) potentially the outputs from other modules.

\gspphot \citep{DR3-DPACP-156} operates on all sources with BP/RP spectra down to $\gmag=19$\,mag. As we expect that more than 99\% of sources down to this brightness are stars, there is a minor overhead of computation time in applying \gspphot\ to every source and \gspspec \citep{DR3-DPACP-186} on all sources which have RVS spectra with {SNR > 20, i.e., $\gmag\la 13-14$\,mag}.

Following these two independent general analyses, \apsis refines the characterization of Gaia sources with specific modules.
\flame operates on a subset of sources with APs of ``sufficient'' precision from \gspphot ($\gmag<18.25$\,mag) and \gspspec ($\gmag<14$\,mag), based on their reported uncertainties.
\msc\ analyses all sources with  $\gmag<18.25$\,mag and treats every source as though it were a system of two unresolved stars.
The remaining modules, specifically \espcs \citep{DR3-DPACP-175}, \esphs, \espels, and \espucd only analyze objects of ``their'' class, i.e., active cool-stars, hot stars, emission-line stars, and ultra-cool dwarfs. Apart from \espucd\ which analyses UCDs fainter than $\gmag = 19$\,mag, the other specific modules only produce results for sources with $G<17.65$\,mag. Finally, \gspphot\ also provides the \azero\ estimates used by \tge\ to produce an all-sky (two-dimensional) map of the total Galactic extinction, meaning the cumulative amount of extinction in front of objects beyond the edge of our Galaxy (see Sect.~\ref{sec:dust} and \citealt{DR3-DPACP-158}).
The various quoted magnitude limits are independent of the stars' physical properties and the quality of the spectra. Instead, these limits came from the \apsis\ processing scheme and processing time limitations.%
\footnote{
We used \gmag\ to divide the entire Gaia data set into chunks of approximately 150 million sources. Some modules (e.g.\ \gspphot) ran faster than others (e.g.\ the ESP modules) and processed fainter chunks of data.
}
In addition to and in contrast with the classifications from some these analysis modules, \apsis comprises two modules dedicated to \textit{empirical} classifications of sources. \dsc classifies sources probabilistically into five classes -- quasar, galaxy, star, white dwarf, physical binary star -- although it is primarily intended to identify extragalactic sources and \oa complement this classification by clustering those sources with the lowest classification probabilities from \dsc. See Sect.~\ref{sec:weird} and details in \citet{DR3-DPACP-157}, and \citet{DR3-DPACP-158}.

\begin{figure*}
    \centering
    \includegraphics[width=2\columnwidth]{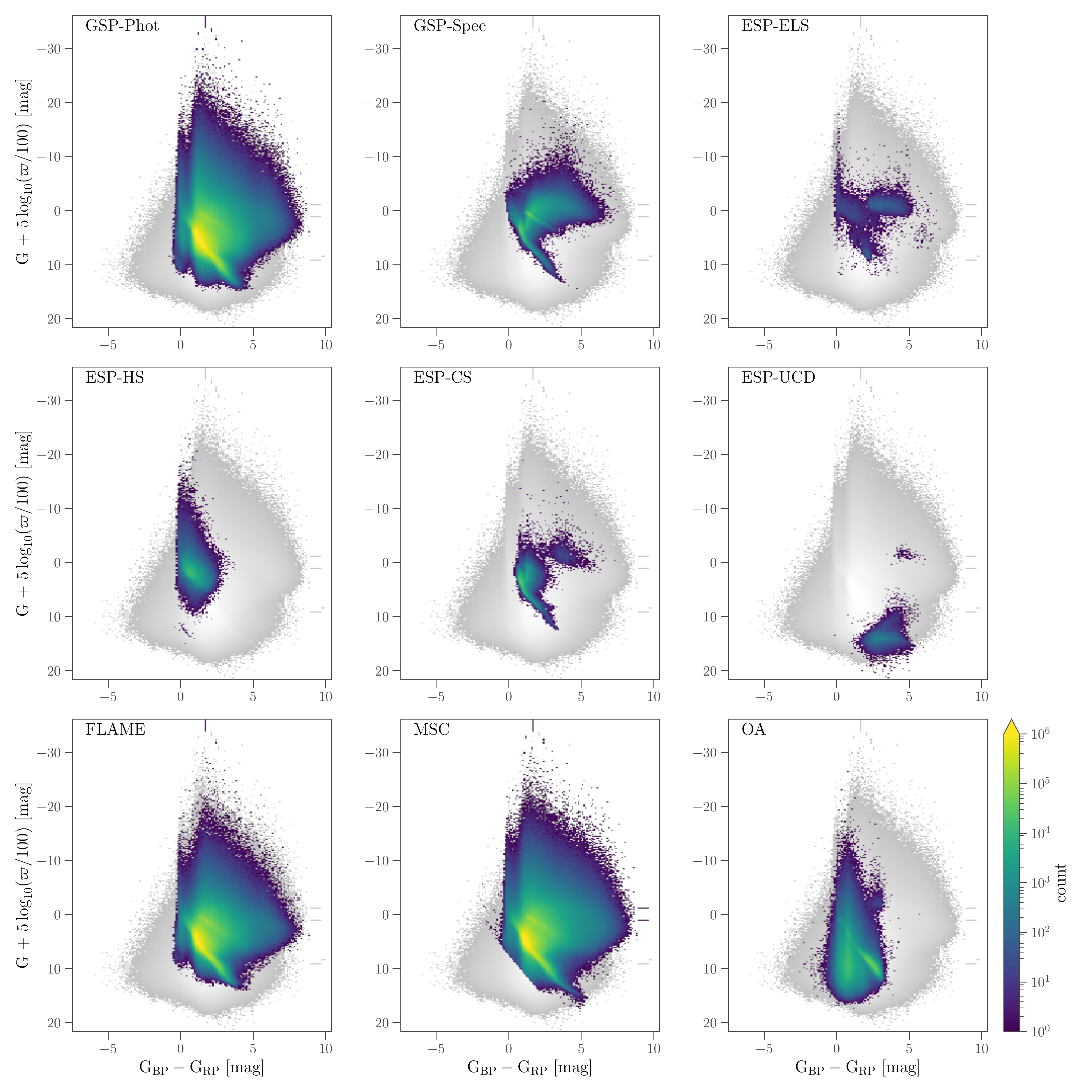}
    \caption{Distributions of the sources processed by \apsis\ in \gdr{3} in CMD space. Each panel highlights the sources a module processed on top of the sources with G, BP, and RP photometry (gray density).}
    \label{fig:cqd_per_module}
\end{figure*}

We summarize the \apsis modules' target selections in Fig.~\ref{fig:cqd_per_module}. We use the inverse parallax as a proxy to emphasize the stellar loci of the targets. Even though we did not explicitly select on \bpmag - \rpmag\ colors, we note that most of the sources with \bpmag - \rpmag < -0.8\,mag in \gdr{3} are not stellar objects according to the \apsis\ processing definitions. This selection translates that stellar evolution models (e.g. PARSEC\footnote{PARSEC isochrones available from \url{http://stev.oapd.inaf.it/cgi-bin/cmd}.}) do not predict bluer stars than \bpmag - \rpmag < -0.6\,mag in the absence of noise in the measurements and within the chemical abundance regime of our analysis.

\subsection{Stellar processing modules \& stellar definition(s)}\label{sec:stellardef}

A principle of \apsis\ in \gdr{3} is to use \textit{only} Gaia data on individual sources when inferring the APs.
We only use non-Gaia observations for validation and calibration.
We define stellar objects as those that remain after removing other kinds of objects: for instance,
extragalactic sources \citep[i.e., galaxies and quasars;][]{DR3-DPACP-101} through dedicated modules such as \dsc and with proper motion, Gaia brightness, and color selections. \apsis presently ignores morphological information \citep{DR3-DPACP-153} and does not take stellar variability \citep{DR3-DPACP-165} into account. As it works with combined epoch spectra (BP, RP, and RVS), some time-variable sources (e.g., Cepheids) received spurious APs from \apsis. \citet{DR3-DPACP-162} summarizes the characterization of variable sources with dedicated pipelines. In the future we plan to investigate using epoch data and whether variability information could improve the quality of our results.

A consequence of our analysis design is that \apsis can assign multiple sets of APs to any given source.
Figure~\ref{fig:apsis_kiel_modules} illustrates the overlap between modules, which for example, leads to four estimates of temperatures for some main-sequence stars. The values we derive not only depend on the data we measure but also on the stellar models we adopt (as embodied in the training data) and other assumptions made, see \citet{DR3-DPACP-157} for a brief overview and the \href{\linktodoc}{online documentation} for details. We can never know a star's ``true'' APs with 100\% confidence. Which estimate to use inevitably remains a decision for the user. For those users who don't want to make this choice, \gspphot estimates APs for all the stars, so there is always a homogeneous set of stellar APs available.

The situation is in the details even more complex because a few of the modules themselves comprise multiple algorithms or multiple sets of assumptions, each providing separate estimates. One reason for this choice is to cross-validate our results: if two or more algorithms give similar results for the same source (and training data), our confidence in the results may increase. For example, \gspspec provides estimates from Matisse-Gauguin \citep{RecioBlanco2016} and a neural-network approach \citep{Manteiga2010} using the same RVS data. Another reason is that we do not use a common set of stellar models: \gspphot operates with four different atmospheric libraries with overlapping parameter spaces but significant differences (see Sect.~\ref{sec:atmosphere-primary}).

Finally, while \gdr{3} reports APs on a wide range of stellar types,
we did not optimize \apsis\ to derive parameters for white dwarfs (WDs), horizontal-branch (HB), and asymptotic giant-branch (AGBs) stars. We did not attempt to model their specific physical conditions (e.g.\ compositional changes due to dredge-up, atomic diffusion, enriched atmosphere, and circumstellar dust).

\subsection{Input data of \apsis\ processing}\label{sec:inputdata}

As \citet{DR3-DPACP-157} describes the \apsis\ input data and their pre-processing exhaustively, here we briefly summarize the most relevant aspects to the stellar APs.
In the context of determining the stellar APs, we used sky positions, the parallaxes, the integrated photometry measurements, and the BP/RP and RVS spectra. However, we note that the classifications by \dsc\ also used proper motions.

Although \apsis\ mainly processed the sources independently (apart from \tge and \oa), their positions on the sky were informative to determine their APs. For instance, we may see a source located near the Galactic center behind a significant amount of extinction, while it would be less likely towards high Galactic latitudes. Therefore, we defined sky position dependent priors, using \citet{2020PASP..132g4501R} as a representative view of the Gaia sky, for instance. The details vary from module to module.

We implemented the parallax zero points from \citet{EDR3-DPACP-132}, which vary with magnitude, color, ecliptic latitude, and astrometric solution type (\linktoparampath{gaia_source}{astrometric_params_solved}). A code is provided with \gdr{3} to compute the parallax zero points.\footnote{\gedr{3} provides the parallax zero point code  \url{https://www.cosmos.esa.int/web/gaia/edr3-code}.}

We used the integrated photometry in the \gmag, \bpmag\, and \rpmag\ bands, in association with the zero-points provided by \citet{EDR3-DPACP-117}.
In addition, we also implemented the correction to the \gmag-band photometry from \citet{EDR3-DPACP-120}, which depends on \gmag, \bpmag-\rpmag\ color, and the astrometric solution type. We emphasize that the parallax zero-point remains calibrated on the original \gmag-band photometry. However, \gdr{3} publishes these corrected values in \linktoparampath{gaia_source}{phot_g_mean_mag}.

\apsis\ derived some of the APs from the analysis of the RVS spectra. The RVS processing pipeline provided us with the time- or epoch-averaged spectra, also called mean spectra, after removing potential cosmic rays and the deblending of overlapping sources. The pipeline delivers the spectra in their stellar rest-frame -- corrected for the star's radial velocity (\linktoparampath{gaia_source}{radial_velocity}) -- and normalized at the local (pseudo-)continuum (\teff\ $\ge$~3\,500~K).
Our analysis used these final spectra re-sampled from $846$ to $870$\,nm, with a constant spacing of $0.01$ nm.  \citet{DR3-DPACP-154} describe in detail the processing of RVS spectra. However, \apsis\ modules rebin the spectra to their optimal use-cases in the perspective of increasing the signal-to-noise ratio of their relevant spectral features \citep[for details]{DR3-DPACP-157}.

Most of the \apsis\ modules produced APs from the analysis of the BP and RP spectra (see examples in Fig.~\ref{fig:seds_teff_modules}).
\gdr{3} provides us with the (epoch) mean BP and RP spectra in a series of coefficients associated with Gauss-Hermite polynomials.
This format results from the complexity of the prism observations. \citet{2021A&A...652A..86C} describes the processing of the spectra. These coefficients contain a flux calibrated (mathematical) continuous representation of the spectra that the \apsis~pipeline internally samples\footnote{\gdr{3} provides GaiaXPy, a Python package to sample the BP and RP spectra; \url{https://www.cosmos.esa.int/web/gaia/gaiaxpy}.} approximately uniformly in pseudo instrumental pixel space, but non-uniform in wavelengths (see Fig.~4 from \citealt{DR3-DPACP-157}).

\subsection{Typical examples and challenges of stellar BP/RP spectra}\label{sec:BPRPexamples}

\begin{figure}
    \begin{center}
        \includegraphics[width=0.95\linewidth]{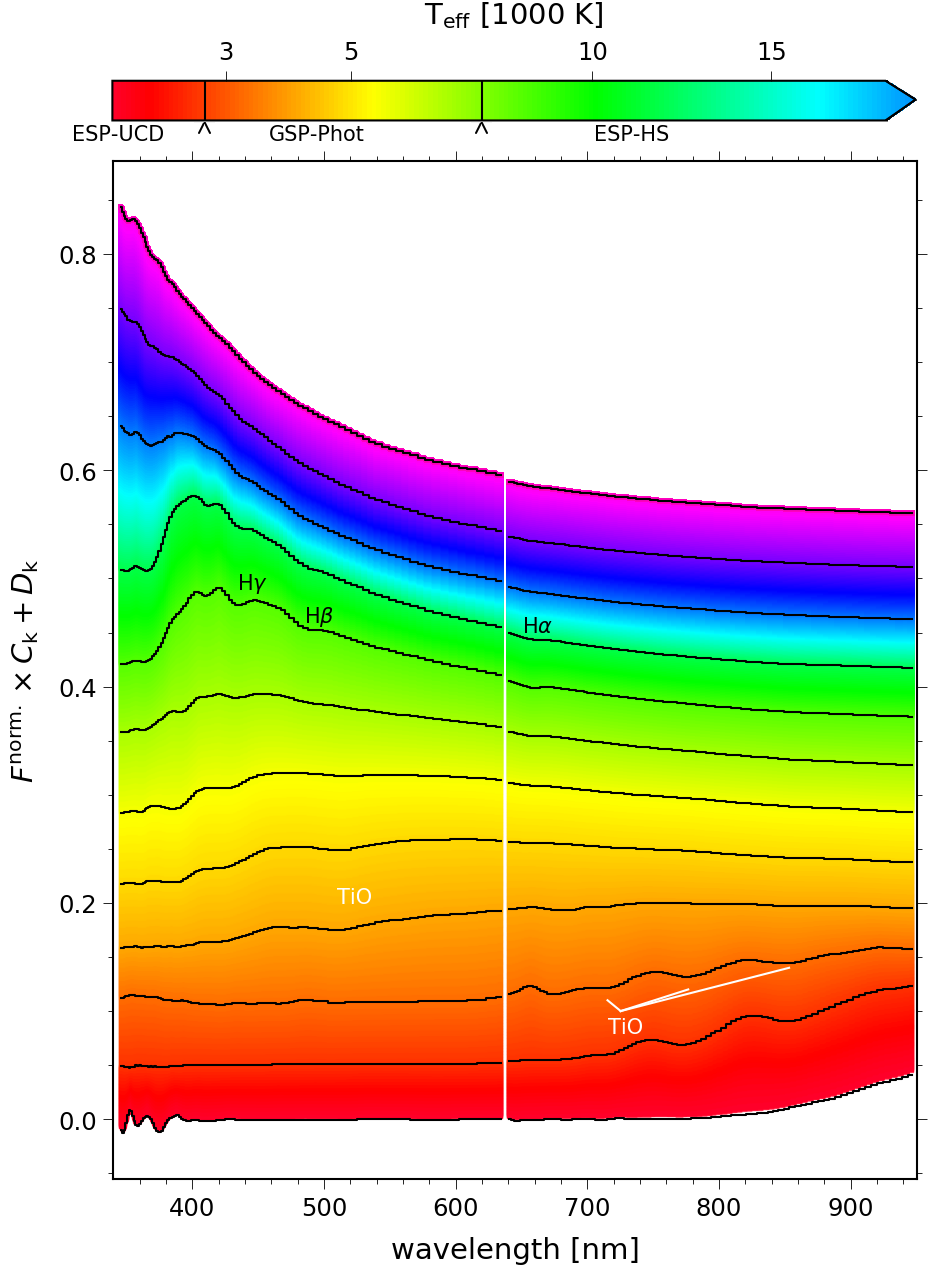}
        \caption{
            Variations of the BP and RP spectra of main-sequence stars with effective temperature. The background color-coding follows the effective temperature scale provided by \espucd, \gspphot, and \esphs (also indicating their optimal \teff\ performance regimes from our validation). We highlighted some spectra for reference and labelled some spectral features. We normalized the spectra to their integrated flux after correcting the BP/RP by the instrument response \citep[see][]{EDR3-DPACP-120}.
            We further stretched and vertically shifted the resulting normalized flux ($F^\mathrm{norm.}$). We restricted our selection to comparable dwarfs:
            \gspphot\ stars with 4 $\leq$ \logg\ $<$ 4.5\,dex, A$_\mathrm{0} < $0.2\,mag and $-$0.1 $\leq$ [Fe/H] $\leq$ $+$0.1\,dex, and \esphs\ stars with 4 $\leq$ \logg $<$ 4.5\,dex, A$_\mathrm{0} < $0.2\,mag.
            (see discussion in Sect.~\ref{sec:BPRPexamples}.)
            \label{fig:seds_teff_modules}
        }
    \end{center}
\end{figure}

\begin{figure}
    \begin{center}
        \includegraphics[width=0.49\textwidth]{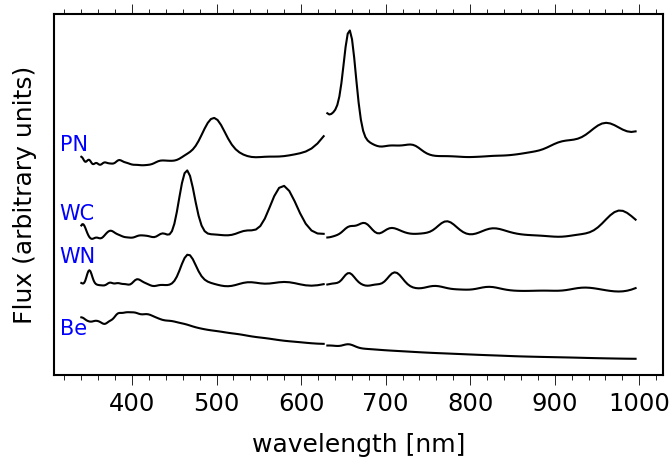}
        \caption{Emission-line features in the BP/RP spectra of various types of objects: planetary nebula (PN), Wolf-Rayet Carbon-rich (WC) and nitrogen-rich (WN) stars, and Be stars. We divided the spectrum flux by the instrument filter responses. We added offsets between the SEDs to place them in the same figure.
        \label{fig:seds_els}
        (see discussion in Sect.~\ref{sec:BPRPexamples}.)
        }
    \end{center}
\end{figure}

The BP and RP spectra reside at the boundary between photometry and spectroscopy.
Due to the low effective spectral resolution from the prisms, these data present only a few noticeable features, as opposed to individual spectral lines in spectroscopy.
On the other hand, where spectroscopy often provides uncertain determination of a stellar continuum, the BP/RP data provide robust determinations with high signal-to-noise ratios similar to photometric measurements.
To illustrate further, Fig.\ref{fig:seds_teff_modules} shows how the spectra of dwarf stars vary with the effective temperature. On this figure, we divided the spectrum fluxes by the instrument filter responses as provided by the  simulation tool internally available to DPAC \citep{EDR3-DPACP-120}. \textit{GaiaXPy} provides the community with a similar tool\footnote{\textit{GaiaXPy}: \url{https://www.cosmos.esa.int/web/gaia/gaiaxpy}}.
Ultra-cool stars mainly emit photons in the RP passband, and their spectra depict strong molecular features. The almost featureless A-, B-, and O-type stars exhibit the Balmer hydrogen lines and jump. And in between, we have the F-, G-, K-, M-type stars characterized by the appearance of TiO bands and metal line blends.
Figure~4 from \citet{DR3-DPACP-157} compares the variation of the BP and RP spectra with effective temperature and extinction using simulations and observational examples.
Based on these data,  we also classify emission-line stars (ELS) by their stellar class by measuring the H$\alpha$ line strength and identifying significant emissions in other wavelength domains.
In Fig.\,\ref{fig:seds_els} we plot the spectral energy distribution (SED) of some of the stellar classes that the \espels\ module estimated.
While one can usually find the strongest features in some planetary nebula and Wolf-Rayet stars, weaker H$\alpha$ emission is more challenging to measure due to the low resolving power of BP and RP spectra.\footnote{The effective resolution of BP and RP spectra decreases towards the red wavelengths, the RP response steeply drops on the blue edge at  $640$\,nm}
The difficulty increases further for the cool ELS stars ($\teff \leq 5\,000$\,K), which spectra show mainly a weak H$\alpha$ emission blended into the local pseudo-continuum shaped by the TiO molecular bands.
Combining the BP and RP data with higher resolution spectra (e.g., RVS, LAMOST, APOGEE) will become an obvious path of choice for the next decades.

\subsection{Typical RVS spectra}\label{sec:RVSexamples}

The RVS spectra share a lot of similarities with RAVE. The RVS have a slightly shorter wavelength window but a higher resolution ($\sim$11\,500): from 846 to 870\,\nm with a resolution element of 0.001\,\nm.

\begin{figure}
    \centering
    \includegraphics[width=0.4\textwidth]{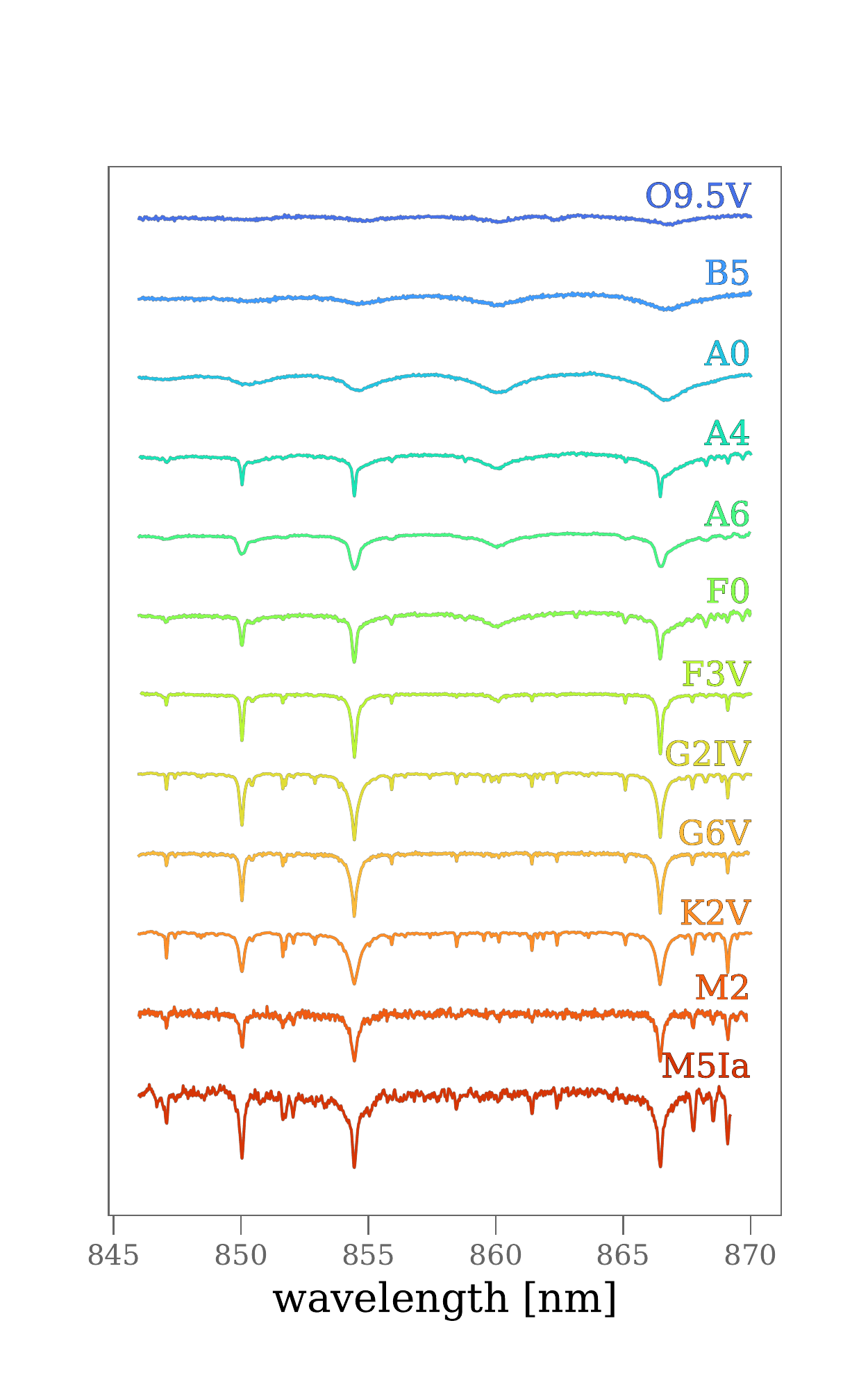}
    \caption{Typical RVS spectra in the OBAFGKM sequence published with the \gaia\ DR3 release. The vertical axis is in arbitrary units for the comparison. The luminosity class V indicates main-sequence stars, and the M5Ia source is a supergiant.
    }
    \label{fig:rvs_examples}
\end{figure}

Figure~\ref{fig:rvs_examples} presents a selection of \gdr{3} typical RVS spectra in the OBAFGKM sequence, a sequence from the hottest (O-type) to the coolest (M-type). Each letter class subdivides itself using numbers with 0 being hottest and 9 being coolest (e.g., A0, A4, A9, and F0 from hotter to cooler). We selected these spectra from their spectroscopic temperatures and surface gravities.

The variations of the RVS spectra with the effective temperature are strong and the spectra of F-, G-, and K-type stars present many atomic lines, but their reliable measurements depends strongly on the temperature and gravity of the star. The \gaia\ \href{https://www.cosmos.esa.int/web/gaia/iow_20210709}{Image of the Week 2021-07-09}
presents an animation of several Gaia RVS stellar spectra and their element abundances. This figure also illustrates the challenge of characterizing O-type stars which present nearly featureless RVS observations.

\section{AP content description and performance}\label{sec:APcontent}

This section describes the AP content of \gdr{3}, their performance, and limitations. We first discuss the object APs individually: their distances in Sect. \ref{sec:distances}, their stellar atmospheric parameters in Sect.~\ref{sec:atmosphere} (i.e., \teff, \logg, metallicity, individual abundances, rotation, and activity), and their evolution parameters in Sect.~\ref{sec:evolution} (i.e., absolute and bolometric luminosities, radius, gravitational redshift, mass, age and evolution stage). These require us to account for dust effects along the line-of-sight summarized in Sect.~\ref{sec:dust} and analyzed in-depth in \citet{DR3-DPACP-158} and \citet{DR3-DPACP-144}. In Sect.~\ref{sec:groups}, we further assess the quality of our APs by focusing on objects in groups (i.e., clusters and binaries). Finally, we discuss the detection of peculiar cases and outliers in Sect.~\ref{sec:weird}.

We emphasize that to avoid repetitions, we summarize only the complete description of some internal precisions of the APs as a function of magnitudes, colors, sky position, and other parameters that appear in other publications (e.g. \citealt{DR3-DPACP-156}, \citealt{DR3-DPACP-157}; \citealt{DR3-DPACP-158}; \citealt{DR3-DPACP-186}; \citealt{DR3-DPACP-175}).

To guide the reader, Appendix \ref{sec:apfields} compiles the various estimates of stellar parameters from \gdr{3} cast into the mentioned categories (corresponding to the following subsections). The compilation indicates which \apsis\ module produces them, and which table and fields store the values in the Gaia catalog. We emphasize that the field names correspond to the catalog in the Gaia Archive but names may differ when using partner data centers.

\subsection{Distances}\label{sec:distances}

Two \apsis\ modules provide distance estimates: \gspphot\ for single stars and \msc\ for unresolved binary stars. Both modules analyze the BP and RP spectra with the Gaia parallaxes to derive distance estimates simultaneously with other astrophysical parameters. We listed the catalog fields related to both modules' distance estimates in Table~\ref{tab:product-module-distance}.

\begin{figure}
    \begin{center}
        \includegraphics[width=0.49\textwidth, angle=0]{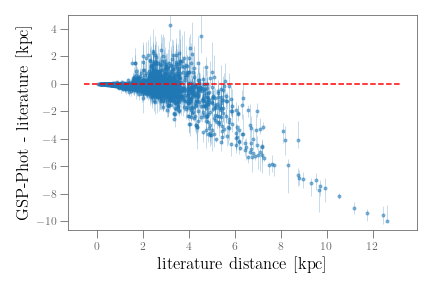}
        \caption{Comparison of \gspphot\ distances in star clusters against literature values. For $2\,019$ star clusters (nearly $200\,000$ stars) spanning distances up to 13\,kpc, we indicate on the y-axis the median offset (dot), and the 50\% quantile (line) of all member star distances w.r.t to cluster distances taken from \citet[][]{2020A&A...640A...1C}. Above $\sim 2.5$\,kpc ($\sim 15$\% of the cluster members), \gspphot\ systematically underestimates distances. (see discussion Sect.~\ref{sec:distances}.)
            \label{fig:GSPPhot-distances-clusters}
        }
    \end{center}
\end{figure}

\begin{figure}
    \begin{center}
        \includegraphics[width=0.49\textwidth, angle=0]{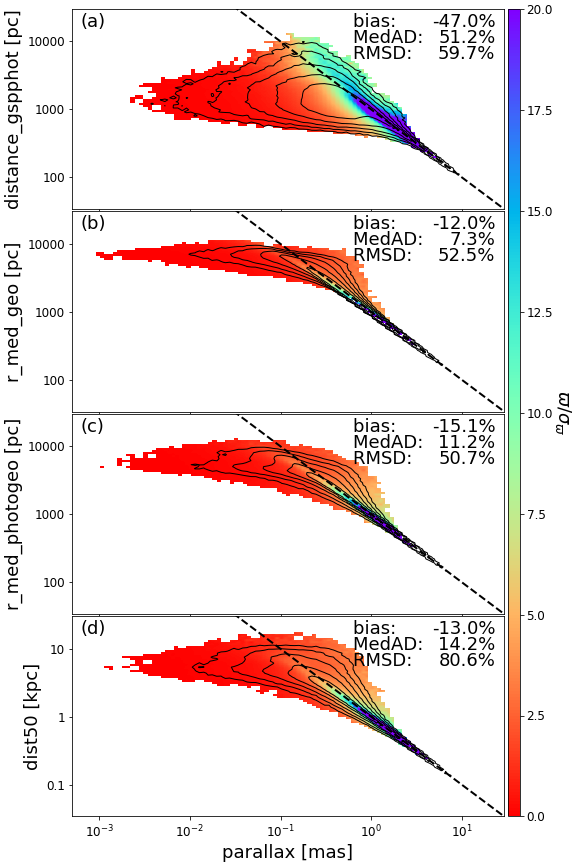}
        \caption{Comparison of distances with parallaxes for a random subset of one million stars, color-coded by parallax signal-to-noise ratio. The black dashed line indicates the expected inverse parallax-distance relation, $\varpi={1}/{d}$. Black contours indicate densities dropping relatively by factors of $3$. From top to bottom, the panels show the \gspphot\ distances from \linktoparampath{gaia_source}{distance_gspphot}, the geometric distances from \citet{BailerJones2021}, the photogeometric distances from \citet{BailerJones2021}, and the distances from \citet{2022A&A...658A..91A}, respectively.
        \label{fig:apsis-GSPPhot-distances-vs-parallax}
        }
    \end{center}
\end{figure}

\begin{figure}
    \centering
    \includegraphics[width=0.49\textwidth]{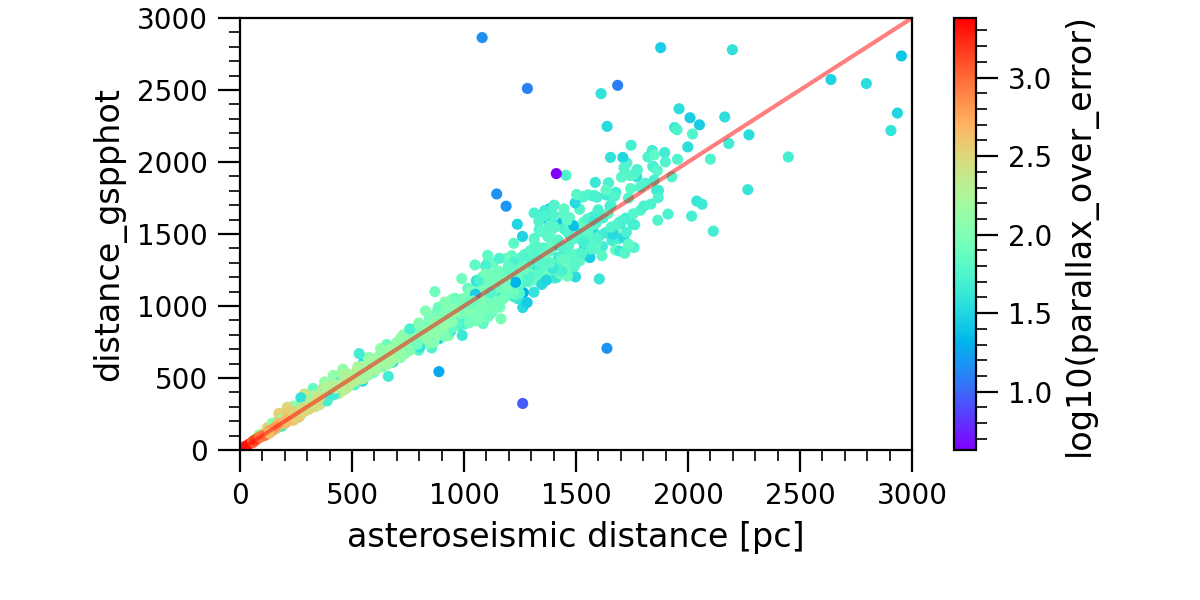}
    \includegraphics[width=0.49\textwidth]{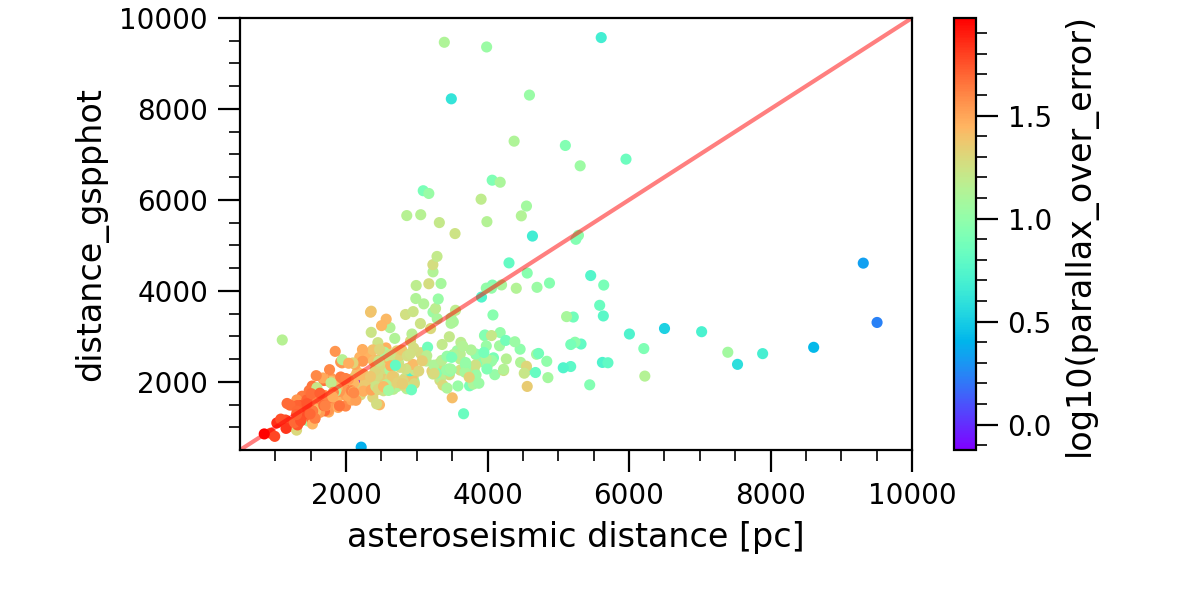}
    \caption{Comparison of \gspphot\ distances (\linktoparampath{gaia_source}{distance_gspphot}) with asteroseismic ones for 2\,236 and 606 stars from \citet{huber2017} (top) and \citet{anders2017} (lower). The continuous red line indicates the bisector for reference in both plots.  We obtain distances in good agreement up to 2 kpc, and some outliers beyond.}
    \label{fig:asteroseismic_distances_vs_gspphot}
\end{figure}

For \gspphot, the distances are reliable out to $\sim$2 kpc. Beyond 2 kpc, \gspphot\ does systematically underestimate distance, as is evident, e.g., from star clusters. Fig.~\ref{fig:GSPPhot-distances-clusters} compares the median \gspphot\ distances of stellar members for each cluster with their literature values by \citet{2020A&A...640A...1C} derived using \gdr{2} data through maximum likelihood. We included the \gdr{3} variable zero point on parallaxes mentioned in Sect. \ref{sec:inputdata}.
We obtain similar results when comparing to the photometric distances by \citet{2013A&A...558A..53K} and in BOCCE \citep{2006AJ....131.1544B,2018A&A...618A..93C} catalogs based on color-magnitude diagram fitting.
However, when the parallax measurement is good (about $\varpi/\sigma_\varpi>10$), the \gspphot distances remain reliable even out to $10$\,kpc, as we show in Fig.~\ref{fig:apsis-GSPPhot-distances-vs-parallax}a.
The reason for this systematic underestimation of distances by \gspphot\ is an overly harsh distance prior. \citet{DR3-DPACP-156} discussed the prior and showed that we could resolve this issue by updating its definition. A prior optimization remains necessary and will be part of further releases.
Figure~\ref{fig:apsis-GSPPhot-distances-vs-parallax} compares also the distances from \citet{BailerJones2021} and \citet{2022A&A...658A..91A} to the \gdr{3} parallaxes and we note that they perform better than \gspphot\ distances.\footnote{\gspphot\ derived distances from parallaxes with the zero-point correction from Sect.~\ref{sec:inputdata}, but Fig.~\ref{fig:apsis-GSPPhot-distances-vs-parallax} uses uncorrected values on the x-axis. Yet, the zero-point alone is unlikely to generate these differences.} For this reason, various DR3 publications chose to not use the \gspphot distances but rather EDR3 distances from \citet{BailerJones2021} \citep[e.g.,][]{DR3-DPACP-75, DR3-DPACP-104, DR3-DPACP-144}.
A further comparison of \gspphot\ distances with those from asteroseismic analyses confirmed a good agreement to 2\,\kpc, and some outliers beyond (see Fig.~\ref{fig:asteroseismic_distances_vs_gspphot}).

\begin{figure}
    \centering
    \includegraphics[width=0.49\textwidth]{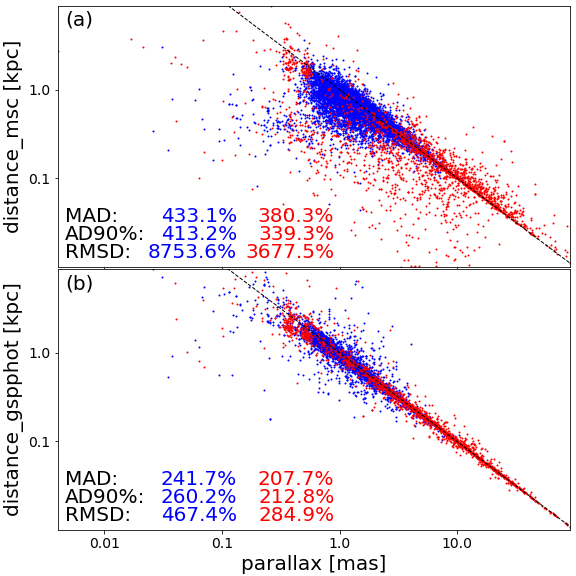}
    \caption{Comparison of Gaia parallaxes with distance estimates from \msc\ (panel a; \linktoparampath{astrophysical_parameters}{distance_msc}) and \gspphot\ (panel b; \linktoparampath{gaia_source}{distance_gspphot}) for 2253 known spectroscopic binaries from \citet{2004A&A...424..727P} (red points) and 10\,407 from \citet{2020A&A...638A.145T} (blue points). We quote the mean absolute differences (MAD), the absolute difference at 90\% confidence (AD90\%) and the RMS differences (RMSD) to the measured parallax for both samples in both panels. Both panels show the same set of stars with estimates from both modules. The anti-diagonal dashed line highlights the parallax as inverse distance.
    }
    \label{fig:MSC-GSPPhot-distances-vs-parallax}
\end{figure}

\msc\ provides distance estimates assuming sources are unresolved binaries with luminosity ratios ranging from $5$ to $1$. At best, \msc's distance estimates would differ from \gspphot's estimates (equivalent to infinite luminosity ratio) by a factor $10$ to $50$\%, respectively. We highlight that distances with luminosity ratios of $5$ significantly differ from single-star assumptions.
Figure~\ref{fig:MSC-GSPPhot-distances-vs-parallax} compares \msc's distance estimates and those from \gspphot\ to the Gaia parallaxes for the spectroscopic binary samples from \citet{2004A&A...424..727P} (mostly $G<10$\,mag) and \citet{2020A&A...638A.145T} (mostly between $G=10$ and $15$\,mag). Overall there is a qualitative agreement between the distances from both modules and the measured parallaxes.  However, \gspphot\ distances exhibit a significantly tighter agreement with the parallaxes than the ones from \msc, despite the single star assumption: their mean absolute differences are only half of those for \msc\ and the RMS differences are more than ten times smaller. However, the RMS differences are dominated by a handful of outliers, whereas the absolute difference at 90\% confidence is more robust yet still much higher for \msc\ than \gspphot.
One source of this mismatch likely comes from the differences in exploiting the BP and RP spectra information: while both \msc\ and \gspphot\ make use of the parallax and the apparent $G$ magnitude, \msc\ normalizes the spectra, whereas \gspphot\ keeps their calibrated amplitudes in their spectra likelihoods \citep[see][for further details]{DR3-DPACP-156}.

Furthermore, interpreting the difference between the two sets of estimates is more complex in practice. Modules adjust their AP sets altogether to fit the observed BP and RP spectra. We emphasize that \msc's double-star assumption allows for more free fit parameters than \gspphot's single-star assumption (8 and 5, respectively). The increased number of fit parameters is likely a source of the more significant dispersion in the \msc\ estimates. We discuss the other APs from \msc\ in Sect.~\ref{sec:unresolved_binaries}.

\subsection{Atmospheric APs}\label{sec:atmosphere}

The atmospheres of stars produce the photons that Gaia collects. Through these photons, we can infer the physical conditions of these layers, which relate to the fundamental stellar parameters. In this section, we characterize the \gdr{3} APs that describe the atmospheric state of the observed stars. We loosely split the APs into three groups: first, the basic static (equilibrium) state of an atmosphere defined by \teff, \logg, metallicity, \mh, and $\alpha$-abundance, \afe\footnote{$\alpha$-elements with respect to iron (\afe), refer to O, Ne, Mg, Si, S, Ar, Ca, and Ti and are considered to vary in lockstep.}; then the dynamic  (departure from equilibrium) state given by the stellar classes, rotation, line emissions, magnetic activity, and mass loss or accretion; and finally the chemical abundances.

\begin{figure}
    \centering
    \includegraphics[width=\columnwidth]{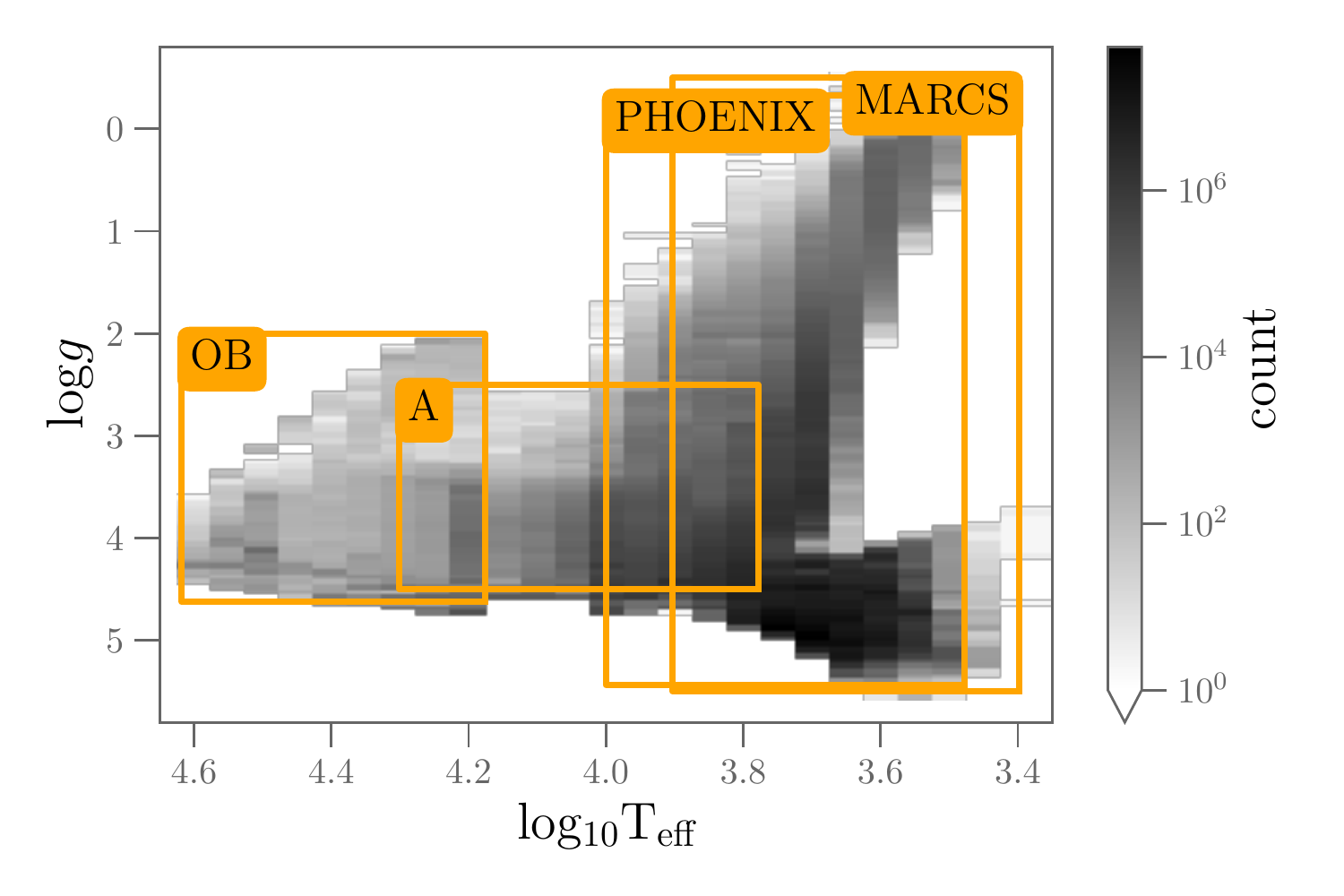}\\
    \caption{Parameter space in the Kiel diagram spanned by the stellar atmosphere libraries used by \gspphot. Boxes indicate the spans of the libraries producing independent estimates.
    The density distribution represents the content from \linktotable{gaiadr3.gaia_source}, which contains only one set of APs per source using the (statistically) ``best'' library (\linktoparam{gaia_source}{libname_gspphot} field) for that one source.}
    \label{fig:gspphot_kiel_libraries}
\end{figure}

The Gaia data set is primarily magnitude-limited and does not select objects on any specific color or class of stars. Consequently, the atmospheric parameters span a great variety of spectral types, from O to M, and even some L-type stars, some of which require target-specific treatment (partly handled by the ESP-modules in \apsis).
Depending on the star (spectral and luminosity) class, we used either empirical or theoretical atmospheric models to estimate the atmospheric parameters of the stars, and sometimes both. The theoretical models try to model the relevant physical processes of the matter-light interaction in stellar atmospheres, while the empirical ones capture some hard-to-model observational effects. The overlap between models and application ranges of \apsis\ modules allows us to check for consistency or the lack thereof (see overlaps in Figs.~\ref{fig:apsis_kiel_modules} and \ref{fig:gspphot_kiel_libraries}).

\subsubsection{Primary atmospheric parameters: \teff, \logg, \mh\ and \afe}\label{sec:atmosphere-primary}

Below we summarized our validation results for the parameters \teff, \logg, \mh, and \afe, which various modules of \apsis\ estimate (see Table~\ref{tab:product-module-atm-1}). We first focus on the FGK-type stars as these constitute the majority of stars in the Gaia data set. Mainly \gspphot\ and \gspspec\ overlap on this stellar-type interval. We emphasize that the application range of the \apsis\ modules varies significantly. To help the reader, we thus organize the description per module.
One way to validate the Gaia-based APs and simultaneously quantify their precision is to compare them with large stellar surveys in the literature. The numbers below serve as a guideline for the global precision of the \gdr{3} results relative to literature works.
Accuracy is harder to quantify globally, but we can assess it in some specific cases, for instance, relative to Gaia benchmark stars \citep[e.g.,][]{Heiter2015} and spectroscopic solar analogs \citep[e.g.,][]{TucciMaia2016}.

\paragraph{\gspphot.} Analyzing BP/RP spectra, \gspphot\ provides multiple sets of APs, one for each of the four supporting theoretical atmospheric libraries: MARCS \citep{2008AA...486..951G}, PHOENIX \citep{2005ESASP.576..565B}, A \citep{2004A&A...428..993S}, and OB \citep{2003ApJS..146..417L, 2007ApJS..169...83L}. Figure~\ref{fig:gspphot_kiel_libraries} shows their parameter space. \gspphot\ analyzes the BP/RP spectra with a Markov-Chain Monte-Carlo approach (MCMC), which also characterizes the uncertainties \citep[method in][]{DR3-DPACP-156}. The reported estimates and uncertainties correspond to the 50th (median) and 16th and 84th percentiles of the (marginalized) MCMC samples, respectively. We also publish the MCMC chains with the catalog through the DataLink protocol \citep{2015ivoa.spec.0617D} implemented by the Gaia Archive.
We compared our APs to those reported in the APOGEE \citep{Abdurrouf2021}, Gaia-ESO \citep{Gilmore2012, 2022GES...WP13}, GALAH \citep{Buder2021}, LAMOST \citep{2011RAA....11..924W, 2014IAUS..306..340W}, and RAVE \citep{Steinmetz2020} catalogs. We characterized the \gspphot\ results by a median absolute error in \teff\ of 119\,K, and a mean absolute error of 180\,K across the various mentioned datasets (details in \citealt{DR3-DPACP-156}). The difference between the two statistics translates the complexity of the distributions. The variations of \logg\ and \mh\ affect the BP and RP spectra only weakly in constrast with the temperature. Although \gspphot\ analyzes the BP/RP spectra with the \gaia\ parallax information and isochrone models, it is this combination that allows \gspphot to determine \logg\ and \mh\ estimates. In \citet{DR3-DPACP-156}, we compared to seismic \logg\ values of solar-like oscillators from \citet{Serenelli2017} and \citet{Yu2018} and we find a median absolute error of $0.2$\,dex for \logg. For \mh, we find that \gspphot\ estimates are typically too low by $0.2$\,dex and exhibit additional systematics. We thus caution against using \mh\ estimates without further investigation. Yet, we find that \mh still encodes some useful information about metallicity, for instance, with empirical calibrations (See \citealt[Sect.~3.5.3]{DR3-DPACP-156} for details).
The comparison of \mh\ to [Fe/H] from APOGEE DR17 \citep{Abdurrouf2021} gives a median absolute deviation of $0.2$\,dex, with globally no offset for stars with $\logg > 2.5$, based on more than 400\,000 FGK stars in common \citep[see][their Fig. 10]{DR3-DPACP-156}.
We assessed the typical precision of \mh\ by measuring the dispersion among FGK members in 187 open clusters with known metallicities, from -0.50 dex to +0.43 dex \citep[see][their Fig. 11]{DR3-DPACP-156}. The residuals show an explicit dependency on the parallax SNR (see Figs.~9 of \citealt{DR3-DPACP-156}). The median absolute deviation is 0.2 dex for $\gmag < 16$\,mag, while for fainter stars, the \gspphot\ metallicities appear to be underestimated by 0.6 dex (median offset) with a dispersion reaching 0.5 dex.
As one can expect, the \gspphot\ performance significantly varies from star to star and depends on the stellar physical conditions encoded in the spectral type, luminosity class, and \mh, for instance (see Fig.~11 from \citealt{DR3-DPACP-156}). For metal-poor stars with  $\mh\,\leq\,-1.5$\,dex, the metallicity sensitivity of the BP and RP spectra diminishes drastically, which leads \gspphot\ to overestimate \mh. This loss of sensitivity is typical of optical photometric metallicity indicators, which is one of the reasons behind dedicated passband designs \citep[e.g.,][]{2010A&A...523A..48J, Starkenburg2017, LopezSanjuan2021} and spectral indices \citep[e.g.,][]{Johansson2010}. \citet{DR3-DPACP-156} interpret this as a consequence of \mh\ having the weakest impact on BP and RP spectra and thus being the parameter that is easiest to compromise.

\begin{figure*}
    \centering
    \includegraphics[width=1.98\columnwidth]{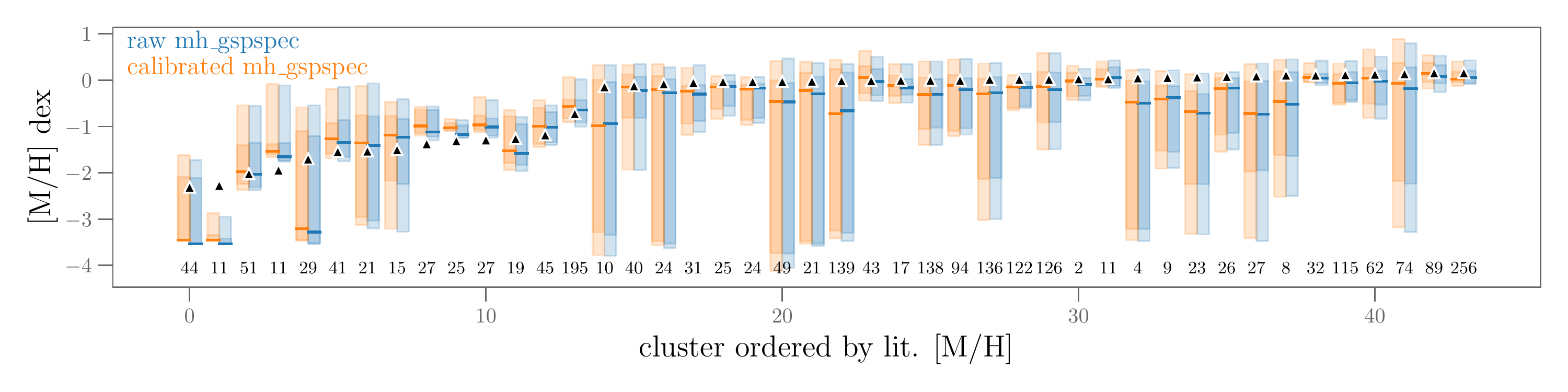}\\
    \caption{\mh\ abundance distributions of member stars per clusters for \gspspec Matisse-Gauguin algorithm (\linktoparampath{astrophysical_parameters}{mh_gspspec}) before and after recommended adjustements (see Sect.~\ref{sec:atmosphere-primary})
    The clusters are ordered by ascending [M/H] literature values (the lowest values to the left).
    These literature values are given by the black triangles. \citet{2020A&A...640A...1C} provided the cluster members.  Marks indicate the median of the distributions, and shaded regions indicate the $68\%$ and $98\%$ intervals.
    Numbers at the bottom indicate how many estimates were available for representing the distribution.
    Ideally, all predictions are within a small interval, which agrees with the triangles.
    We did not filter the estimates using the flags to keep enough stars per cluster, but nevertheless, the agreement is remarkable.}
    \label{fig:gspspec-cluster-mh}
\end{figure*}

\paragraph{\gspspec.} Analyzing RVS spectra with primarily SNR > 20 (i.e, $\gmag \lesssim 16$\,mag), \gspspec\ estimates the stellar APs using synthetic spectra based on MARCS models and with two different algorithms (``Matisse-Gauguin'' and ``ANN''; see \citealt{Manteiga2010, RecioBlanco2016, DR3-DPACP-186} for details.). Unlike \gspphot, \gspspec\ does not exploit additional information like parallax or photometric measurements.
\gspspec\ estimates uncertainties per star from the ensemble of APs from 50 Monte-Carlo realizations of the spectra: for each, \gspspec\ draws a spectrum from the noise (i.e., spectral flux covariances estimated by \citealt{DR3-DPACP-154}) and derives a set of atmospheric parameters and chemical abundances (see Sect.~\ref{sec:abundances}).
The reported lower and upper confidence values correspond to the 16th and 84th percentiles of the MC results per star, respectively. In addition, we provide quality flags to identify estimates potentially suffering from bad pixels, low signal-to-noise ratio, significant line broadening due, for instance, to stellar rotation ($v\sin i$), poor radial velocity (RV) correction, and grid border effects.
We discuss below the results from the Matisse-Gauguin and ANN algorithms, available in the \linktotable{astrophysical_parameters} table and \linktotable{astrophysical_parameters_supp}, respectively.


We validated and quantified the accuracy of the Matisse-Gauguin parameters for FGK stars against literature data. We selected results with corresponding AP flags equal to zero and compared our estimates with APOGEE DR17 \citep{Abdurrouf2021}, GALAH-DR3 \citep[]{Buder2021} and RAVE-DR6 \citep{Steinmetz2020}.
We find with a comparison with APOGEE-DR17 a median offsets and MAD of ($-32$; $58$)\,K, ($-0.32, 0.12$)\,dex and ($+0.04,0.08$)\,dex, for $\teff$, \logg\ and \mh. The spectra from RAVE and RVS share very similar wavelength coverage, which led \citet{DR3-DPACP-186} to extensively compare the \gspspec\ performance against those stellar parameters. We find similar statistics when comparing with the other catalogs \citep[see details in][, esp. their Fig.~11]{DR3-DPACP-186}.

However, we found for giant stars a bias in the \logg\ and smaller biases in the \mh\ and \afe\ values from Matisse-Gauguin. \citet{DR3-DPACP-186} provide corrective prescriptions in the form of a polynomial function of \logg and suggests an analogous (\logg-dependent) correction for \mh and \afe\ to reduce this issue between dwarfs and giants.  We calibrated the \logg\ and \mh corrections on the AP values from APOGEE-DR17, GALAH-DR3, and RAVE-DR6 simultaneously. Applying those corrections leads to \logg\ and \mh median offsets and MAD of APOGEE-DR17 to $(-0.005; 0.15)$\,dex and $(0.06; 0.12)$\,dex, respectively.
However, we calibrated the $\afe$ correction on a sample of solar-like stars (in terms of metallicity, galactocentric position, and velocity, see Sect.~\ref{sec:abundances}). This correction reconciles dwarf and giant on the same \afe\ scale. We found in the solar-like sample on average $\afe=0$ for all stellar types after calibration.
We further assessed the typical precision of the non-corrected \mh\ by measuring the dispersion in stellar clusters of known metallicity, similarly to what we did for \gspphot. Figure~\ref{fig:gspspec-cluster-mh} compares the dispersion of the \mh\ abundance distributions of member stars per clusters for \gspspec Matisse-Gauguin algorithm before and after the recommended adjustments. Even though the corrections did not affect the overall agreement, we note that we did not apply filters based on the associated flags.
We further restricted ourselves to the FGK members in $162$ open clusters of \citet{2020A&A...640A...1C}, and we found an average MAD of $0.11$\,dex per cluster. We noted a larger dispersion and a negative offset ($-0.12$\,dex) for dwarfs. For $64$ globular clusters ($\mh \leq -0.50$\,dex), the typical dispersion per cluster is $0.20$\,dex with an median offset of $+0.12$\,dex. However, these statistics describe the data regardless of the quality flags. If we require unset \mh\ flag bit zero \citep[see details in][]{DR3-DPACP-186}, the metallicities agree better with the literature, with absolute offsets values lower than $0.10$\,dex, and with typical dispersions of $0.075$\,dex for open clusters and $0.05$\,dex for globular clusters. Note, however, that the filtering also reduces the number of stars significantly, leaving us with 40\% of the $2\,271$ members of open clusters and only 4\% of $1\,224$ members in globular clusters. These sources are primarily removed for low-SNR spectra mostly due to GCs being far away. These settings also remove fast rotators, hot stars, and some K-, and M-giants in OCs. Finally, stars nearby the model grid borders, predominantly hot dwarfs, and cool giants in the case of the OC and GC, respectively. However from this test, we should not conclude to a metallicity dependent performance as metal-poor stars are rare and predominantly known in GCs.

\begin{figure*}
    \centering
    \includegraphics[width=1.98\columnwidth]{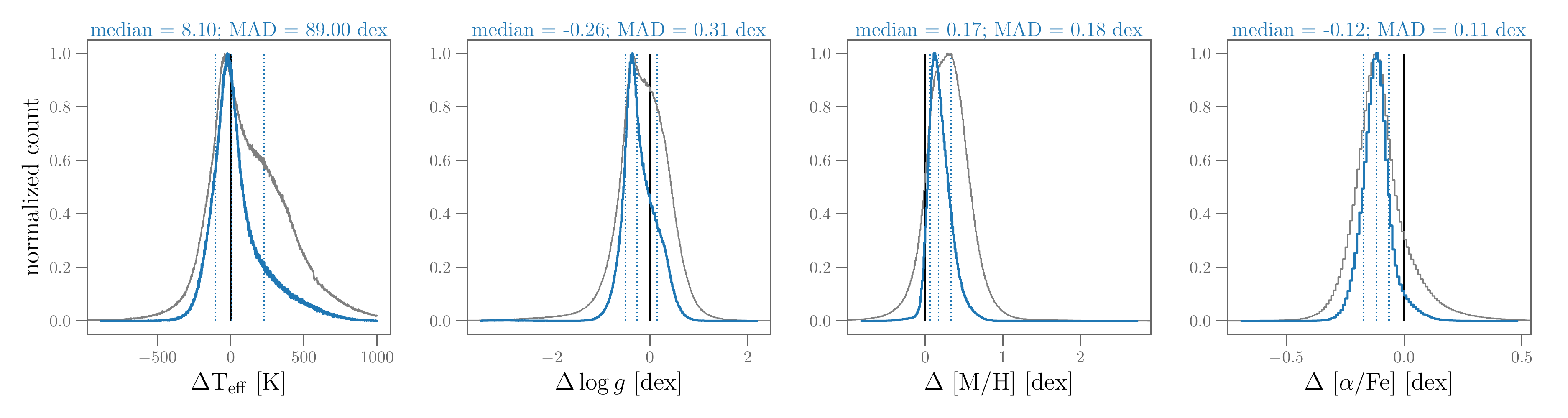}
    \caption{Comparison of the APs from \gspspec-Matisse Gauguin and ANN for the \gdr{3} sample with the first thirteen and eight values in \linktoparampath{astrophysical_parameters}{flags_gspspec}) and (\linktoparampath{astrophysical_parameters_supp}{flags_gspspec_ann} equal to zero. This represents $1\,084\,427$ stars in the \gdr{3} catalog. For reference, we also indicated the distribution without the flag filtering in gray.}
    \label{fig:gspspec-algo-deltas}
\end{figure*}


The artificial neural networks algorithm (ANN) in \gspspec ANN provides a different parametrization of the RVS spectra, independent from the Matisse-Gauguin approach. In contrast with Matisse-Gauguin's forward modeling approach, ANN projects the RVS spectra onto the AP label space. We trained the network on the same grid of synthetic spectra as the Matisse-Gauguin algorithm, in this case adding noise according to different signal-to-noise scales in the observed spectra \citep{Manteiga2010}.
ANN's internal errors are of the order of a fraction of the model-grid resolution and show no significant bias, confirming the ANN projections' consistency of the synthetic spectra grid.
In \citet{DR3-DPACP-186}, we compared the ANN results with the literature values and found similar biases to those of Matisse. Equivalently, we provide also calibration relations for \teff, \logg, \mh\ and \afe\ to correct these biases.

Figure~\ref{fig:gspspec-algo-deltas} compares the APs from both algorithms of \gspspec on a sample of $1\,084\,427$ in \gdr{3} with respective estimates. We also restricted this comparison to the good flag status: the first thirteen and eight values in (\linktoparampath{astrophysical_parameters}{flags_gspspec}) and (\linktoparampath{astrophysical_parameters_supp}{flags_gspspec_ann}) equals to zero.
Overall, the algorithms agree with each other. Once we apply the calibration relations to both algorithms estimates, we found for spectra with SNR $\ge{150}$ deviations with median values of  $-94$\,K,  $-0.05$\,dex , $0.1$\,dex, and $0.04$\,dex for \teff, \logg, \mh\ and \afe, respectively. For the sample sample, we found MAD values of $93$\,K, $0.11$\,dex, $0.10$\,dex, and $0.05$\,dex, respectively.


\paragraph{\gspphot\ and \gspspec\ overlaps.}

\begin{figure}
    \centering
    \includegraphics[width=0.9\columnwidth, clip, trim= 11.5cm 0 0 0]{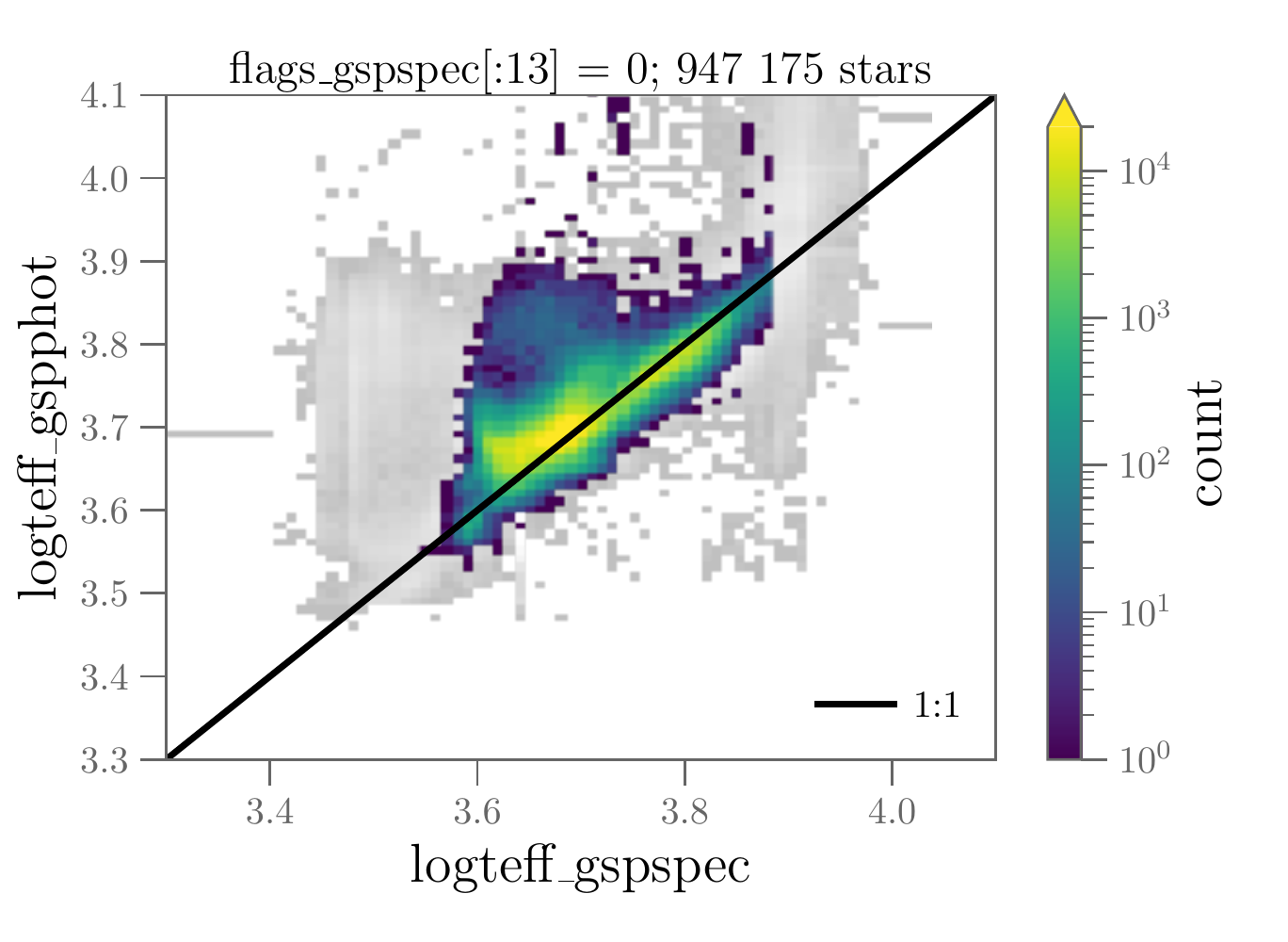}\\
    \includegraphics[width=0.9\columnwidth, clip, trim= 11.5cm 0 0 0]{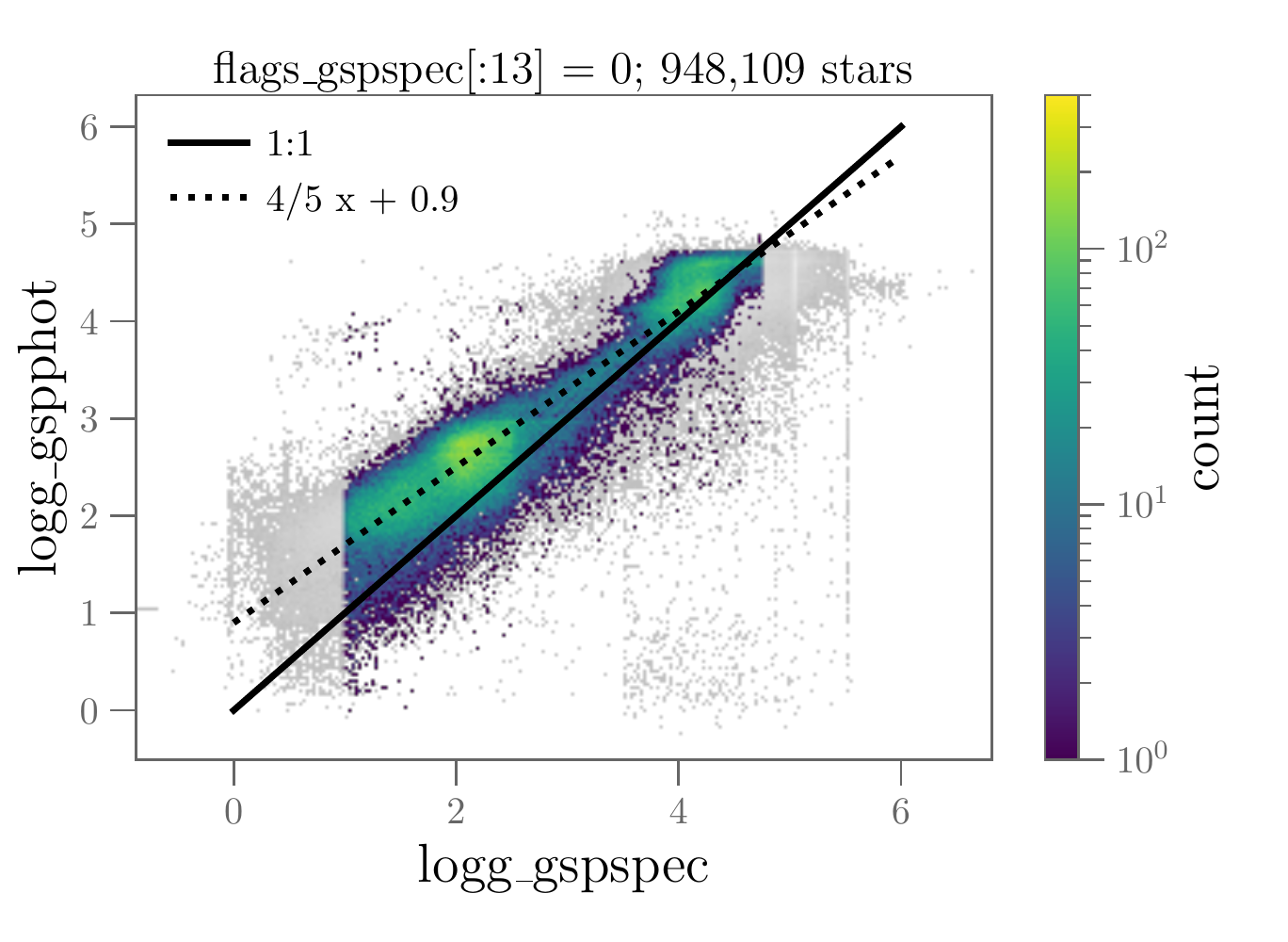}
    \caption{Comparison of the temperatures (top) and surface gravity (bottom) estimates from \gspphot\ and \gspspec. The \gspphot\ and \gspspec\ algorithms are calibrated to the literature values. We plotted in gray the $\sim 3.2$ million sources in \gdr{3} with both \linktoparampath{astrophysical_parameters}{teff_gspphot}, \linktoparam{astrophysical_parameters}{teff_gspspec}  and \linktoparampath{astrophysical_parameters}{logg_gspphot}, \linktoparam{astrophysical_parameters}{logg_gspspec}. The highlighted distribution corresponds to those with the first thirteen values in \linktoparam{astrophysical_parameters}{flags_gspspec} equals to zero ($\sim 1$ million sources). We indicated the identity lines and the identified divergence in \logg\ between the modules. We note that the \gspspec\ recommended calibration of \logg\ does not affect significantly this comparison.}
    \label{fig:gspphot-gspspec-aps1}
\end{figure}

Figure~\ref{fig:gspphot-gspspec-aps1} compares the temperatures and gravity estimates from \gspphot\ and \gspspec. The \teff\ estimates strongly agree overall but some outliers remain visible on the plot, most likely originating from \gspphot\ sensitivity to low-quality parallaxes. In particular, we traced back to variable stars the plume at $\log_{10}$\linktoparam{astrophysical_parameters}{teff_gspphot} $\sim 3.8$ (see \citealt{DR3-DPACP-156} for details). On this sample, we found a median offset of $98$\,K, an MAD of 246\,K.
It is very apparent that the \logg\ estimates systematically differ strongly between the modules. The recalibration prescription from \citet{DR3-DPACP-186} mitigates the differences, but does not remove them completely. We found a median offset of $0.35$\,dex, and an MAD of $0.34$\,dex. \citet{DR3-DPACP-186} identified a similar trend in the \gspspec\ \logg values when comparing to those of the literature (see their Fig.~10).

Solar analog stars are stars closest to the Sun in temperature, gravity, and metallicity. We selected 200+ spectroscopic solar analogs from the literature (mostly from \citealt{Datson2015} and \citealt{TucciMaia2016}) with relative \teff\ within $\pm100$\,K, \logg and [Fe/H] within $\pm0.1$\,dex to those of the Sun values. We compare the biases and dispersion of the \gspphot\ and \gspspec\ Matisse-Gauguin APs on this sample of stars. We note that solar analogs are dwarf stars, which are little to not affected by the Matisse-Gauguin corrections mentioned above.
We find that \gspphot\ underestimates \teff\ by between $30$\,K (PHOENIX) and $90$\,K (MARCS), with a standard deviation $\sigma \sim 100$\,K in both cases. In contrast, \gspspec\ estimates have essentially no \teff\ bias ($+10$\,K) but slightly larger dispersion ($\sigma \sim  130$\,K).
Irrespective of the atmosphere library (\linktoparam{gaia_source}{libname_gspphot}), \gspphot\ underestimated the \logg\ values by 0.12\,dex, but with a standard deviation of $\sigma \sim 0.14$\,dex they remain statistically compatible with the solar value. \gspspec\ results are as accurate as those from \gspphot\ around the solar locus, but they present a larger dispersion of $0.42$\,dex (calibration of \logg\ does not change this value). We recall that \gspspec\ uses only the RVS spectra as input, while \gspphot\ also uses parallaxes and constraints from isochrones.
\mh\ values are nearly solar for \gspspec\ with an offset of $0.1$\,dex and $\sigma \sim 0.05$ (again, without significant impact of the recommended corrections), but we found larger offsets for \gspphot\ when using PHOENIX ($-0.4 \pm 0.2$\,dex) and MARCS models ($-0.2 \pm 0.2$\,dex).
\cite{DR3-DPACP-156} discussed the systematic and significant discrepancies between APs based on the PHOENIX and MARCS libraries. For solar-like stars, they found substantial differences in the original atmosphere models that are still under investigation at the time of writing this manuscript.

Ideally, \gspphot\ and \gspspec\ would return results in perfect agreement with each other.
In practice, they don't, but rather complement each other. The two modules analyze data with different spectroscopic resolutions and wavelength ranges. To first order, \gspphot\ relies on the stellar continuum over the whole optical range from the BP/RP low-resolution spectra (from $330$ to $680$\,\nm). In contrast, \gspspec\ investigates atomic and molecular lines in the continuum-normalized medium-resolution spectra in the narrow infrared window of RVS (from $846$ to $870$\,nm). Hence the modules analyze different aspects of the light emitted from stars. Additionally, interstellar extinction significantly affects the BP and RP spectra, but RVS data only in the region of the diffuse interstellar band around 860\,nm  \citep[e.g.,][]{DR3-DPACP-144}. Therefore, \gspphot's AP determination significantly depends on determining the amount of extinction correctly, while it has little impact on \gspspec's AP inference (see Sect.~\ref{sec:dust}).

\begin{figure}
    \begin{center}
        \includegraphics[width=1\linewidth, angle=0]{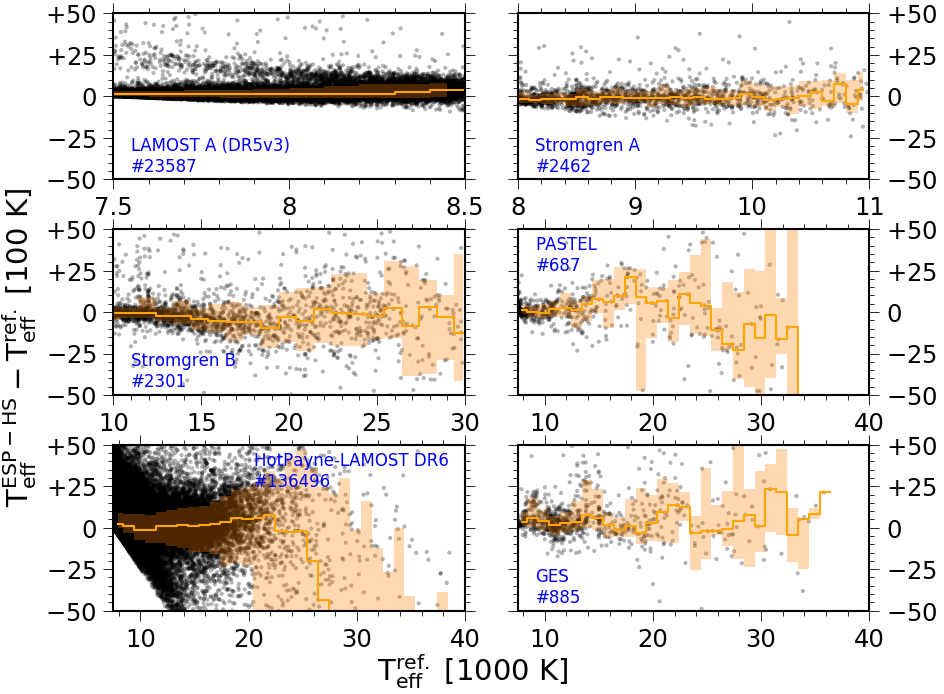}
        \caption{Effective temperature difference between values obtained by \esphs\ and those from literature catalogs as a function of the literature \teff. We indicate the running median with the solid orange lines, and the interquantile dispersions by the orange shaded regions. The blue numbers on each panel indicate how many stars are compared. We emphasize the scales on the x and y axes are scaled by 1\,000 and 100\,K, respectively. Reference estimates are: from the LAMOST (DR5) A-type stars \citep[LAMOST A (DR5v3),][]{Luo2019}, derived from the Stromgren photometry (Stromgren A \& Stromgren B) by adopting the updated calibration of \citet{1993A&A...268..653N}, from the Pastel catalog \citep[PASTEL,][]{2016A&A...591A.118S}, from the LAMOST OBA (DR6) catalog \citep[HotPayne LAMOST DR6,][]{2021arXiv210802878X}, and from the Gaia ESO survey \citep[GES,][]{2022GES...WP13}.
            \label{fig:esphs_teff_offset}
        }
    \end{center}
\end{figure}

\begin{figure}
    \begin{center}
        \includegraphics[width=1\linewidth, angle=0]{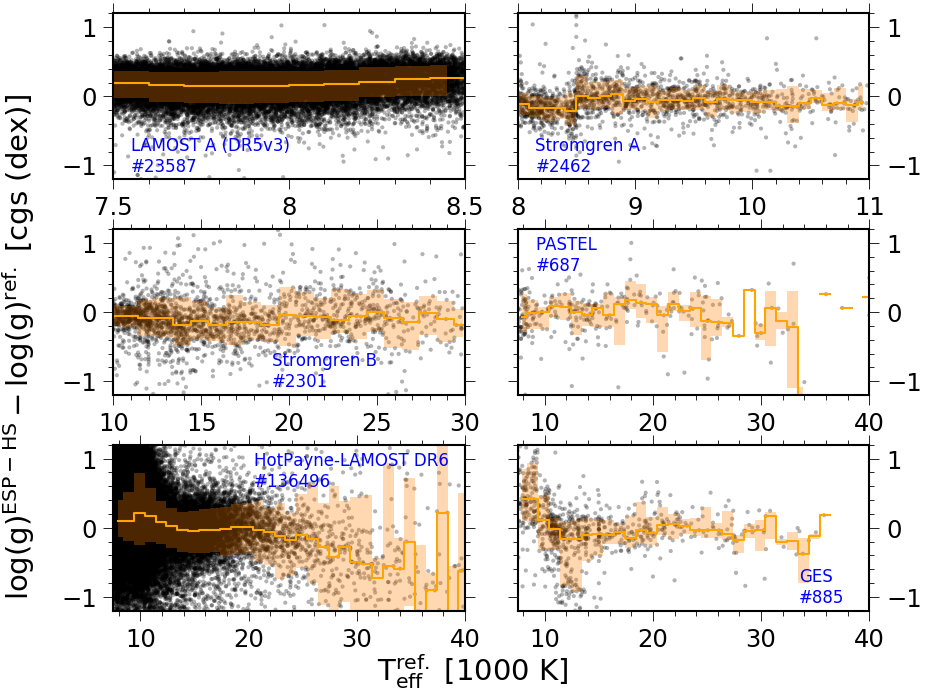}
        \caption{Surface gravity difference between values obtained by \esphs\ and those from literature catalogs. Same conventions as Fig.\ref{fig:esphs_teff_offset}.
            \label{fig:esphs_logg_offset}
        }
    \end{center}
\end{figure}

\paragraph{\esphs.}
Stars hotter than $7\,500$\,K (O, B, and A-type stars) undergo a specific analysis by the \esphs\ module. It operates in two modes: simultaneous analysis of BP, RP, and RVS spectra (``BP/RP+RVS''), or BP and RP only.
\esphs\ first estimates the star's spectral type\footnote{Originally produced by \esphs, the spectral type classification procedure moved to the \espels module for practical reasons.} from its BP and RP spectra to further analyze O, B, and A-type stars only. (\linktoparampath{astrophysical_parameters}{spectraltype_esphs}: CSTAR, M, K, G, F, A, B, and O). Hot stars of these spectral types are inherently massive, short-lived according to stellar evolution, and consequently these are young stars \footnote{We assume our data dominated by disk stars, and therefore ignoring horizontal branch stars from the Halo and \esphs  does not include models for white dwarf atmospheres}. Hence, \esphs\ assumes a Solar chemical composition, and therefore it does not provide any metallicity estimate. See module details in \linksec{ssec:cu8par_apsis_esphs}{the \gdr{3} online documentation, Sect.~3.3.8}.
For the stars hotter than 7\,500~K, the overlap between \gspphot\ and \esphs\ allows us to cross-validate our effective temperature estimates.  We find that \esphs\ tend to provide \teff\ greater than the \gspphot\ values due to different internal ingredients.
We quantify further the potential systematics from \esphs with respect to catalogs in the literature. Figures~\ref{fig:esphs_teff_offset} and \ref{fig:esphs_logg_offset} show the residuals relative to literature compilations for \teff and \logg, respectively. Below 25\,000~K, we obtain reasonable agreement of \esphs's temperatures with the catalogs estimates.
Overall, the dispersion in \teff\ increases with temperature from $\sim 300$\,K for the A-type stars to $500-2\,000$\,K for B-type stars. Above $25\,000$\,K, we find, relative to the \teff\ vs. spectral type scale of \citet{2010A&A...524A..98W}, a systematic underestimation of our temperatures by 1\,000~K to 5\,000\,K for the Galactic O-type stars, while it can be up to 10\,000~K for their LMC target samples. However, we also recall that this particular LMC sample is sub-solar metallicity, i.e, outside the model limits of \esphs. Similarly the dispersion in \logg\ increases from about 0.2~dex in the A-type stars temperature range to $\sim$0.4~dex for the O-type stars. More detailed numbers for the offset and dispersion of \teff\ and \logg\ relative to the catalogs considered in Fig.\,\ref{fig:esphs_teff_offset} and Fig.\,\ref{fig:esphs_logg_offset} are available in \citet{onlinedocdr3}.
We found that \esphs\ underestimated uncertainties by a factor of 5 to 10 in the BP/RP+RVS mode while reporting the correct order of magnitude in the BP/RP-only mode. We did not inflate the reported uncertainties in the \gdr{3} catalog accordingly. The first digit of \linktoparampath{astrophysical_parameters}{flags_esphs} reports which mode \esphs\ estimates come from (i.e. 0: ``BP/RP+RVS'', 1: ``BP/RP-only'').
We emphasize that we filtered out a significant number of bad fits of \esphs, but known outliers remain present (e.g. \teff\ $>$ 50\,000~K). In addition, \esphs\ processed white dwarfs (WD) despite not using a suitable library. Finally, some classes of stars intrinsically cooler than $7\,500$~K (e.g., RR Lyrae stars) were misclassified as O, B, or A-type stars and \esphs\ analyzed and reported on them assuming a correct classification.

\begin{figure*}
    \centering
    \includegraphics[width=\textwidth, clip, trim=0 0.6cm 0 0]{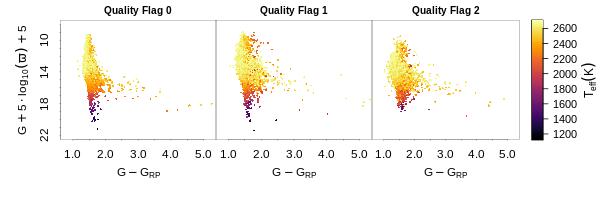}
    \caption{Color-absolute magnitude diagram of the UCD candidates from the \espucd\ analysis. The y axis uses the inverse parallax as distance estimate and we assumed negligible extinction. The color code reflects the \espucd\ \teff estimate according to the scale on the right hand side.}
    \label{fig:espucd_camd}
\end{figure*}

\begin{figure}
    \centering
    \includegraphics[width=0.47\textwidth]{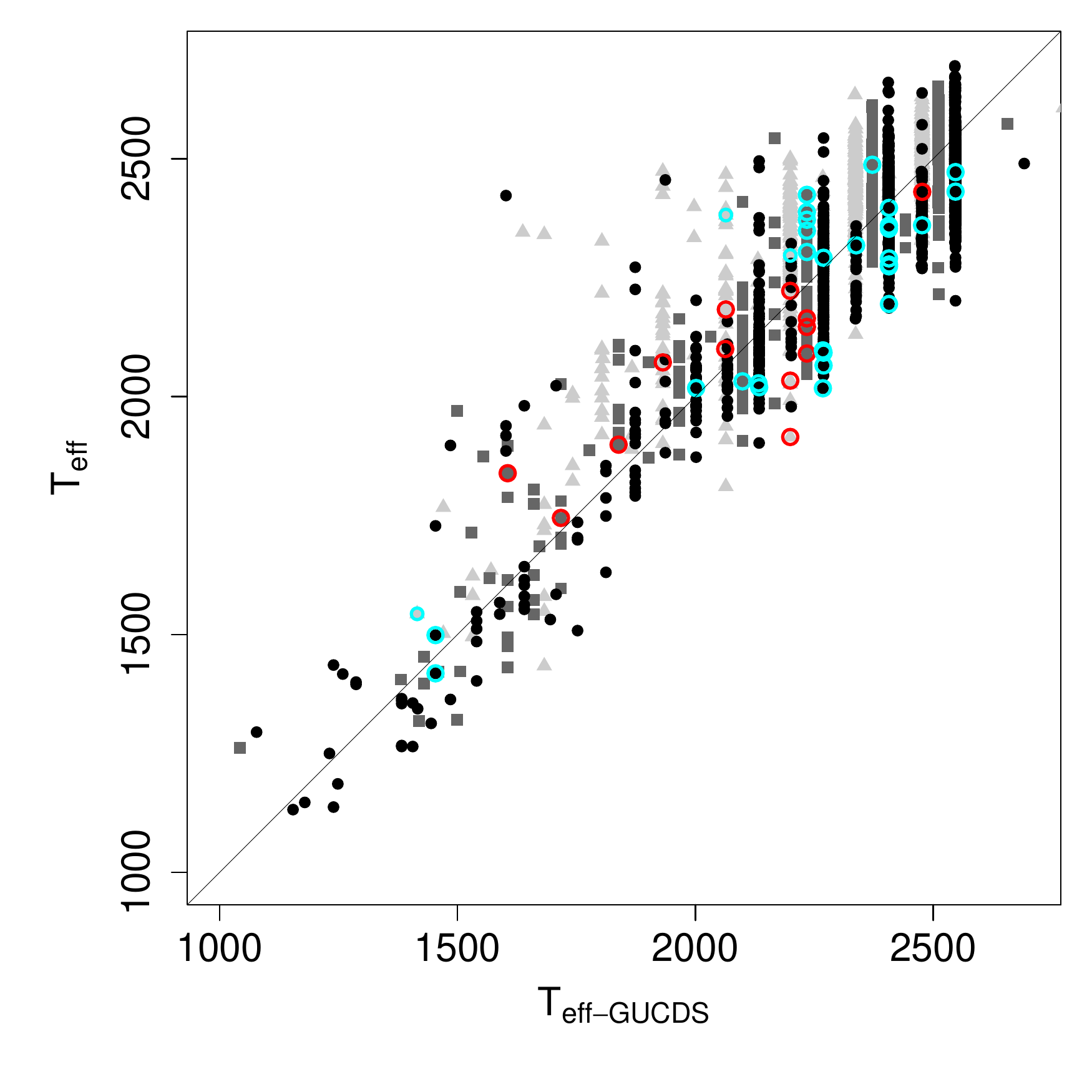}
    \caption{Comparison of the effective temperatures (in Kelvin) between \espucd\ estimates and those obtained by converting the GUCDS spectral types using the calibration by \citet{2009ApJ...702..154S}. Black circles correspond to quality 0, dark gray squares to quality 1, and light gray triangles to quality 2. Cyan symbols denote low-metallicity sources and red symbols denote young sources.}
    \label{fig:espucd_gucds}
\end{figure}

\begin{figure}
    \centering
    \includegraphics[width=0.47\textwidth]{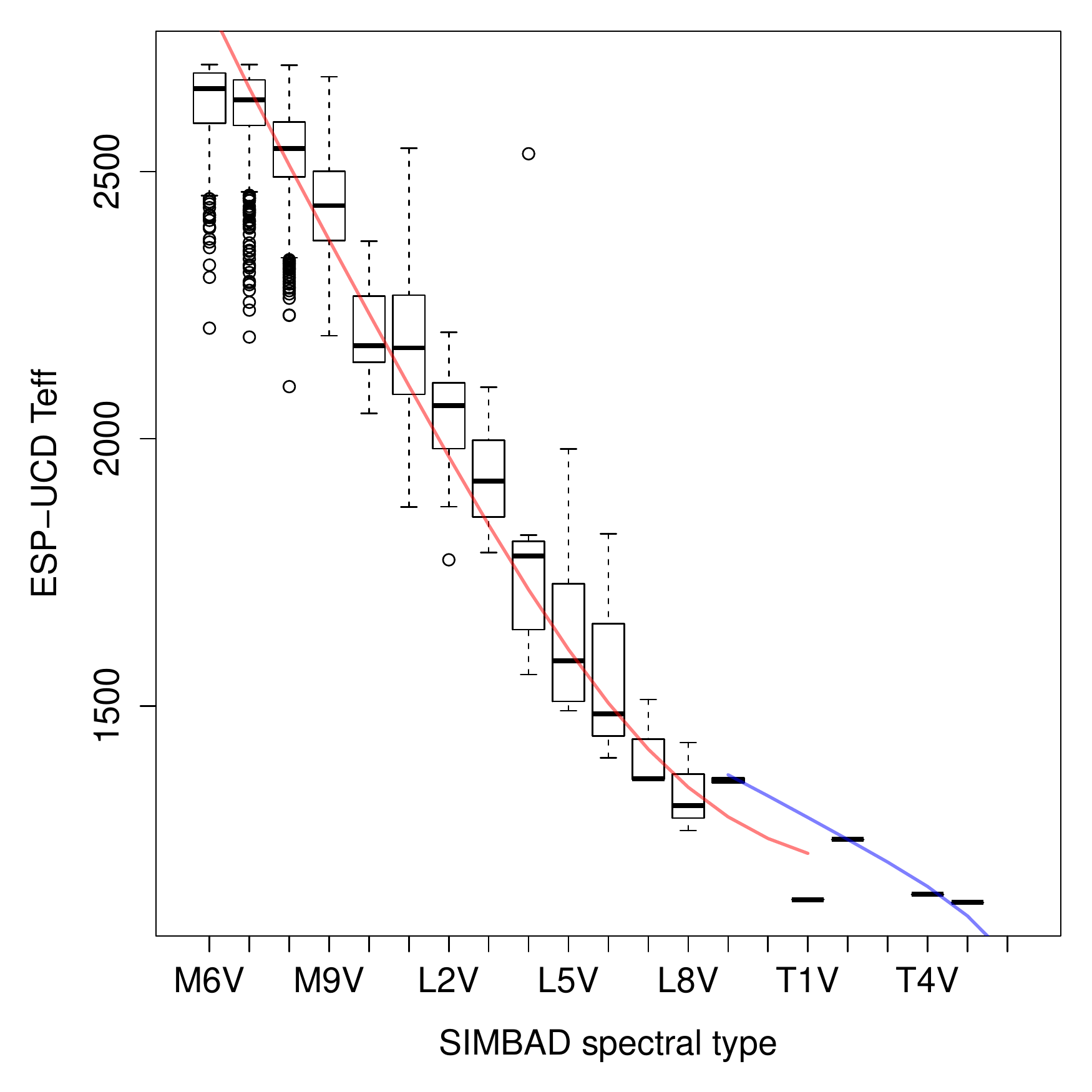}
    \caption{Comparison of the SIMBAD spectral types with the \espucd effective temperatures. The boxplots represent each spectral type's medians, interquartile ranges, and outliers. The red and blue solid lines represent the calibrations from \citet{2009ApJ...702..154S} for the M and L optical spectral types and the infrared spectral type T sources, respectively. }
    \label{fig:espucd_simbad}
\end{figure}

\paragraph{\espucd.}
At the faint end of the luminosity distribution, we transition between the ``standard'' stars burning hydrogen and the brown dwarfs not massive enough to sustain nuclear fusion.
We define ultracool dwarfs (UCDs) as sources of spectral type M7 or later \citep{1997AJ....113.1421K} which corresponds to \teff $\leq 2\,656$\,K according to the calibration by \citet{2009ApJ...702..154S}.
Using a combination of parallaxes, color indices, and RP spectra, we identified 94\,158 UCD candidates in \gdr{3} with \teff $< 2,700\,K$ despite the \gaia\ instruments being suboptimal to observe these intrinsically faint sources. We note that unsurprisingly the flux in the BP band is negligible (or even absent) for these very red and faint sources. The adopted threshold (2\,700\, K) is slightly hotter and more inclusive than the quoted 2\,656\, K to take into account the \teff\ estimate uncertainties.
\citet{DR3-DPACP-157} detail our characterization module, the complete UCD selection criteria, our quality filters, and our training set definition. \espucd\ produced effective temperatures for 94\,158 UCD candidates in \gdr{3}, the vast majority of them (78\,108) having $\teff > 2\,500$\,K.
However, \gdr{3} provides temperature estimates from \espucd\ (\linktoparampath{astrophysical_parameters}{teff_espucd}) but it does not include the corresponding \logg or \mh estimates due to the poor performance of \espucd\ on these properties and a severe lack of literature reference in this regime. We plan to publish them in \gdr{4}.
\espucd provides a flag (\linktoparampath{astrophysical_parameters}{flags_espucd}) to encode the quality of the data in one of three categories based on the Euclidean distance between a given RP spectrum and the closest template in the training set and the signal to noise ratio of the integrated RP flux. Quality Flag 0 corresponds to the best RP spectra distance below 0.005; quality 1 corresponds to sources with distances between 0.005 and 0.01 and SNR > 30 relative uncertainties $\sigma_{RP}/f_{RP} <= 0.03$; and finally quality flag 2 corresponds to sources with distances between 0.005 and 0.01 but SNR < 30. (The \linksec{ssec:cu8par_apsis_espucd}{\gdr{3} online documentation} provides a more detailed description of the quality flags.)

Figure \ref{fig:espucd_camd} shows the color-absolute magnitude diagram (CAMD) for all the UCD candidates we detected for the three \espucd\ quality categories. We find good consistency in CAMD positions and the inferred effective temperatures: as expected for these stars, their temperatures strongly correlate with \mg. We note that Fig.~\ref{fig:espucd_camd} uses the inverse parallax as a good distance proxy to approximate \mg, because 95\% of the sources have SNR $\varpi/\sigma_{\varpi}>5$ (the median parallax SNR in the three quality categories, 0, 1 and 2, are 25, 11, and 7.5 respectively). Overall, as the quality degrades, the vertical sequence spreads and becomes noisier w.r.t. the temperature scale.

More quantitatively, we compare our inferred temperatures with those of the Gaia UltraCool Dwarf Sample \citep[GUCDS; ][]{2017MNRAS.469..401S,2019MNRAS.485.4423S}. We translated the GUCDS spectral types using the calibration by \citet{2009ApJ...702..154S}, and we found an RMS of 103 K and a MAD of 88 K for the entire sample (see Fig.~\ref{fig:espucd_gucds}). We note that these statistics include the low-metallicity and young sources.
%
Figure~\ref{fig:espucd_simbad} compares the \espucd effective temperatures with SIMBAD spectral types when available, which sample includes and extends the GUCDS. We indicate the two spectral type-\teff calibration relations by \citet{2009ApJ...702..154S} for optical and infrared spectral types to provide a comparison reference.
These two relations are those we used to define the empirical training set of the \espucd module.
We note that the spectral type M6 corresponds to an effective temperature $\sim 2\,800$ K.
This temperature is hotter than the \espucd parameter space limit. However, \espucd\ attributed cooler \teff values to some of these stars, which we published but led to the apparent negative bias for the M6V bin in Fig.~\ref{fig:espucd_simbad}.
\citet{DR3-DPACP-123} (Sect.~7) further explore the stellar population of UCDs in the Galaxy, and their properties.

\subsubsection{Secondary atmospheric estimates: stellar classes, rotation, emission, activity}\label{sec:atmosphere-secondary}

\paragraph{Classification.}
There are four main stellar classifications from \apsis (see fields in Table~\ref{tab:product-module-atm-2}). First, \dsc\ primarily distinguishes between extragalactic sources (quasars and galaxies) and stars (single stars, physical binaries, and white dwarfs).
Users can classify sources using \dsc's probabilities of a source to belong to a given class. However, 99\% of \gdr{3} sources processed by \apsis\ are most certainly stars (or binaries). Hence \dsc's classification is not the most relevant for stellar objects \citep[see][]{BailerJones2021,DR3-DPACP-157}.

\oa\ measures similarities between observed BP and RP spectra of different sources to produce an unsupervised classification using self-organizing maps \citep[SOMs;][]{Kohonen2001}. One can use these maps to find similar groups of stars once labeled (details in \citealt{DR3-DPACP-157}) and peculiar or outlier sources (see Sect.~\ref{sec:weird}).
Finally, the user might prefer using the spectral types from \esphs and the classification of \espels\ for emission-line star types of stellar sources. This section focuses on the \esphs and \espels\ classification tailored to stellar objects.

\esphs\ estimates the spectral type of a source from its BP/RP spectra. While primarily focused on hot stars, it provides the following main classes:  CSTAR, M, K, G, F, A, B, and O).
We find from a cross-match with the LAMOST OBA catalog of \citep{2021arXiv210802878X} that \esphs\ obtained {62\%} of the Galactic A- \& B-stars (assuming the other catalog is complete). Conversely, we find only $186$ (30\%) of the $612$ Galactic O-type stars published in the Galactic O-type Stars catalog \citep[GOSC,][]{2013msao.confE.198M}. This low fraction reflects the persisting difficulties of deriving reliable hot star APs from Gaia BP and RP spectra.

\espels\ identifies the BP and RP spectra that present emission features and classifies the corresponding target into one of the seven ELS classes listed in Table~\ref{tab:espels_class_distribution}.
We recall that \espels\ processed stars brighter than $\gmag<17.65$\,mag (see Sect.~\ref{sec:overview}).
The \espels\ classification as ELS relies on detecting line-emission and primarily on measuring the H$\alpha$ pseudo-equivalent width (see below).
We tagged particular failure modes with the quality flag (\linktoparampath{astrophysical_parameters}{classlabel_espels_flag}; see Table~\ref{tab:espels_class_distribution}).
Primarily, this flag takes values ranging from 0 (best) to 4 (worst) depending on the relative strength of the two most probable classes (i.e., \espels\ published random forest classifier class probability estimates in \linktoparampath{astrophysical_parameters}{classprob_espels_wcstar}, \linktoparam{astrophysical_parameters}{classprob_espels_wnstar}, etc.).
In addition, \linktoparampath{astrophysical_parameters} indicates the \gspphot\ AP values we used to make the classification was removed by the final \gdr{3} filtering or when those APs disagreed with the spectral type estimated by \espels. These two modes correspond to \linktoparam{astrophysical_parameters}{classlabel_espels_flag} first bit  $1$ and $2$, respectively (Table~\ref{tab:espels_class_distribution}).
We emphasize that the identification of Wolf-Rayet stars (WC \& WN) and planetary nebula does not depend on any APs. All but five of the 136 detected WC stars have typical spectroscopic features. The missed Galactic WC stars taken from \citet{2015MNRAS.447.2322R}\footnote{\url{http://pacrowther.staff.shef.ac.uk/WRcat/}} are usually fainter than the processing limit of $\gmag = 17.65$\,mag, or a low-quality H$\alpha$ pEW estimate  (e.g., weaker emission lines with type WC8 or WC9).
Half of the $431$ WN star candidates do not show any typical emission line. Most of the known WNs in the literature are fainter than the processing limit of \espels.


\begin{table}
    \caption{Number of identified candidates per ELS class and per quality flag (QF) value. QF $\le$ 2 implies a probability larger than 50\%.}
    \label{tab:espels_class_distribution}
    \centering
    \begin{tabular}{@{}l@{\hspace{1.5\tabcolsep}}r@{\hspace{1.5\tabcolsep}}r@{\hspace{1.5\tabcolsep}}r@{\hspace{1.5\tabcolsep}}r@{\hspace{1.5\tabcolsep}}r@{\hspace{1.5\tabcolsep}}r@{\hspace{1.5\tabcolsep}}r@{}}
        \hline
        Class   & \multicolumn{7}{c}{{\linktoparam{astrophysical_parameters}{classlabel_espels_flag}} (QF)}                                                        \\
                & 0      & 1     & 2     & 3     & 4      & 10-14  & 20-24   \\
        \hline\hline
        Be      & 3\,210  & 3\,118 & 2\,815 & 2\,332 & 1\,475 & 122    & 3\,879  \\ 
        Ae/Be   & 35      & 94     & 231    & 519    & 972    & 299    & 1\,754  \\ 
        T Tauri & 914     & 2\,052 & 1\,594 & 1\,083 & 740    & 1\,180 & 27\,594 \\ 
        dMe     & 0       & 1      & 5      & 54     & 178    & 43     & 380     \\ 
        PN      & 37      & 52     & 83     & 85     & 16     & 0      & 0       \\ 
        WC      & 106     & 13     & 8      & 7      & 2      & 0      & 0       \\ 
        WN      & 173     & 38     & 29     & 142    & 47     & 0      & 0       \\ 
        \hline
    \end{tabular}
\end{table}

\paragraph{Stellar Rotation}
While deriving the astrophysical parameters,  \esphs\ also measures the line broadening on the RVS spectrum by adopting a rotation kernel. This by-product of the \esphs\ processing corresponds to a projected rotational velocity (\vsini; \linktoparampath{astrophysical_parameters}{vsini_esphs}) obtained on co-added mean RVS spectra \citep{DR3-DPACP-154}. It therefore differs from \linktoparampath{gaia_source}{vbroad} obtained on epoch data by the radial velocity determination pipeline \citep{DR3-DPACP-149}. The \esphs\ estimate suffers from the same limitations as \linktoparam{gaia_source}{vbroad} -- mostly the limited resolving power of the RVS -- increased by the poor \vsini-related information for OBA stars in this wavelength domain. In addition, the determination of \linktoparam{astrophysical_parameters}{vsini_esphs} is affected by the higher uncertainty of the epoch RV determination expected for stars hotter than 10\,000~K \citep{DR3-DPACP-151}, and by the use of a Gaussian {\it mean} ALong-scan LSF with a resolving power of 11\,500 \citep[][Sect.2.2]{DR3-DPACP-157}.

\begin{figure}
    \begin{center}
        \includegraphics[width=1\linewidth]{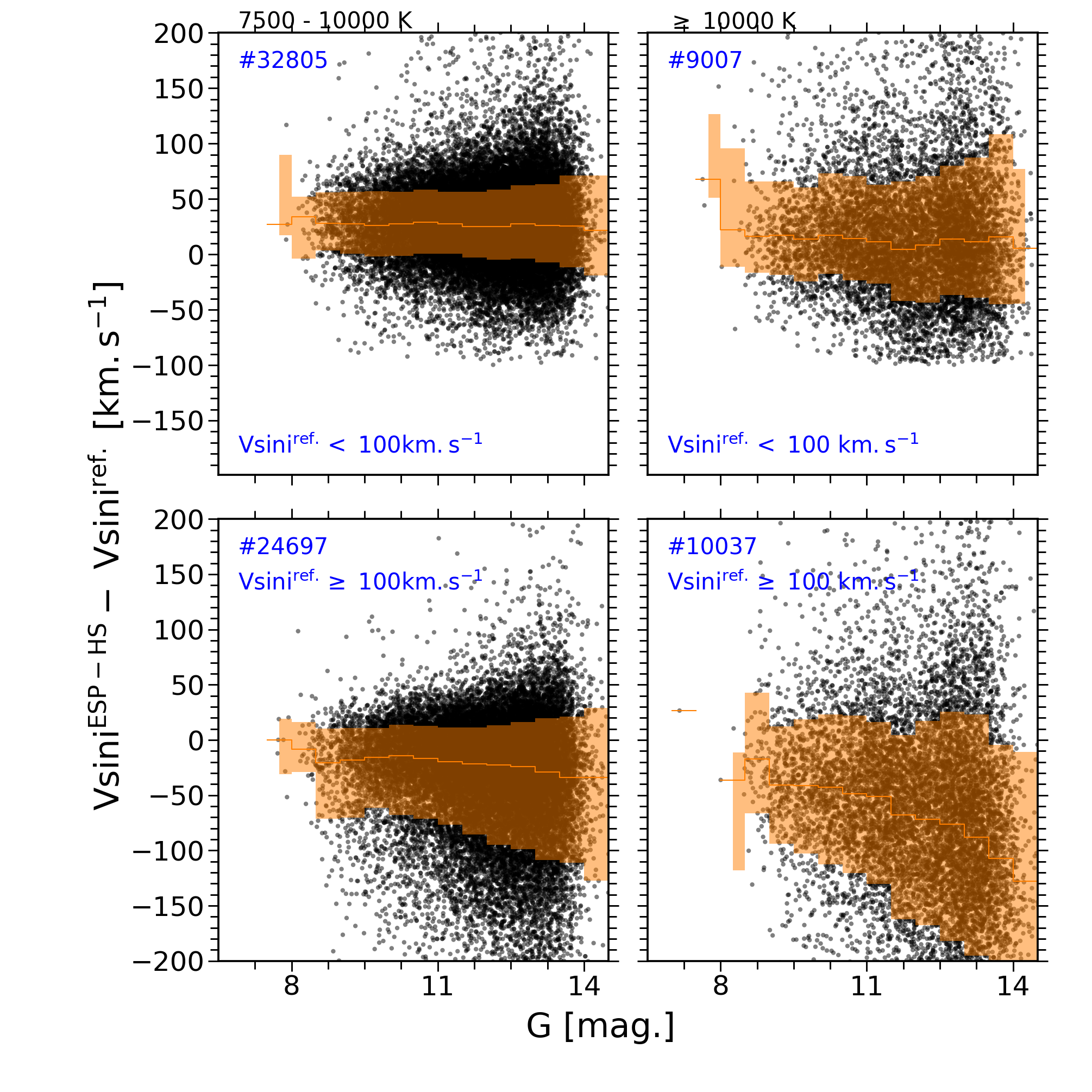}
        \caption{Distribution with the G magnitude of the differences between the LAMOST OBA results (\vsini$^\mathrm{ref.}$) and \esphs\ \vsini\ measurement. Stars cooler and hotter than 10\,000~K are plotted in the left and right panels, respectively. A distinction is also made between slow (upper panel) and rapid (lower panel) rotators. The running median is shown in orange, while the interquartile dispersion (at 14.85\% and 85.15\%) is represented by the orange shades.
        }
        \label{fig:cu8par_apsis_qa_esphs_vsini_lamost}
    \end{center}
\end{figure}

In Fig.\,\ref{fig:cu8par_apsis_qa_esphs_vsini_lamost} we present a comparison between the \vsini\ measurements by \esphs\ to those obtained in the framework of the LAMOST survey for OBA stars which presents the largest overlap with the results of \esphs\ compared to other surveys. The agreement rapidly decreases with magnitude, and effective temperature, while the most sensitive features to rotational broadening disappears from the RVS domain. The half inter-quantile dispersion (i.e. 14.85 \% - 15.15 \%) varies from 25 $\mathrm{km.s}^{\mathrm{-1}}$ to 40 $\mathrm{km.s}^{\mathrm{-1}}$ in the A-type \teff\ domain when the magnitude $G$ ranges from 8 to 12, respectively. At hotter temperatures, it varies from 60 $\mathrm{km.s}^{\mathrm{-1}}$ to 75 $\mathrm{km.s}^{\mathrm{-1}}$ at $G=8$ and $G=12$, respectively.

\begin{figure}
    \begin{center}
        \includegraphics[width=1\linewidth, angle=0]{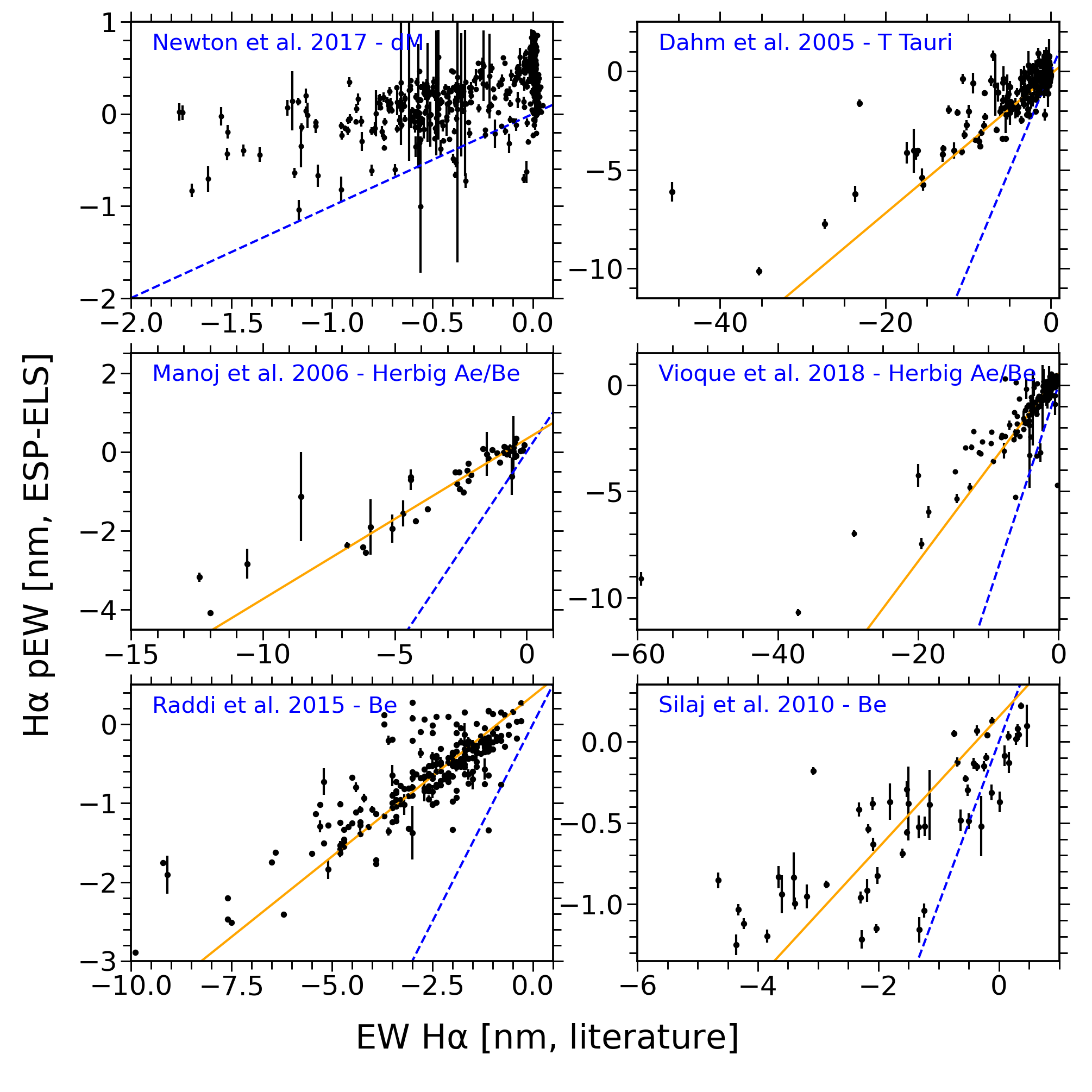}
        \caption{Comparison of the H$\alpha$ pseudo-equivalent width measured by \espels\ to the equivalent width published for different types of ELS classes. Each panel compares our estimates with the reference indicated in blue assuming their stellar class. The identity relation is given by the broken blue line and compared to a linear fit (orange line) through the data (Eq.\,\ref{eq:espels.line.ha} and Table\,\ref{tab:espels.linearfit.ha}).
        \label{fig:cu8par_apsis_espels_halpha_litt}
        }
        \nocite{2015MNRAS.446..274R, 2005AJ....129..829D, 2018AA...620A.128V, 2017ApJ...834...85N, 2010ApJS..187..228S, 2006ApJ...653..657M}
    \end{center}
\end{figure}

\paragraph{H$\alpha$ emission}
The \espels\ classification of a star as ELS relies primarily on measuring the H$\alpha$ pseudo-equivalent width (pEW; \linktoparampath{astropysical_parameters}{ew_espels_halpha}).
However, measuring the H$\alpha$ emission line is challenging due to the low resolving power of BP and RP spectra and the steep loss of transmission at that wavelength (blue side). Figure~\ref{fig:cu8par_apsis_espels_halpha_litt} compares our H$\alpha$ pEW estimates to the values provided by various authors \citep{2015MNRAS.446..274R, 2005AJ....129..829D, 2018AA...620A.128V, 2017ApJ...834...85N, 2010ApJS..187..228S, 2006ApJ...653..657M}. We found a general consistency between the estimates, except for stars cooler than $4\,000$\,K, for which overlapping spectral molecular bands significantly alter the local continuum. We mitigated this effect using synthetic spectra and the \gspphot's APs. However, the mismatches between the observed and theoretical spectra and some systematics in the APs we used to select the synthetic spectra led us to misclassify active M dwarf and T Tauri stars. For the hotter targets, we attempted to link the \espels\ estimate, pEW(H$\alpha$), and the published measurements presented in Fig.\,\ref{fig:cu8par_apsis_espels_halpha_litt} with the following linear relation:
\begin{equation}
EW^\mathrm{ref.}(H\alpha) = \alpha + \beta \times pEW(H\alpha),
\label{eq:espels.line.ha}
\end{equation}
\noindent where Table~\ref{tab:espels.linearfit.ha} provides the coefficients, $\alpha$ and $\beta$, with their uncertainty. We indicated by the orange lines the fitted relations n Fig.\,\ref{fig:cu8par_apsis_espels_halpha_litt}.

\begin{table}
    \centering
    \caption{Coefficients of the line fitted (Eq.\,\ref{eq:espels.line.ha}) through the data points presented in Fig.\,\ref{fig:cu8par_apsis_espels_halpha_litt}\label{tab:espels.linearfit.ha}.
    }
    \begin{tabular}{lcc}
        \hline
        Ref. for $EW^\mathrm{ref.}(H\alpha)$ & {$\alpha$} & {$\beta$} \\
        \hline\hline
        \citet{2006ApJ...653..657M} & $-$0.811 $\pm$ 0.165 & $+$2.464 $\pm$ 0.198 \\
        \citet{2005AJ....129..829D} & $+$0.407 $\pm$ 0.075 & $+$2.835 $\pm$ 0.087 \\
        \citet{2018AA...620A.128V}  & $-$1.196 $\pm$ 0.133 & $+$2.266 $\pm$ 0.141 \\
        \citet{2015MNRAS.446..274R} & $-$0.886 $\pm$ 0.085 & $+$2.454 $\pm$ 0.117 \\
        \citet{2010ApJS..187..228S} & $-$0.380 $\pm$ 0.241 & $+$2.480 $\pm$ 0.371 \\
        \hline
    \end{tabular}
\end{table}

\paragraph{Chromospheric activity Index}
The \espcs\ module computed an activity index \linktoparam{astrophysical_parameters}{activityindex_espcs} from the analysis of the Ca\,II\,IRT (calcium infrared triplet) in the RVS spectra for $2\,141\,640$ stars in \gdr{3}.
\espcs\ defines cool stars as stars with $G \lesssim 15$\,mag, $\teff \in[3000, 7000]$\,K, $\logg \in [3.0, 5.5]$\,dex, and $\mh \in [-0.5, 1.0]$\,dex. Stars with APs from \gspspec\ within these intervals undergo the analysis by \espcs.
The activity index is the excess of the Ca\,II\,IRT lines from comparing the observed RVS spectrum with a purely photospheric model (assuming radiative equilibrium). The latter depends on a set of \teff, \logg, and \mh\ from either \gspspec\ or \gspphot\ (\linktoparam{astrophysical_parameters}{activityindex_espcs_input} set to "M1" or "M2", respectively), and a line broadening estimate \linktoparampath{gaia_source}{vbroad} when available.
We measure the excess equivalent width in the core of the Ca\,II\,IRT lines by computing the observed-to-template ratio spectrum in a $\pm \Delta \lambda = 0.15 \mathrm{nm}$ interval around the core of each of the triplet lines. This measurement translates the stellar chromospheric activity and, in more extreme cases, the mass accretion rate in pre-main-sequence stars.
\cite{DR3-DPACP-175} detail the \espcs\ module, method, and scientific validation.

\subsubsection{Chemical Abundances}\label{sec:abundances}

\begin{figure}
    \centering
    \includegraphics[width=0.39\textwidth]{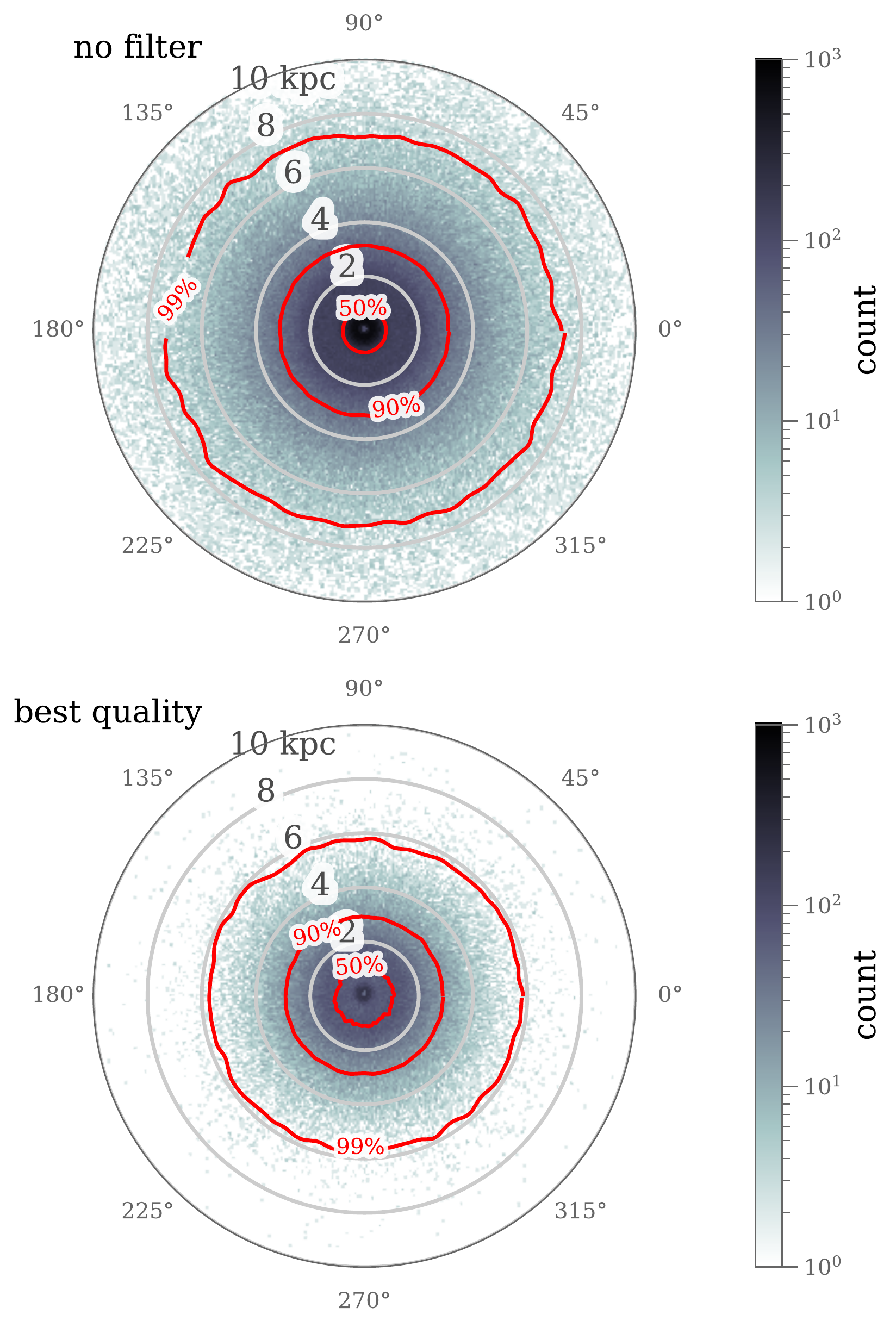}
    \caption{Galactic top-down view of the processing coverage of \gspspec. The top and bottom panels show the entire and best quality samples, respectively. We projected the sources for which \gspspec provides abundance estimates using the distances from \gspphot. The distribution centers around the Sun with the Galactic center at (0, 8\,kpc). The contours indicate the 50, 90, and 99\% quantiles of the distribution, corresponding to $\sim$ 1, 3, and 6\,kpc, respectively. }
    \label{fig:gspspec_xy_coverage}
\end{figure}

\begin{figure}
    \includegraphics[width=\columnwidth]{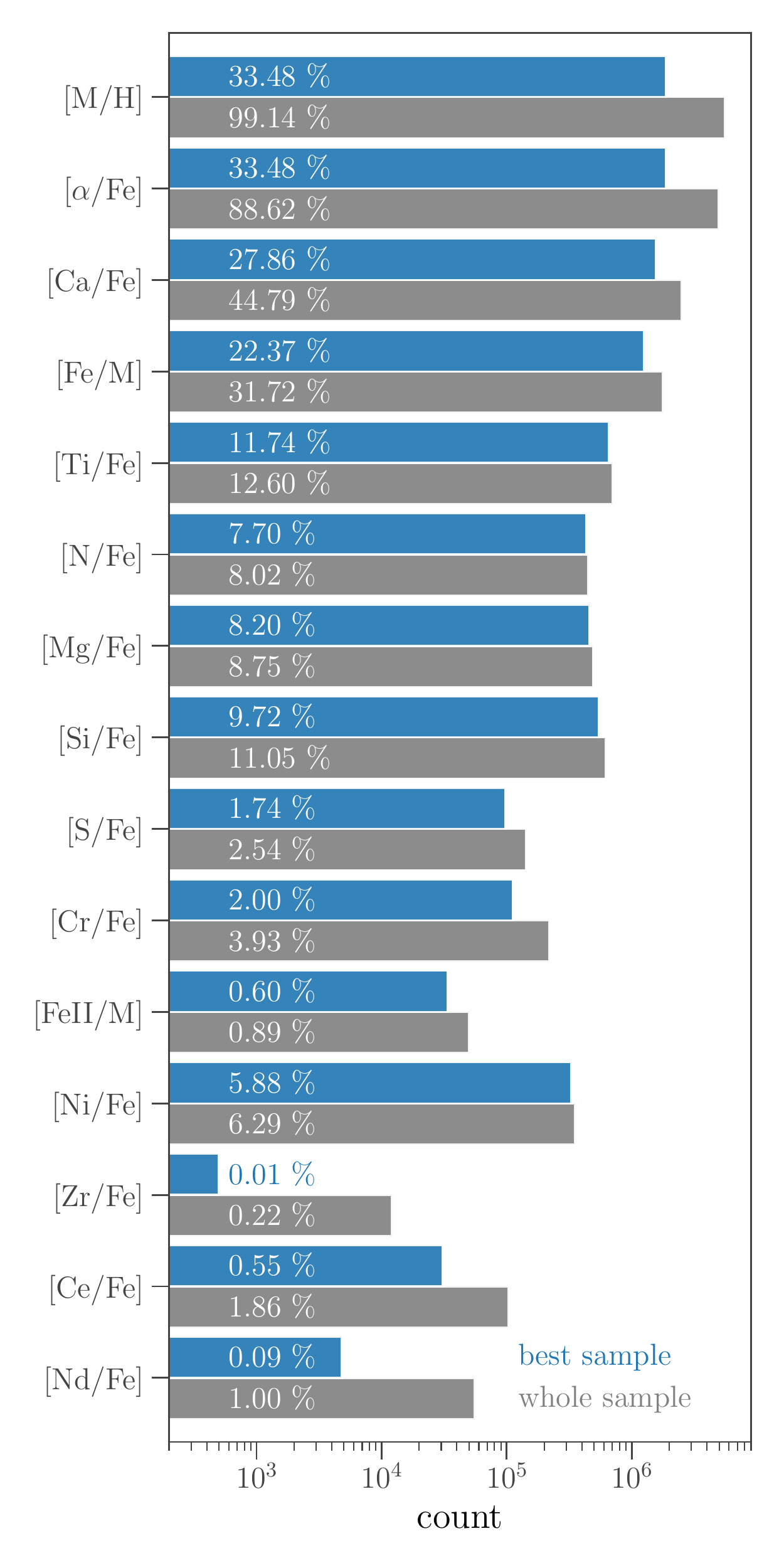}
    \caption{Number of stars with individual abundance ratio estimates from \gspspec. The two sets represent the whole sample and the best quality sample (quality flags equal to zero) in gray and blue, respectively. We indicate at the top \mh\ and \afe\ for reference (discussed in Sect.\,\ref{sec:atmosphere-primary}). The percentages correspond to the fraction of estimates with respect to the $5\,594\,205$ stars processed by \gspspec. Note that on that scale, 1\% corresponds to $40\,000$ stars.
    \label{fig:abundance_counts}
    }
\end{figure}

In \gdr{3}, \gspspec\ -- most specifically the Matisse-Gauguin algorithm -- provides us with $13$ chemical abundance ratios from $12$ individual elements (N, Mg, Si, S, Ca, Ti, Cr, Fe, Ni, Zr, Ce, and Nd; with the FeI and FeII species) as well as equivalent-width estimates of the CN line at $862.9$\,nm. These chemical indexes rely on the line-list and models from \citet{Contursi21} and \citet{DR3-DPACP-186}, respectively.
For each of the $13$ abundance estimates, \gspspec\ reports two quality flag bits, a confidence interval, the number of used spectral lines, and the line-to-line scatter (when more than one line).
We note that the \gdr{3} catalog contains the [FeI/M] and [FeII/M] as all the other ratios, dictated by the parameterization of the synthetic model grids in \gspspec. To obtain [FeI/H] and [FeII/H], one has to add the star's [M/H].  \citet{DR3-DPACP-186} describe in detail the definition and measurements of these chemical abundances, as well as their quality flags.
Figure~\ref{fig:gspspec_xy_coverage} shows the spatial extent of the abundance estimates in a top-down Galactic view. The coverage indicates that \gdr{3} provides abundance estimates for a significant fraction of the stars observed by \gaia\ within 4\,\kpc as indicated by the 99\% quantile contour.
Figure~\ref{fig:abundance_counts} decompose the \gdr{3} catalog content into the individual abundance ratios for the best quality and whole sample. We note that \citet{DR3-DPACP-186} also provide intermediate selections.
\citet{DR3-DPACP-104} analyzes the chemical abundance estimates in the context of the chemistry and Milky Way structure, stellar kinematics, and orbital
parameters.

The validation of individual abundances is challenging as no fundamental standards exist for stars other than the Sun. One needs particular attention when comparing with literature data, which suffers from different zero points and underlying assumptions (e.g., assumed solar-scaled composition).

We expect our derived abundances to have the usual limitations discussed in the literature stemming from model assumptions (e.g., 1D or 3D model atmospheres, hydrostatic, local thermodynamic equilibrium,
the atomic line list) to observational effects (e.g., possible line blends, limited resolution of RVS, instrumental noise). These effects can lead to systematic offsets in the abundance determinations that depend on the atmospheric parameters.
%
However, we could estimate (and correct) these systematic offsets using the \gspspec\ outputs alone and specific samples of stars. For instance, we selected stars from the immediate solar neighborhood ($\pm250$\pc from the Sun), with metallicities close to solar ($\pm 0.25$) and velocities close to the local standard of rest ($\pm25$ km/s). In this sample, any ratio of abundances (i.e.,[$X_1$/$X_2$] for two elements $X_1$ and $X_2$) deviating from zero (i.e., solar value) indicates systematics independent of the atmospheric parameters. In \citet{DR3-DPACP-186}, we detail our samples and analysis, and we provide \logg-dependent calibration relations for $10$ chemical abundances, out of the $13$,  in the form of polynomials (of the third or fourth-order).
In particular, {Table~3} of \citet{DR3-DPACP-186} lists the coefficients values as well as the \logg\ intervals over which the calibration is applicable (and comparison statistics with the literature).
 For instance, we selected sample stars from APOGEE DR17 \citep{Abdurrouf2021} and GALAH-DR3 \citep{Buder2021} with \gspspec\ quality flags all equal to zero and literature uncertainties smaller than $500$\,K, $0.5$\,dex and $0.2$\,dex for \teff, \logg and [Mg/Fe] or [FeI/H], respectively. This sample contains 1\,100 stars with [Mg/Fe] and 92\,000 with [FeI/H] estimates. When comparing these with \gspspec\ abundances, we found a median abundance offset of $-0.15$ dropping to $0.0$\,dex for [Mg/Fe] and $-0.15$ to $0.05$\,dex for [FeI/H], before and after applying those calibration relations, respectively \citep[see][for further details]{DR3-DPACP-186}.

\subsection{Evolutionary APs}\label{sec:evolution}

\gdr{3} provides several parameters describing the evolution of a star that we group in two sets.
\gspphot\ and \flame\ produce these parameters (see Table~\ref{tab:product-module-abundances}). We emphasize that \flame\ produces two sets of estimates: one using \gspphot's APs and one using \gspspec's obtained from the BP/RP and RVS spectra analysis, respectively, in addition to using photometry and distance (or parallax).

We first discuss in Sect.~\ref{ssec:evolution_rl} the ``observed'' parameters: luminosity \lum, absolute magnitude \mg, radius \radius, and gravitational redshift \gravshift.
These are relatively model-independent in contrast with the mass \mass, age \age, and evolutionary stage \evolstage, which strongly depend on evolution models. We discuss these in Sect.~\ref{ssec:evolution_ma}.

\subsubsection{Radius, luminosity, absolute magnitude, and gravitational redshift}\label{ssec:evolution_rl}

\begin{figure}
    \begin{center}
        \includegraphics[width=0.47\textwidth, angle=0]{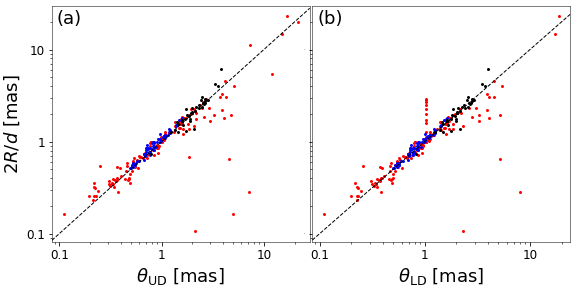}
        \caption{Using \gspphot\ radius and distance estimates to predict measurements of angular diameters from ground-based interferometry by \citet{Boyajian2012a,Boyajian2012b, Boyajian2013} (blue points), \citet{Duvert2016} (red points) and \citet{2021arXiv210709205V} (black points). Panel a: Comparison of $2R/d$ to estimates assuming uniform disk. Panel b: Comparison of $2R/d$ to estimates accounting for limb-darkened angular diameter.
            \label{fig:gspphot_angular_diameters}
        }
    \end{center}
\end{figure}

\begin{figure*}
    \centering
    \includegraphics[width=0.49\textwidth]{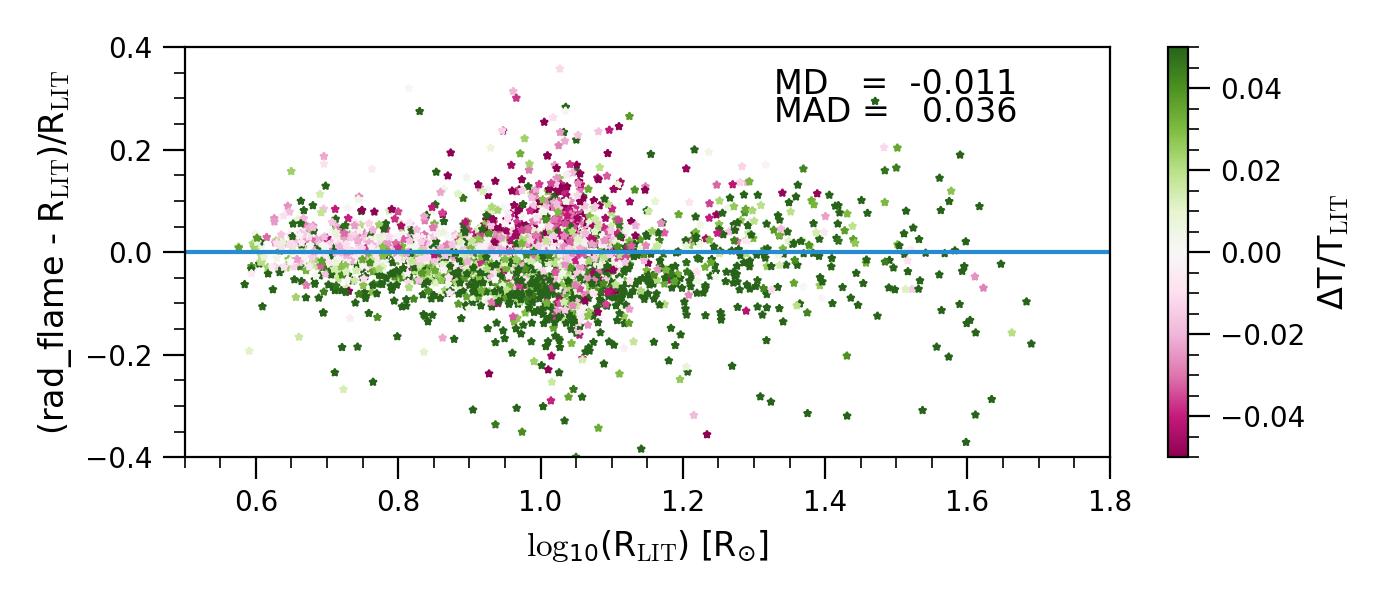}
    \includegraphics[width=0.49\textwidth]{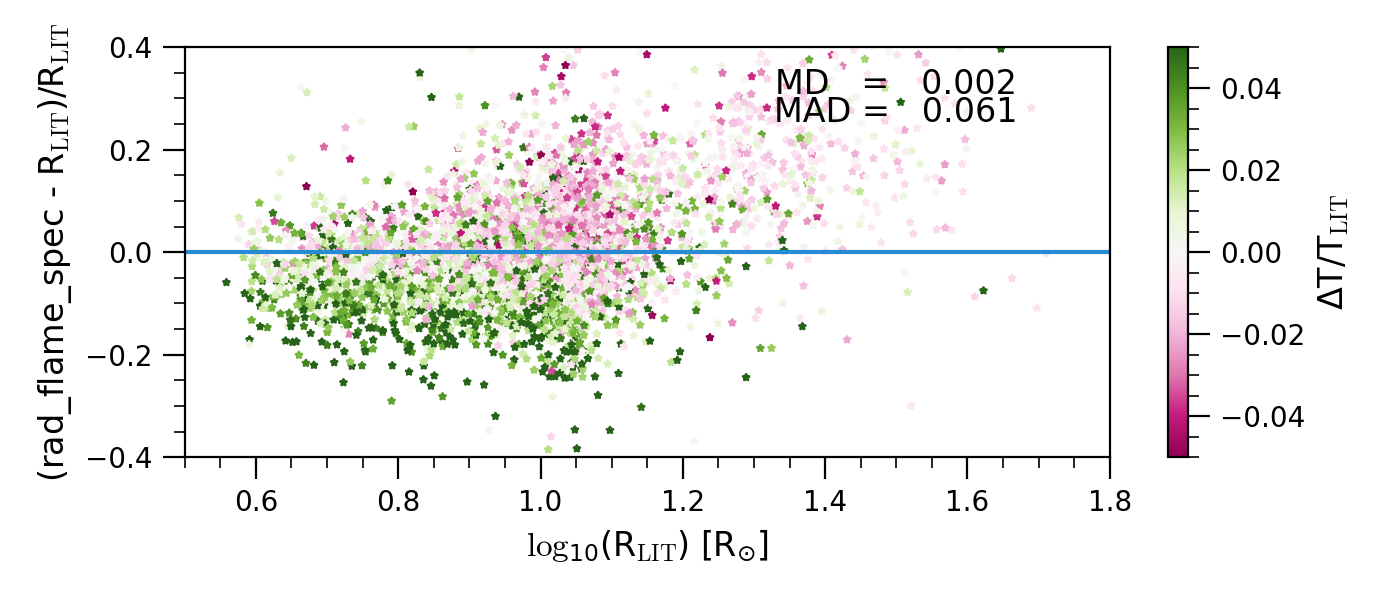}\\
    \includegraphics[width=0.49\textwidth]{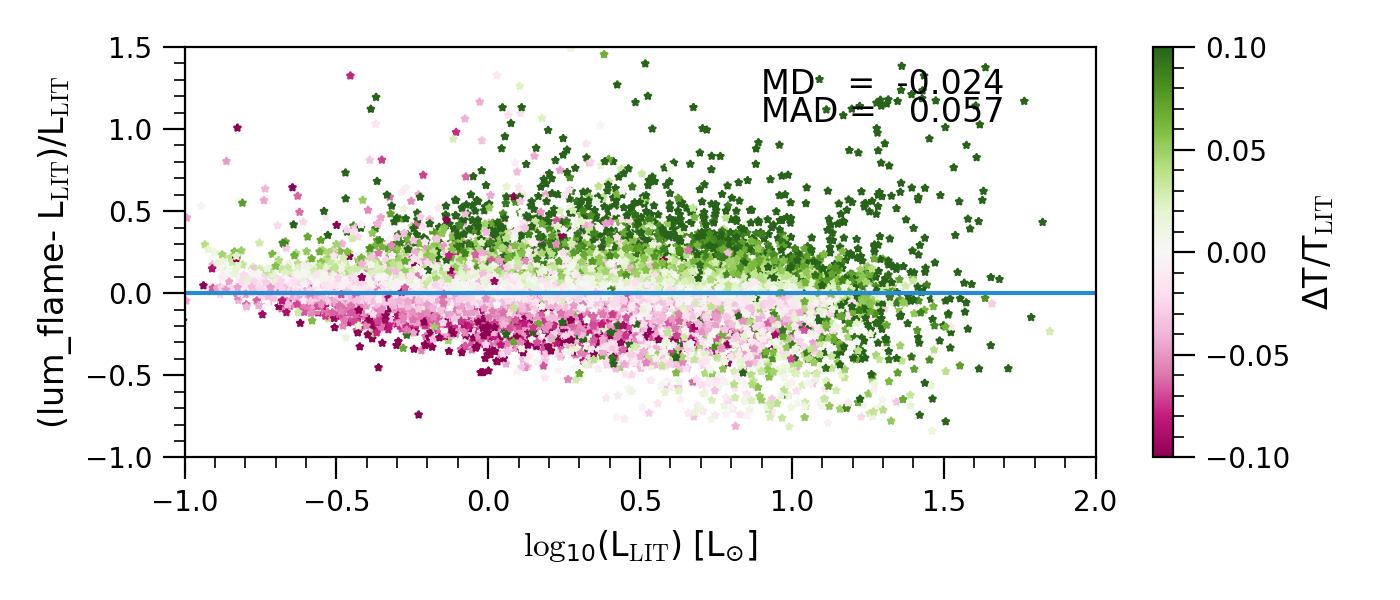}
    \includegraphics[width=0.49\textwidth]{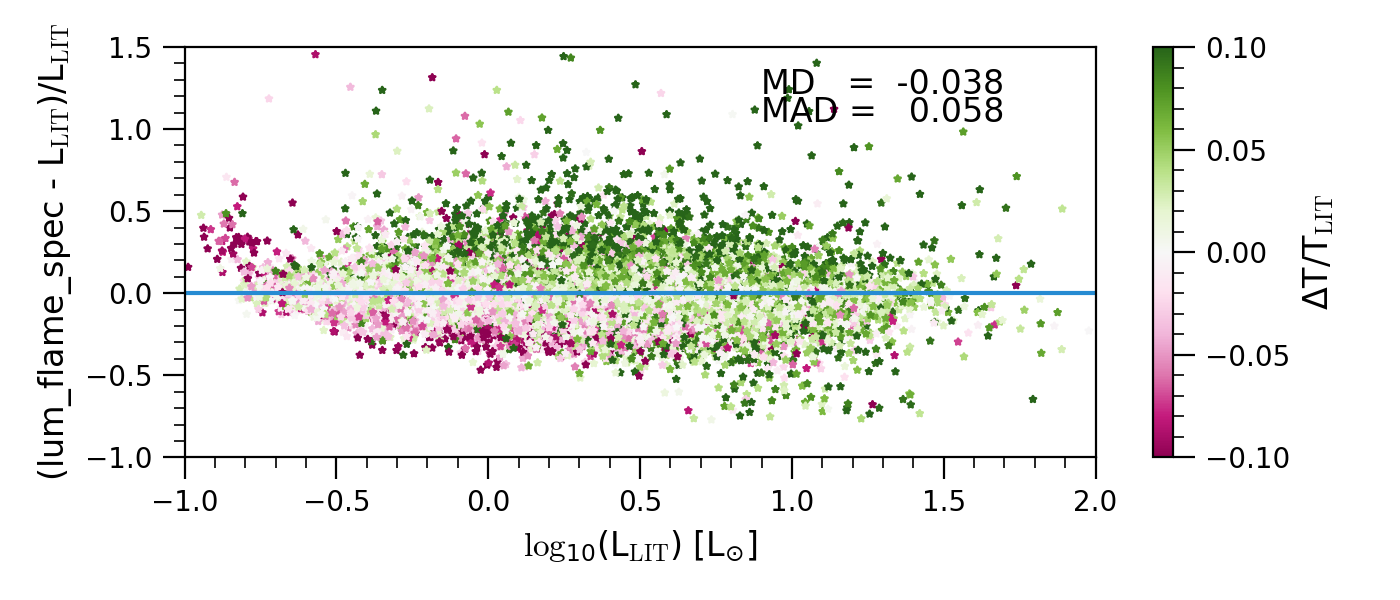}\\
    \includegraphics[width=0.49\textwidth]{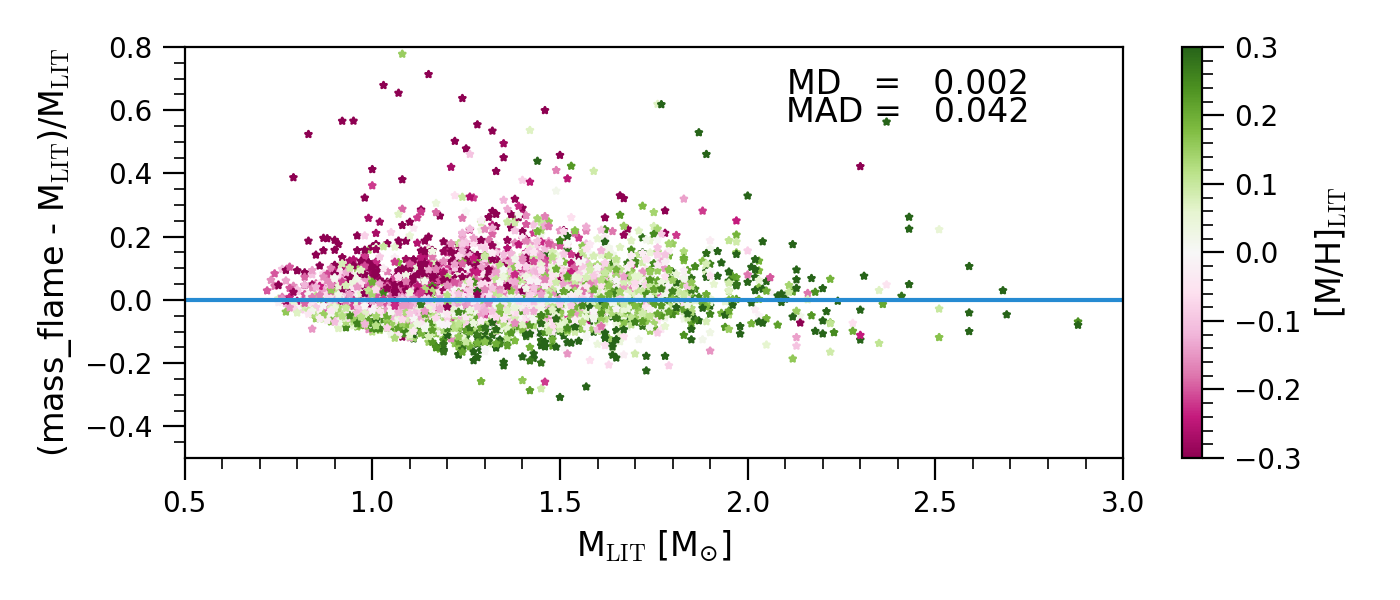}
    \includegraphics[width=0.49\textwidth]{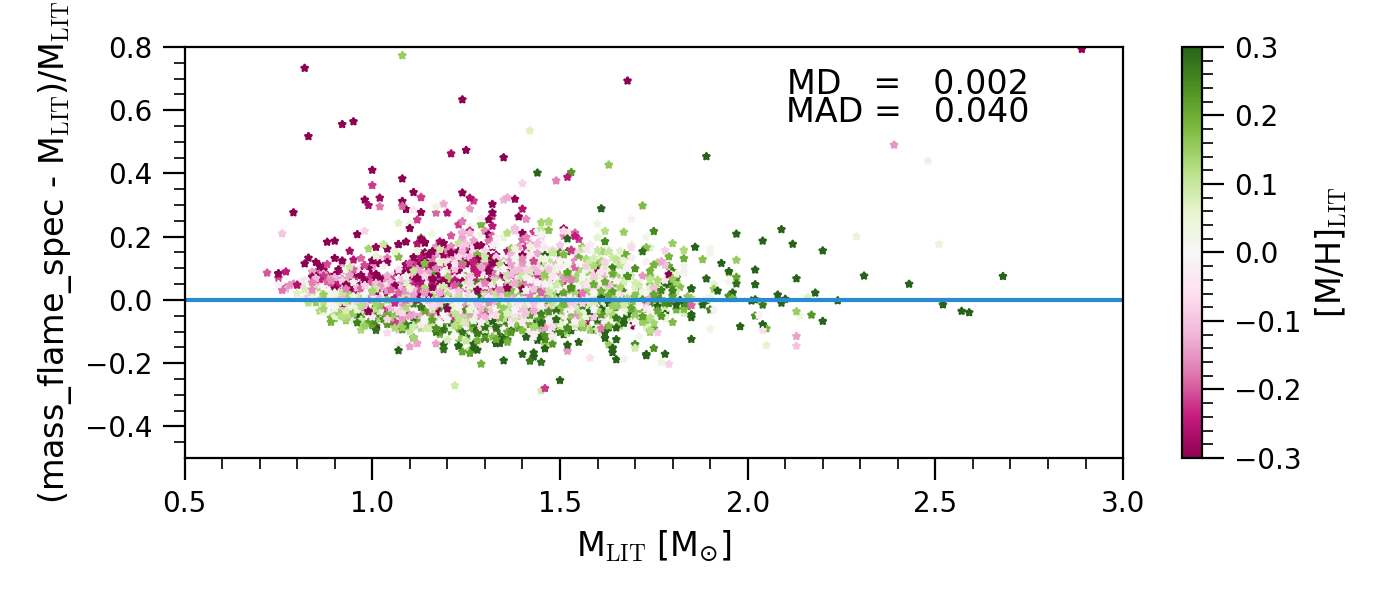}
    \includegraphics[width=0.49\textwidth]{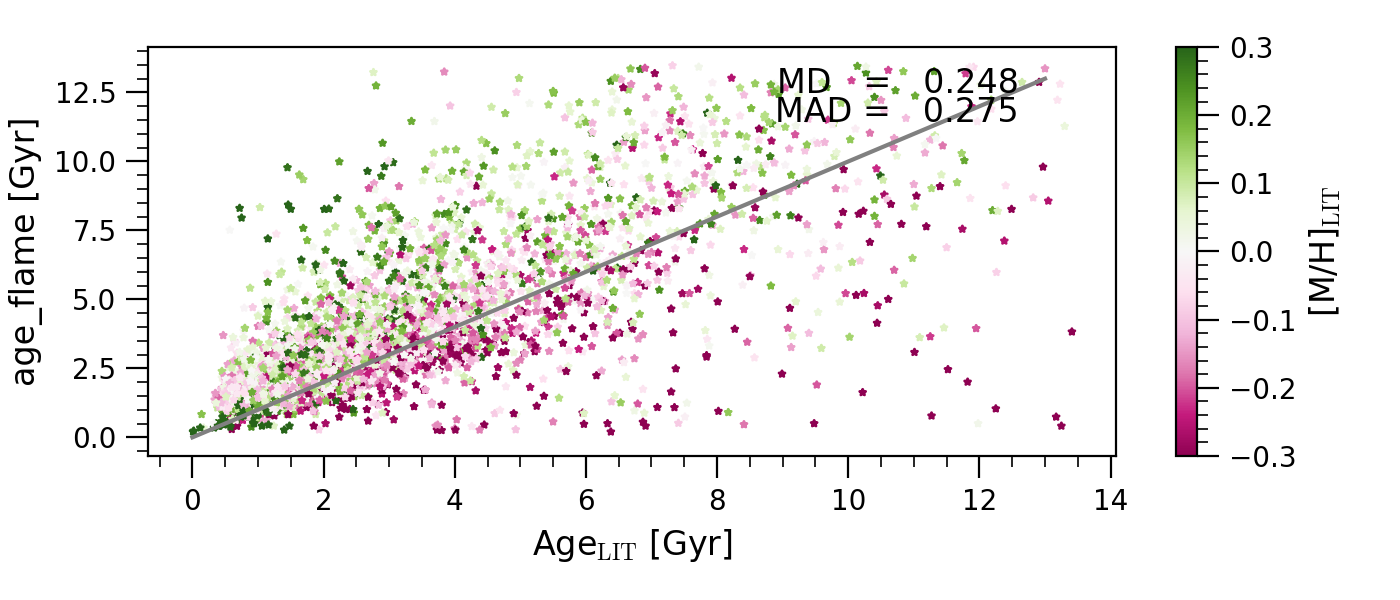}
    \includegraphics[width=0.49\textwidth]{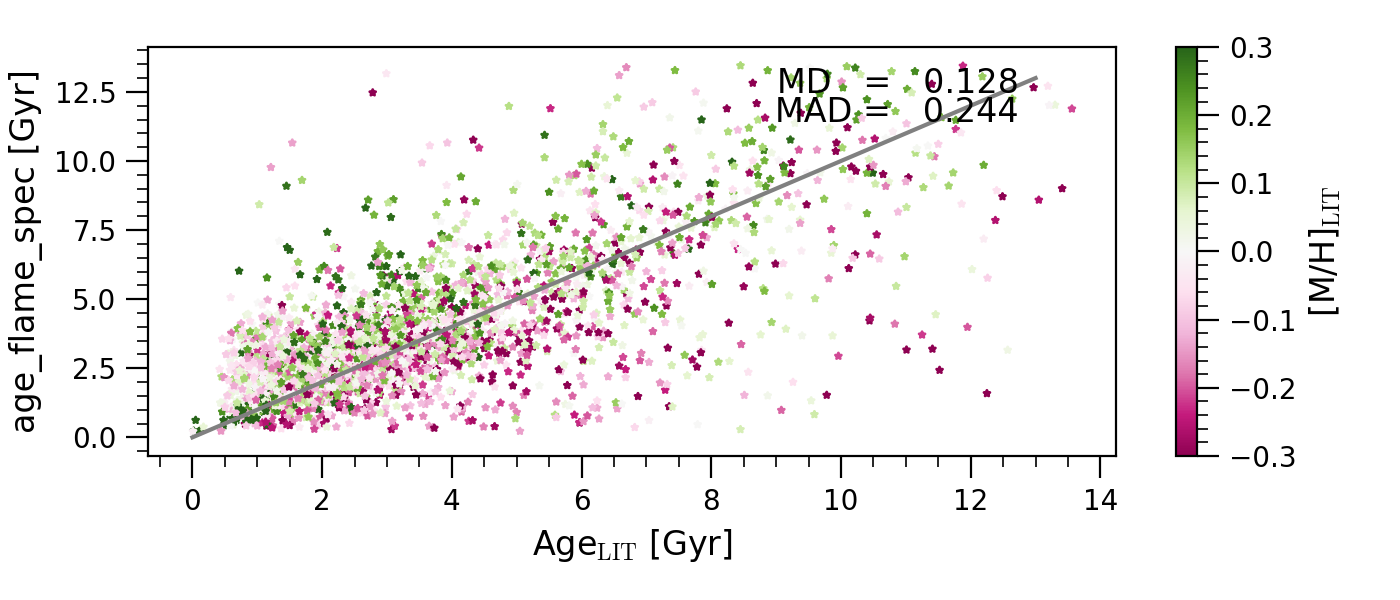}
    \caption{Comparison of \radius, \lum, \mass, and age from \flame\ to literature values. The left and right panels compares the estimates based on \gspphot\ and \gspspec from the \linktotable{astrophysical_parameters} and \linktotable{astrophysical_parameters_supp}, respectively.  The top panel compares \linktoparam{astrophysical_parameters}{radius_flame} and \linktoparam{astrophysical_parameters_supp}{radius_flame_spec} for giants with asteroseismic radii from \citet{2018ApJS..239...32P}. The second panel compares main sequence luminosities \linktoparam{astrophysical_parameters}{lum_flame} and \linktoparam{astrophysical_parameters_supp}{lum_flame_spec} with those from \citet{2017AJ....154..259S} using a random selection of 90\,000 stars. 
    The third panel compares \linktoparam{astrophysical_parameters}{mass_flame} and \linktoparam{astrophysical_parameters_supp}{mass_flame_spec} with masses from \citet{2011yCat..35300138C}, and the bottom panel compares
    \linktoparam{astrophysical_parameters}{age_flame} and \linktoparam{astrophysical_parameters_supp}{age_flame_spec} from that same catalog.}
    \label{fig:lumflame_stevens}
\end{figure*}

\paragraph{Stellar radius.}
From the analysis of the BP and RP spectra, \gspphot\ estimates the stellar radii \linktoparampath{astrophysical_parameters}{radius_gspphot} and the distances \linktoparampath{astrophysical_parameters}{distance_gspphot}.
We validate the ratio of twice the estimated radius to the estimated distance, $2R/d$, by comparing them with interferometric measurements of angular diameters.
Figure~\ref{fig:gspphot_angular_diameters} presents the excellent agreement with the samples from \citet{Boyajian2012a,Boyajian2012b, Boyajian2013, Duvert2016}, and \citet{2021arXiv210709205V}. We note that all of these targets are brighter than $G<9.6$ and more than 90\% of them have high-quality parallaxes with $\frac{\varpi}{\sigma_\varpi}>20$ such that \gspphot\ results should be very reliable \citep{DR3-DPACP-156}.

\flame also provides radii estimates with a different approach based on the APs from either \gspphot or \gspspec combined with the Gaia photometry, and parallaxes.
The top panels in Fig.~\ref{fig:lumflame_stevens} compare \linktoparampath{astrophysical_parameters}{radius_flame} and
\linktoparampath{astrophysical_parameters}{radius_flame_spec} with asteroseismic radii for giants from \citet{2018ApJS..239...32P}. The agreement is at the 1\% level with a scatter of 4\%.  Comparisons with other similar catalogs show agreement at the 1 -- 2\% level, see further comparisons in \linksec{}{online documentation}.

\paragraph{Bolometric luminosity.}
\flame\ estimates the bolometric luminosities, \lum, using bolometric corrections based on \gspphot's  and \gspspec\ APs.
We compared the \lum\ estimates with bolometric fluxes from \citet{2017AJ....154..259S}. We selected a random subset of 90\,000 main-sequence sources with \gdr{3} parallaxes (panels from the second row in Fig.~\ref{fig:lumflame_stevens}).
We found that \linktoparampath{astropphysical_parameters}{lum_flame} and
\linktoparampath{astropphysical_parameters}{lum_flame_spec}  agree well with the literature with a median offset of 2--3\% and a dispersion of around 5--6\%.  We also compared our estimates with other catalogs, such as \citet{2011yCat..35300138C},  with a median offset of  $+0.01$\,\Lsun and similar dispersion.

\begin{figure*}
    \begin{center}
        \includegraphics[width=0.97\textwidth, angle=0]{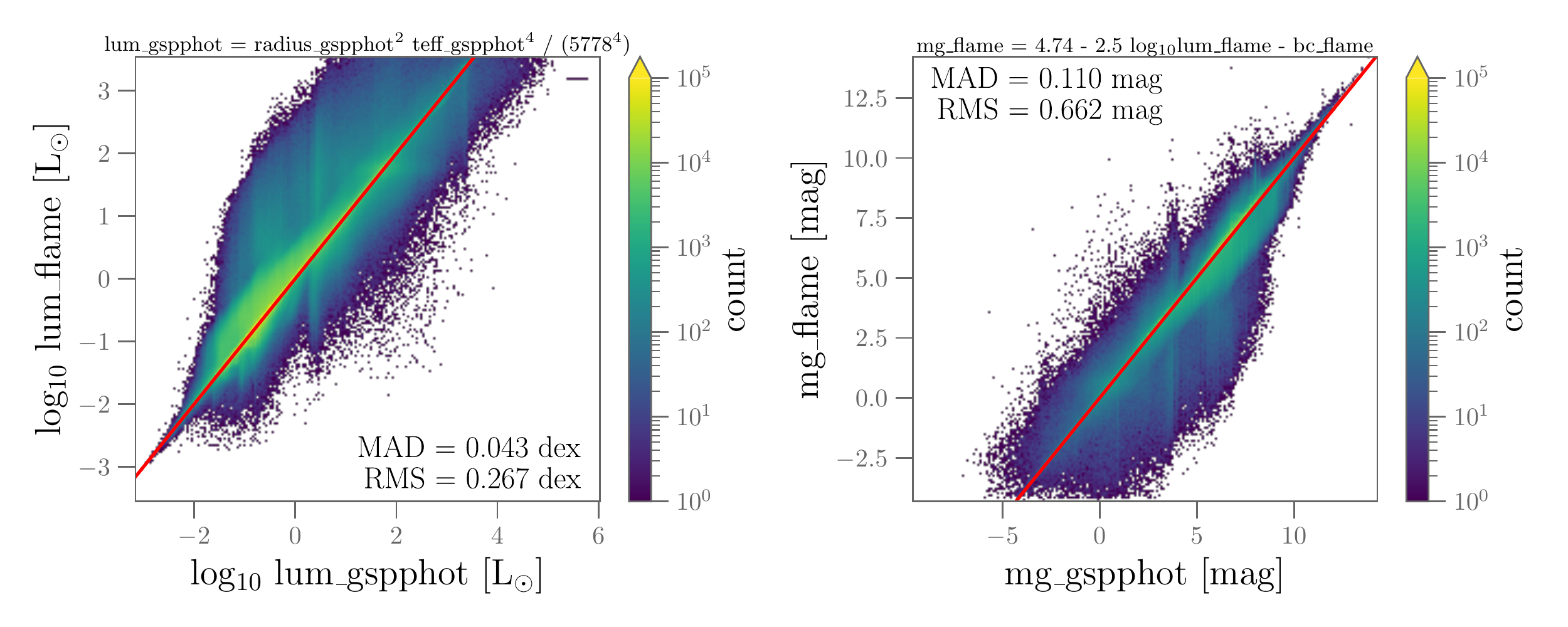}
        \caption{Comparison of luminosities (left) and absolute magnitudes (right) from \gspphot\ and \flame\ for all \gdr{3} sources with estimates from both modules. Numbers quote the median absolute difference (MAD) and the root mean squared error (RMS).
        We indicated the equations we used to construct the luminosities from \gspphot\ from the radius and temperatures, and the absolute magnitudes from \flame\ from the luminosities and bolometric corrections.}
        \label{fig:apsis-FLAME-vs-GSPPhot-MG}
    \end{center}
\end{figure*}

\paragraph{Absolute magnitude \mg.}
\apsis\ provides two sets of absolute magnitude: one from \gspphot\ obtained from the direct analysis of the BP and RP spectra, \gmag\ magnitude (and parallax); one from \flame if we use its luminosity \lum and the bolometric correction as follow:
$M_G = 4.74 -2.5 \log_{10}(L/L_\odot) - BC_{G}$.
Figure~\ref{fig:apsis-FLAME-vs-GSPPhot-MG} compares these two magnitude estimates. We find that most of the stars follow the bisector indicating consistent results. However, we find a median absolute deviation of the order of $0.1$\,mag, and some artifacts. For instance, there are a couple of vertical stripes (e.g., \fieldName{mg_gspphot} = 3\,mag), which could indicate anomalies due to \gspphot's models. In general, we find that \flame\ tends to overestimate luminosity, leading to underestimating \mg, when using parallaxes when fractional uncertainties are on the order of 15-20\%. In contrast, but not surprisingly, we find a stronger agreement when \flame\ use \fieldName{distance_gspphot} than when it uses \fieldName{parallax} as a distance proxy. \linktoparam{astrophysical_parameters}{flags_flame} indicates which distances proxy led to the luminosity estimates.

\begin{figure}
    \centering
    \includegraphics[width=0.45\textwidth]{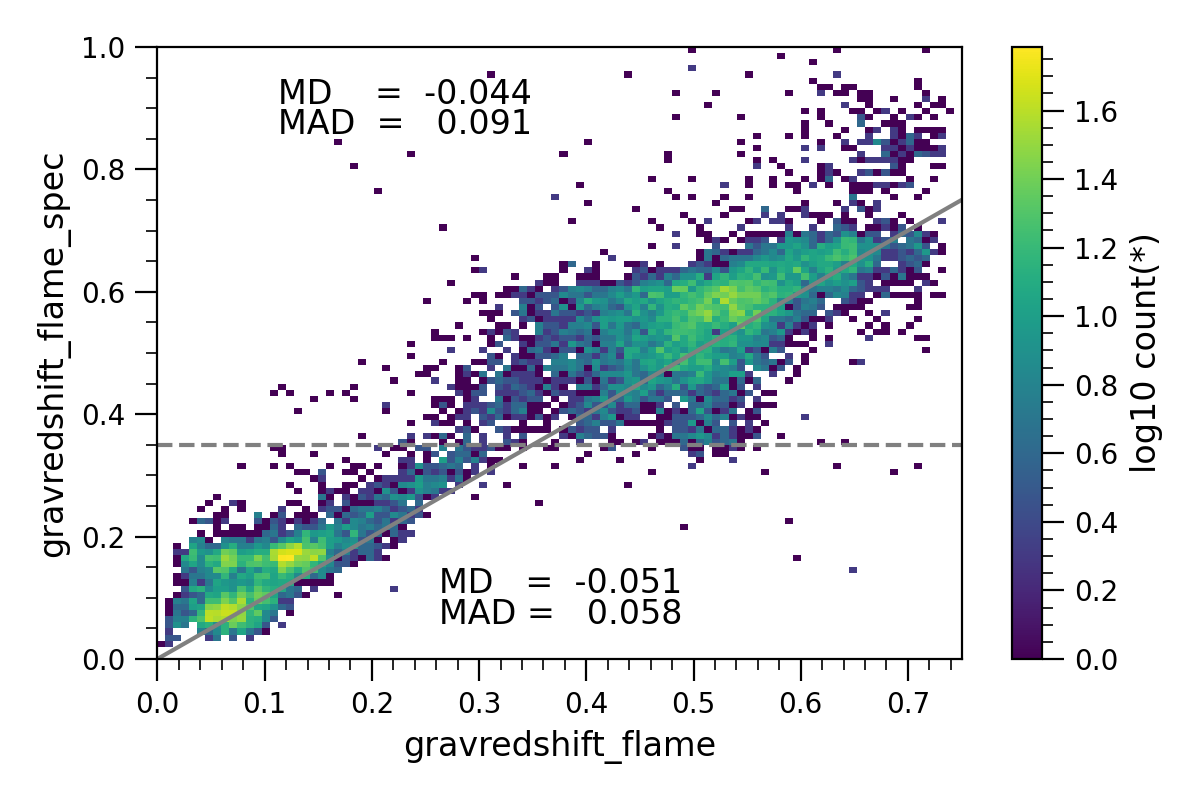}
    \caption{Comparison between \linktoparam{astrophysical_parameters}{gravredshift_flame} and
    \linktoparam{astrophysical_parameters_supp}{gravredshift_flame_spec} expressed in \kms.  The bisector is indicated by the grey solid line.  We indicated the mean offsets (MD) and absolute deviations (MAD) between the two clusters of values above and below $0.35$ \kms.   The differences reflect the different \logg\ values from \gspspec\ and \gspphot\ along with the different radii derived by \flame.}
    \label{fig:gravshift}
\end{figure}

\paragraph{Gravitational Redshift.}
\flame\ produces another model-independent parameter that is the gravitational redshift \gravshift (\linktoparampath{astrophysical_parameters}{gravredshift_flame} and \linktoparampath{astrophysical_parameters_supp}{gravredshift_flame_spec}). Typical values range from 0.05 to 0.8 \kms.
Figure~\ref{fig:gravshift} compares \linktoparam{astrophysical_parameters}{gravredshift_flame} and
\linktoparam{astrophysical_parameters_supp}{gravredshift_flame_spec}
We found a good consistency between the two flavors, with median offset values of $-0.05$ \kms.  This disagreement is a direct reflection of the different input data used to produce the value: \logg\ and \teff\ from \gspspec\ and \gspphot, and \radius\ from \flame.
Additionally, we selected solar-analog stars from a random subset of $2$ million stars from \gdr{3}, those for which \gspphot\ gave \teff\ within $100$\,K, and \logg\ within $0.2$\,dex of the Sun's values. This selection contained $46\,667$ stars, with a mean \gravshift\ of $588 \pm 15$ m~s$^{-1}$, in agreement with the expected value of $600.4 \pm 0.8$ m~s$^{-1}$ for the Sun \citep{2014MNRAS.443.1837R}. We repeated this test for the \gspspec\ based result, and we obtained a mean \gravshift\ of $590 \pm 8$ m~s$^{-1}$. Although the second sample contained only $386$ sources, we also found a good agreement with the known Sun's value.

\subsubsection{Mass, age and evolution stage}\label{ssec:evolution_ma}

\begin{figure}
    \centering
    \includegraphics[width=0.47\textwidth]{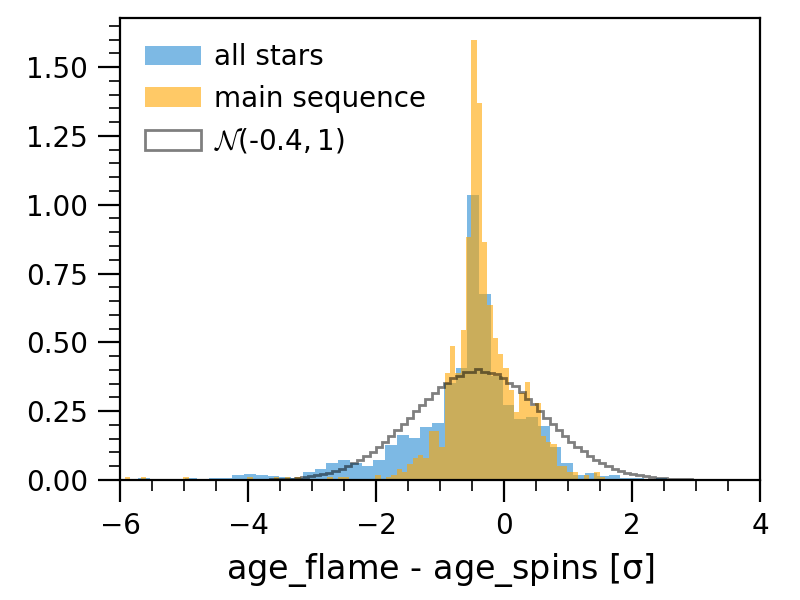}
    \caption{Difference between \linktoparampath{astrophysical_parameters}{age_flame} with the age derived using the SPinS code normalized by their joint uncertainties.  The Gaussian represents the ideal case but centered on the peak difference (--0.4$\sigma$) of the results using all stars irrespective of their evolutionary status.
    The input data are identical, and we assumed a solar-metallicity prior for both codes. We highlighted the sample of MS stars discussed in Sect.~\ref{ssec:evolution_ma}.}
    \label{fig:comparespins_flame}
\end{figure}

\begin{figure*}
    \centering
    \includegraphics[width=0.99\textwidth]{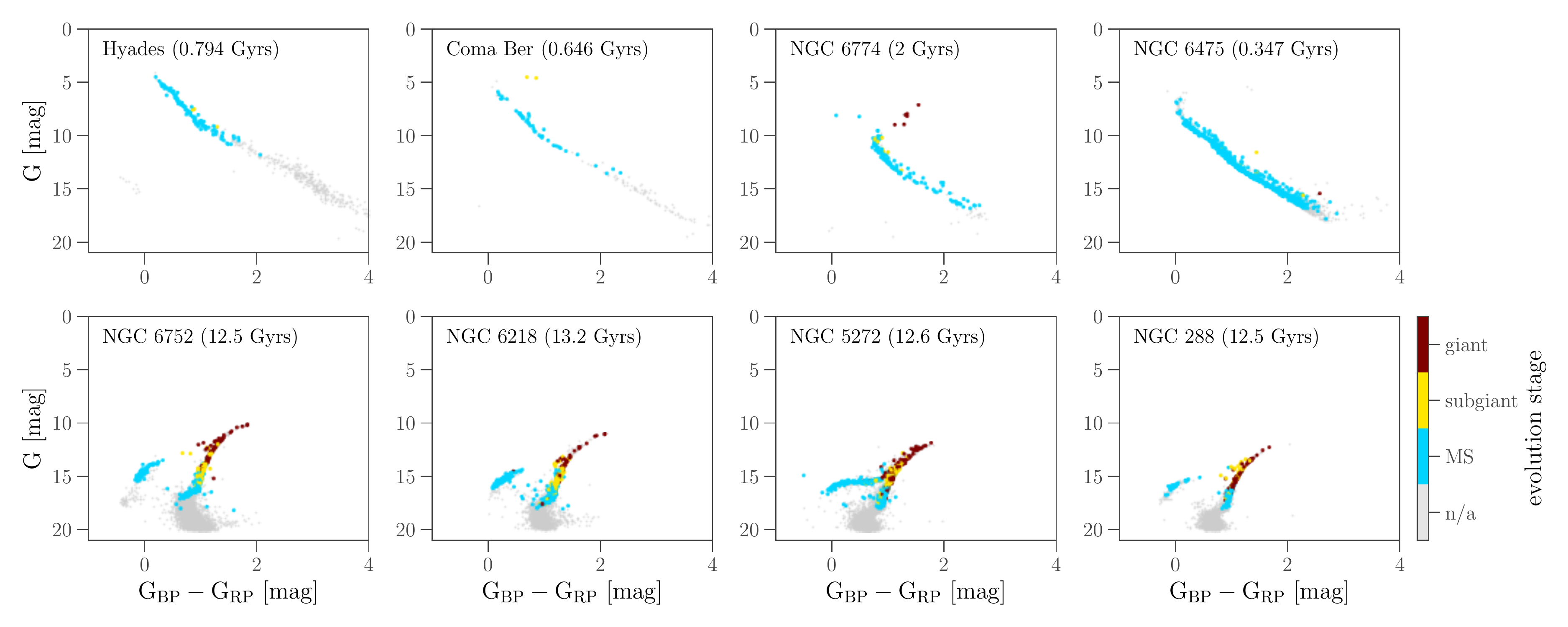}
    \caption{
        Evolution stage distribution from \flame in the CMDs of some selected clusters. The top panels represent the CMDs of four open clusters near solar metallicity, and the bottom panels are those of 4 low-metallicity globular clusters. \citet{2018A&A...616A..10G} provided us with the cluster members. We parsed the values of \linktoparampath{astrophysical_parameters}{evolstage_flame} into the three stages: main-sequence, sub-giant, and giant color-coded according to the scale on the right-hand side. We also indicated in gray other members without phase estimates.}
    \label{fig:cluster_evolstage_flame}

\end{figure*}

This section focuses on the most intrinsic evolution parameters: the mass \mass, age \age, and evolution stage \evolstage. These are unique products of \flame\ (with both \gspphot- and \gspspec-based flavors). These parameters are strongly model-dependent as they directly relate to the stellar evolution models, here the BASTI models \citep{Hidalgo2018}.
In addition, we emphasize that \flame\ assumes solar metallicity during the determination of those parameters. Hence, we recommend using those estimates cautiously for stars with $\mh < -0.5$\,dex.

\paragraph{Stellar masses.} We compared \flame's masses with those from \citet{2011yCat..35300138C} for main-sequence stars (see third panel in Fig.~\ref{fig:lumflame_stevens}). Although we do not expect a significant influence, we note that \citet{2011yCat..35300138C} also used the BASTI models in their analysis, but they used an older version from \citep{basti2004}. We find excellent agreement between the two estimates with a MAD of $0.002$\,\Msun\ with a scatter of $0.042$\,\Msun. Overall, \flame\ produces results comparable to literature results, with some outliers or disagreement with other catalogs that we traced back to the different input \teff or \logg estimates.
In particular one can reduce these outliers for giants if their retrict the \mass estimates (\linktoparam{astrophysical_parameters}{mass_flame}, \linktoparam{astrophysical_parameters}{mass_flame_spec}) to only when (i) $1.0 < \mass < 2.0\,\Msun$ and  (ii) $\age >1.5$\,Gyr.

\paragraph{Stellar ages.}
Overall, we find an agreement between the ages from \flame and the literature for non-evolved stars (i.e., main-sequence stars).
The bottom panel of Fig.~\ref{fig:lumflame_stevens} compares the \linktoparampath{astrophysical_parameters}{age_flame} and \linktoparampath{astrophysical_parameters_supp}{age_flame_spec} with ages from \cite{2011yCat..35300138C}.
In this comparison, we found a mean offset on the order of $0.1$ to $0.3$\,Gyr with a dispersion around $0.25$\,Gyr.
However, it is more delicate to estimate ages for the giant stars reliably because their ages are very dependent on their fitted mass.  In addition, \flame\ only relies on \lum\ and \teff\, to obtain ages and masses, which suffers from significant degeneracies. In addition, ages rely heavily on the solar abundance assumption in the \flame\ processing.

One can trace most differences compared with the literature to the different input \teff\ and \lum estimates. To support this statement, we compared \flame's ages to the ones we obtained with the SPinS public code \citep{spins2020}. We generated random sets of $600$ stars with the SPinS code using the same \gdr{3} APs that \flame\ uses and compared the output ages in four different magnitude intervals.  Figure~\ref{fig:comparespins_flame} compares the estimates with \linktoparampath{astrophysical_parameters}{age_flame}. The agreement for the main sequence stars is always to 1-$\sigma$. The agreement is poorer for the evolved stars, but it remains within 3-$\sigma$ (see \citealt{DR3-TN-OLC-035} for more details).

Section~\ref{sec:groups} presents further analysis of the masses and ages using clusters and further comparisons of mass and age with external data. We also present the analysis of the turn-off ages of some clusters in the online documentation, see {\linksec{}{online documentation}}.

\paragraph{Evolution stage.} The \evolstage\ parameter is an integer that takes values between $100$ and $1\,300$, representing the time step along a stellar evolutionary sequence. To first order, we tag main-sequence stars with values between $100$ and $420$, subgiant stars those between $420$ and $490$, and the giants above as defined in the BASTI models \citep{Hidalgo2018}.
Figure~\ref{fig:cluster_evolstage_flame} represents the evolution stage for members of four open star clusters (top panels; roughly solar metallicity) and four metal-poor globular clusters (bottom panels). We took the system members from \citet{2018A&A...616A..10G}. These clusters were selected to contain a statistically significant number of stars in the three evolution phases estimated from \flame. Overall, the main-sequence and giants evolution stages cover the expected color-magnitude space. Although less numerous, the subgiant evolution stages are consistent with the expected color-magnitude space.
However, we also find discrepancies with expectations due to the stellar models only covering the Zero-Age-Main-Sequence (ZAMS) to the tip of the red giant branch. The bottom panels in Fig.~\ref{fig:cluster_evolstage_flame} clearly show horizontal giant branch stars incorrectly labeled as main-sequence stars. Outside the ZAMS to the tip of the red-giant branch phases, \flame\ labels any star incorrectly. Again, the assumption of solar abundance in \flame\ is challenged in those metal-poorer globular clusters.

As no other module produces \mass, age, or $\epsilon$ parameters, the only other method to assess their quality is to determine their consistency within other open clusters or wide binaries, which we discuss in Sect.~\ref{sec:groups}.

\subsection{Extinction, Dust \& ISM}\label{sec:dust}

\begin{figure*}
    \centering
    \includegraphics[width=0.90\textwidth, clip, trim=0 0 2cm 0]{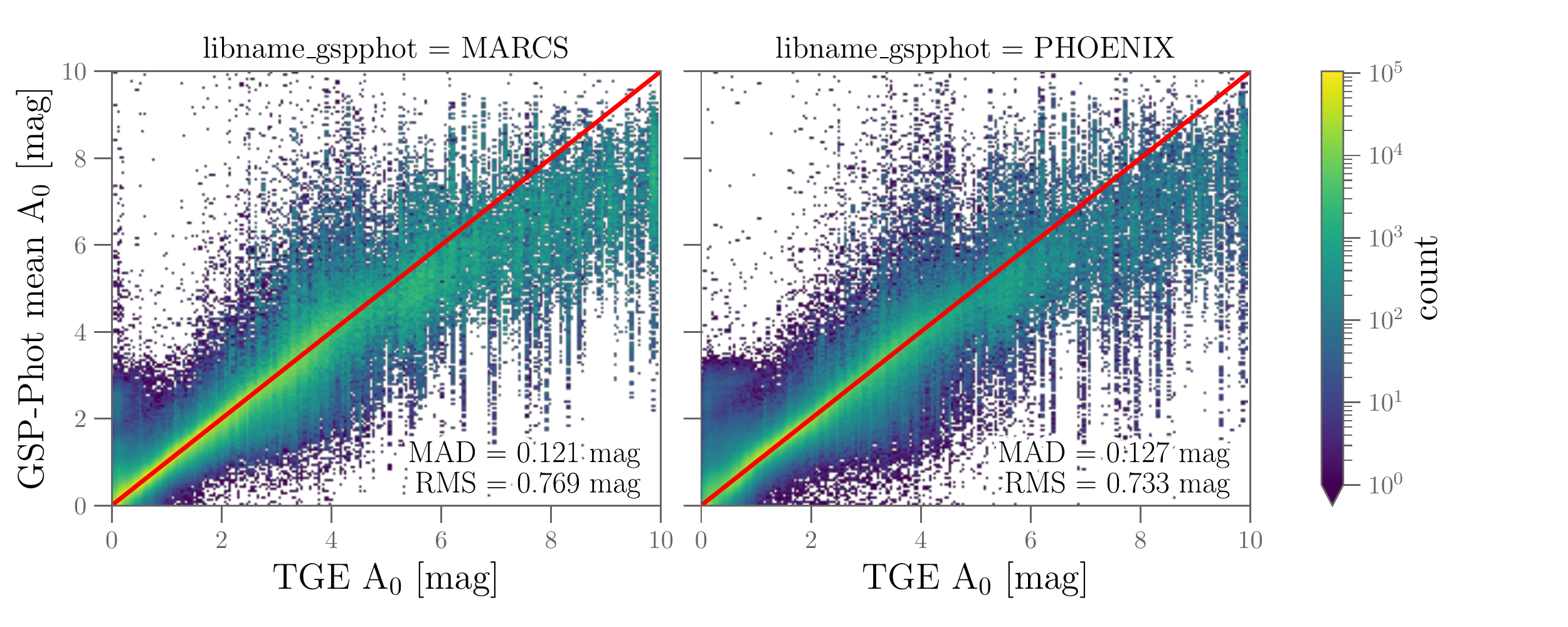}
    \caption{Comparison of \tge\ and \gspphot\ extinction estimates \azero\ limited to giant stars. We calculated the mean extinction \linktoparampath{astrophysical_parameters}{azero_gspphot} per healpix level 9 to compare to \tge\ optimized map \linktoparampath{total_galactic_extinction_map}{a0}. We partially included the \tge\ tracer selection: $3\,000 < $\linktoparam{astrophysical_parameters}{teff_gspphot}$<5\,700$\,Kand $-10 < $\linktoparam{astrophysical_parameters}{mg_gspphot}$<4$\,mag (we did not filter on distances). This represents $21\,244\,458$ and $9\,271\,775$ stars for MARCS and PHOENIX library, respectively.}
    \label{fig:extinction_gspphot_tge}
\end{figure*}

When estimating the intrinsic stellar APs, it is also necessary to consider the effect of interstellar extinction on the observed SED, resulting in an estimation of the line-of-sight extinction for each star. We thus have extinction estimates from \gspphot, \esphs (for hot stars), and \msc (for double stars) as one of the spectroscopic parameters estimated from BP/RP spectra (\azero, \ag, \abp, \arp, \ebpminrp). We also have an independent extinction estimate by \gspspec\ based on the analysis of the diffuse interstellar bands (DIB) (see field details in Table~\ref{tab:product-module-extinction}).

\paragraph{\gspphot.}
For all processed sources, \gspphot\ primarily estimates the monochromatic extinction \azero\ at $541.4$\,nm (\linktoparampath{astrophysical_parameters}{azero_gspphot}) by fitting the observed BP and RP spectra, parallax and apparent $G$ magnitude. However, \gspphot\ also estimates the broadband extinctions \ag, \abp, and \arp, as well as \ebpminrp\ obtained from the models (\linktoparampath{astrophysical_parameters}{ag_gspphot},
\linktoparam{astrophysical_parameters}{abp_gspphot}, \linktoparam{astrophysical_parameters}{arp_gspphot}, and \linktoparam{astrophysical_parameters}{ebpminrp_gspphot} respectively).
Extinction is a positive quantity, thus \gspphot\ imposes a non-negativity constraint on all estimates. Consequently, it can lead to a small systematic overestimation of extinction in truly low-extinction regions ($\azero<0.1$\,mag)\footnote{The mean or median of a positive distribution is always strictly positive, but never null}. \citet{DR3-DPACP-156} demonstrated this effect for the Local Bubble where \gspphot\ estimates a mean extinction of $\azero=0.07$\,mag instead of zero. Yet, a decreasing exponential approximates reasonably well the distribution of \gspphot's \azero\ in the Local Bubble, and it is also the maximum-entropy distribution of a non-negative random variate with a true value of zero. In other words, the exponential is equivalent to a Gaussian noise in more common contexts. Consequently, the exponential's standard deviation (identical to the mean value) of $0.07$\,mag provides an error estimate for \azero. Similarly, \citet{DR3-DPACP-156} reported similar values of $0.07$\,mag for \abp, $0.06$\,mag for $A_G$, and $0.05$\,mag for \arp\ within the Local Bubble. These values are in agreement with \citet{2019A&A...631A..32L} finding a $0.02$\,mag.
While one could allow small values of negative extinctions such that results for low-extinction stars may scatter symmetrically around zero, \citet{DR3-DPACP-156} showed that this is not sufficient in the case of StarHorse2021 \citep{2022A&A...658A..91A}, whose \texttt{av50} in the Local Bubble peaks around $0.2$\,mag twice as much as \gspphot. We found that StarHorse2021 extinction \texttt{av50} estimates appear globally larger than \azero\ from \gspphot\ by $0.1$\,mag, which is likely a bias in the StarHorse2021 catalog \citep[see ][their Fig.~15]{2022A&A...658A..91A}. \citet{DR3-DPACP-156} also observed that in high-extinction regions \texttt{av50} can become significantly larger than \azero. It is currently unclear whether this is an overestimation by StarHorse2021 or an underestimation by \gspphot\ (or both).

\begin{figure}
    \centering
    \includegraphics[width=1\linewidth]{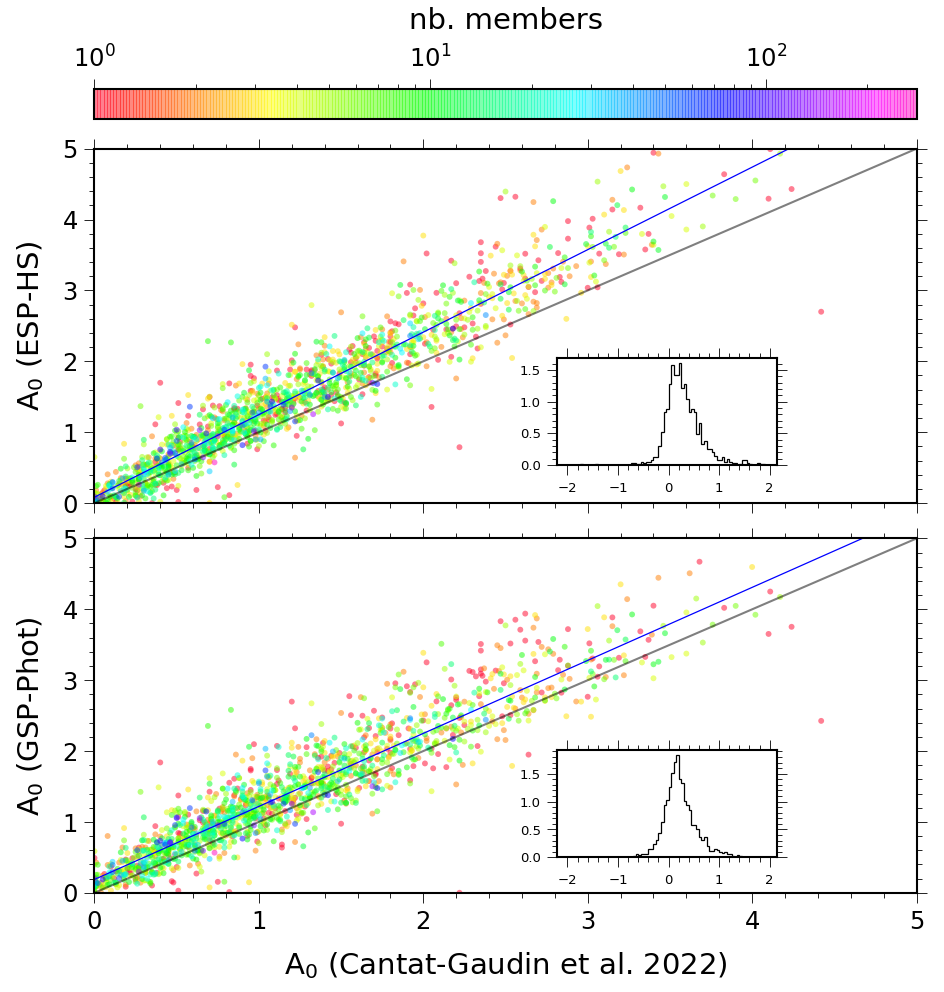}
    \caption{Comparison of \esphs\ (top) and \gspphot\ (bottom) extinction estimates \azero\ for hot stars in star clusters. We used the cluster members and mean extinctions from \citet{2020A&A...640A...1C}.We compute the \gspphot\ and \esphs\ median estimates using stars with $\teff > 7\,500$\,K only.
    We color-coded the data by the number of hot star members with estimates we found in the cluster (w.r.t. colorbar at the top).
    On both panels, the gray lines represent the identity relation, and the blue lines a linear regression through the data points. The insets show the normalized distribution of the differences, \azero(\gspphot or \esphs) - \azero(literature).
}
    \label{fig:extinction_clusters}
\end{figure}

Using Solar-like stars, \citet{DR3-DPACP-123} investigated the $\bpmag-W_2$ color, which uses the \gaia\ and AllWise passbands for two reasons: (i) a color is a quantity independent of distance, and (ii) as the extinction in the AllWISE $W_2$ band is negligible, we can safely associate any correlation to  $\bpmag$ (i.e., a proxy for \abp). We find that the $\bpmag-W_2$ color agrees closely to a linear trend with \gspphot's \abp\ estimate to within $0.087$\,mag RMS scatter, which is consistent with the $0.07$\,mag obtained for \abp\ in the Local Bubble. We also found that the linear relation holds from the low-extinction regimes to high extinctions ones. Additionally, Fig.~\ref{fig:extinction_clusters} shows also good agreement of our \azero estimates with our expectations in open clusters with only a mild overestimation of $\sim 0.1$\,mag (see Sect.~\ref{sec:clusters}).

\paragraph{\tge.}
\gspphot\ also provides the \azero\ estimates used by \tge\ to produce an all-sky (two-dimensional) map of the total Galactic extinction, meaning the cumulative amount of extinction in front of objects beyond the edge of our Galaxy.
\tge\ selects giant star ``tracers'' at the edge of the Milky Way, more specifically, stars with \linktoparampath{gaia_source}{classprob_dsc_combmod_star} $> 0.5$, \linktoparampath{gaia_source}{teff_gspphot} between $3\,000$ and $5\,700$\,K, \linktoparampath{gaia_source}{mg_gspphot} between $-10$ and $4$\,mag, and distances from the galactic plane beyond $300$\,\pc using the \linktoparampath{gaia_source}{distance_gspphot}. Once selected, \tge\ groups the tracers per HEALpix with levels adapting from 6 ($\sim$ 0.08 deg$^2$) to 9 ($\sim$ 0.01 deg$^2$) to have at least 3 stars per group. Finally,  \tge\ estimates \azero\ from the median and standard deviation of the ensemble of \linktoparampath{gaia_source}{azero_gspphot} values per defined HEALpix. We emphasize that \tge\ provides two tables: \linktotable{total_galactic_extinction_map}, which contains the map with a variable HEALpix resolution (\linktoparam{total_galactic_extinction_map}{healpix_level}) and \linktotable{total_galactic_extinction_map_opt}, which contains the resampled information at HEALpix level 9.
It is important to remark that \tge\ primarily uses \linktoparampath{gaia_source}{azero_gspphot}, which contains estimates with a mixture of atmosphere libraries, so-called ``best fit'' estimates.
Figure~\ref{fig:extinction_gspphot_tge} compares the \tge\ estimates to those of \gspphot, for the MARCS and PHOENIX atmosphere libraries providing APs for the giant stars. Although one could expect some AP variations from a set of atmosphere models to another, we find statistically no significant differences between the two libraries and \tge\ estimates. The large dispersion along the y-axis mostly reflects the low numbers of stars beyond 16\,\kpc from the Galactic center, especially with high extinction values. \citet{DR3-DPACP-158} provides a more detailed description of the methodology and performance assessment of the \tge\ maps, especially comparisons with non-stellar tracers (e.g., Planck).

\paragraph{\esphs.} For hot stars with $G<17.65$\,mag, \esphs\ also estimates \azero by fitting the observed BP and RP spectra (\linktoparam{astrophysical_parameters}{azero_espels}). And likewise \gspphot, \esphs\ also provides \ag, and \ebpminrp.
We compared the extinction \azero\ from \gspphot\ and \esphs using star clusters for the hotter stars (Fig.\, \ref{fig:extinction_clusters}). Both modules find consistent \azero\ estimates when deriving extinctions greater than {0.3}\,mag. However, over this hot star sample, we find that \gspphot\ tend to overestimate extinction by about $0.1$\,mag constantly, and \esphs\ overestimate by a factor $1.2$. Overall, for all stars with \gspphot\ and \esphs\ estimates, we found a MAD of $0.120$\,mag, and RMS of $0.380$\,mag. However, we emphasize that these differences, esp. RMS statistics, also vary with the spectral libraries (\linktoparampath{gaia_source}{libname_gspphot} or \linktoparampath{astrophysical_parameters}{libname_gspphot}). If we restrict the comparison to the OB star library that best describes this temperature regime, we found an improved RMS of $0.170$\,mag. Hence this illustrates the importance of choosing or exploring which spectral library is appropriate for the sources of interest.

\paragraph{\msc.} \msc also estimates the \azero\ parameter by assuming that the BP and RP spectra represent a composite of an unresolved binary: two blended coeval stars at the same distance (\linktoparam{astrophysical_parameters}{azero_msc}). \msc's performance is similar to \gspphot (see Sect.~\ref{sec:unresolved_binaries}).

\paragraph{\gspspec-DIBs} In addition to the stellar APs, \gspspec\ estimated the equivalent width of diffuse interstellar bands (DIBs) in the RVS spectra for $476\,117$
stars in \gdr{3}. The DIB spectral feature arises from largely unidentified molecules ubiquitously present in the interstellar medium (ISM).
\gspspec measures the DIB profile parameters: the equivalent width (\linktoparampath{astrophysical_parameters}{dibew_gspspec}) and characteristic central wavelength (\linktoparampath{astrophysical_parameters}{dib_gspspec_lambda}) using a Gaussian profile fit for cool stars and a Gaussian process for hot stars. We described in detail the DIB measurements procedure in \citet{DR3-DPACP-186} (Sect.~6.5) and further assessed the performance of those in \citet{DR3-DPACP-144}. We emphasize that one should restrict themselves to using the DIB estimates with quality flags \linktoparampath{astrophysical_parameters}{dibqf_gspspec} $\leq 2$ (Definition in Table~2 of \citealt{DR3-DPACP-144}).
Although one can question the standard analysis in this field, we applied the approach to compare our results with the literature. We estimated a linear relation between \linktoparam{astrophysical_parameters}{dibew_gspspec} and \linktoparam{astrophysical_parameters}{ebpminrp_gspphot} as
\begin{equation}
\ebpminrp = 4.508 (\pm 0.137) \times EW_{862} - 0.027 (\pm 0.047).
\end{equation}
We identified the strong outliers to this relation as having an overestimated \ebpminrp\ from \gspphot (linked to an incorrect temperature estimate; see \citealt{DR3-DPACP-144}).
\gspspec\ also measured DIBs for hot stars ($\rm T_{eff} > 7500\,K$), providing us with a total of $1\,142$ high-quality DIB measurements. We compared these with the extinction estimates from \esphs (\linktoparampath{astrophysical_parameters}{ebpminrp_epshs}) and found an excellent agreement with the relation we obtained above (see Fig.~9 of \citealt{DR3-DPACP-144}).
We further compared the DIB EW with the \azero values of the \tge\ HEALPix level 5 map (\linktotable{total_galactic_extinction_map}) where we found a strong linear correlation given by  $\rm EW = 0.07 \times \azero + 0.03$ up to $\azero \sim 1.5$\,mag after which we found a shallower trend. We suspect the slope change originates from \tge\ providing total extinction far beyond the distance of stars with DIB\,$\lambda$862 measurements.

Finally, we estimated the standard quantity E(B--V)/EW of $\rm 3.105\pm0.048$, which lies in the range of the derived ratios in the literature (Compilation in {Table~3 in \citealt{DR3-DPACP-144}}).

\subsection{Groups of stars}\label{sec:groups}

\subsubsection{Clusters}\label{sec:clusters}

\begin{table}
    \caption{Parameter residuals to reference values in star clusters. \label{tab:cu8par_Clusters}}
    \begin{center}
        \begin{tabular}{ccccc}
            \hline
            Parameter   & Module   & M$^{(1)}$     & MAD$^{(2)}$  & Units \\
            \hline\hline
            \teff       & \gspphot & 34    & 400  & K     \\
            \teff       & \gspspec & 6     & 160  & K     \\
            \logg       & \gspphot & 0.01  & 0.22 & dex   \\
            \logg       & \gspspec & -0.30 & 0.44 & dex   \\
            \ag         & \gspphot & 0.12  & 0.10 & mag   \\
            \mass(spec) & \flame   & -0.02 & 0.10 & \Msun \\
            \mass(phot) & \flame   & -0.11 & 0.14 & \Msun \\
            \age(spec)  & \flame   & -0.40 & 0.60 & Gyr   \\
            \age(phot)  & \flame   & 0.30  & 0.40 & Gyr   \\
            \hline
        \end{tabular}
    \end{center}
    $^{(1)}$: median estimates of the residuals;\\
    $^{(2)}$: mean absolute deviation (MAD) of the residuals.\\
    \linktoparam{astrophysical_parameters}{flags_gspspec} with \texttt{f1,f2,f4,f5,f8}=0.
\end{table}

\begin{figure*}
    \begin{center}
        \includegraphics[width=0.9\textwidth]{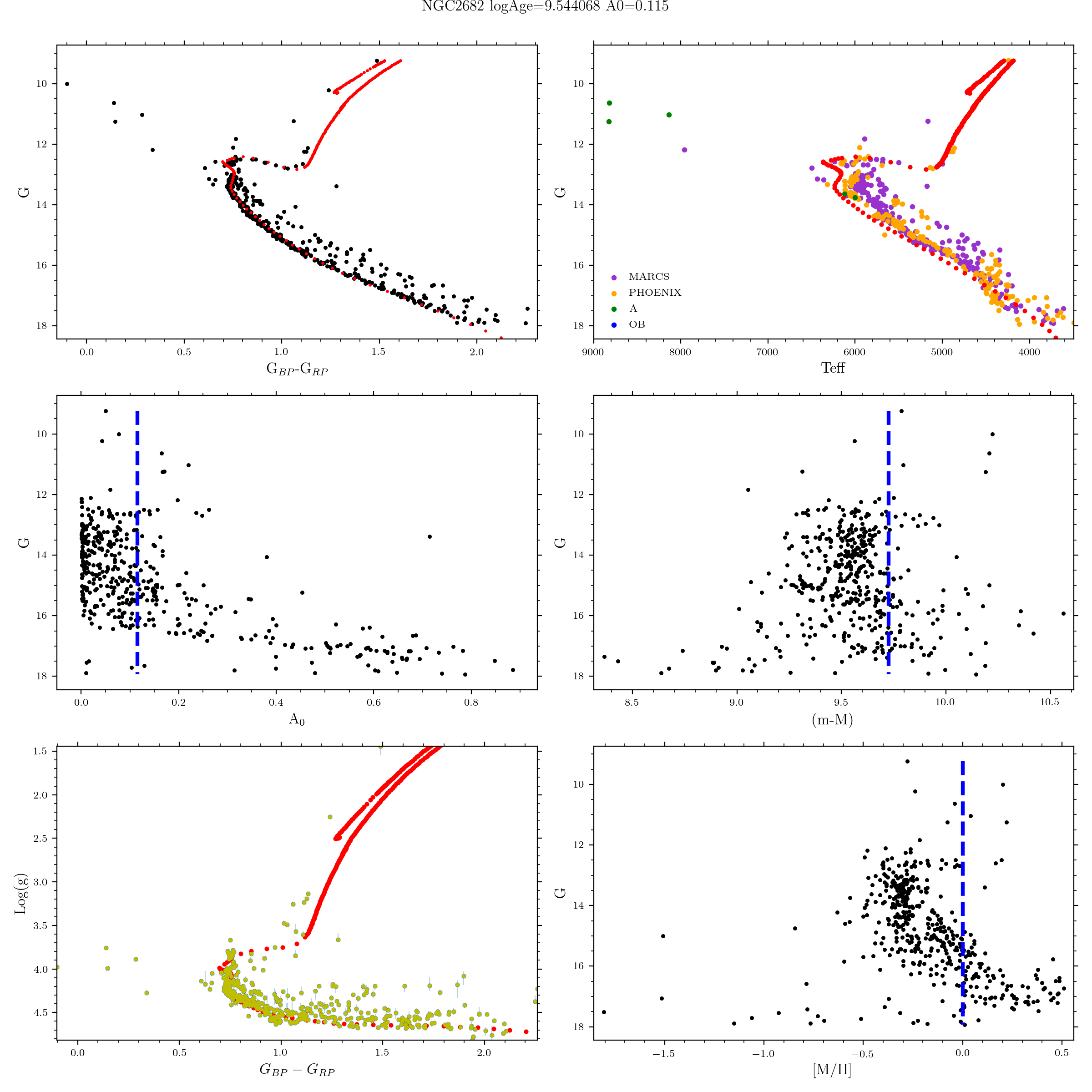}
        \caption{Illustration of the performances of \gspphot in the Messier\,67 cluster.
        The PARSEC isochrone (indicated by the red dots in the panels) indicates the reference of $\log_{10}(Age/yr)=9.5$, $\azero=0.11$\,mag, and a solar metallicity.
        We also indicate with the vertical dashed blue lines the reference \azero, distance modulus, and \mh.
        Top-left panel: the Gaia CMD of M67.
        Top-right panel: \gmag\ vs. \linktoparampath{gaia_source}{teff_gspphot}. The colors indicate the different ``best'' libraries (\linktoparampath{gaia_source}{libname_gspphot}).
        Middle-left panel: \gmag\ vs. \linktoparampath{gaia_source}{azero_gspphot}.
        Middle-right panel: \gmag\ vs. the distance modulus derived from \linktoparampath{gaia_source}{distance_gspsphot}.
        Bottom-panel left: \linktoparampath{gaia_source}{logg_gspphot} vs. \bpmag-\rpmag (yellow dots).
        Bottom-right panel: \gmag\ vs. \linktoparampath{gaia_source}{mh_gspphot}. }
        \label{fig:M67}
    \end{center}
\end{figure*}

Star clusters are very effective in assessing the stellar parameters' qualities, as proven in previous Gaia data releases. Open star clusters are coeval populations: same age, same metallicity, about the same extinction, and distance.

\apsis\ processed all the stars independently and, in particular, did not exploit the coevolution of stars. This section presents some of the key results concerning the global quality of the APs in star clusters. We provide additional validation, known issues, some calibration relations, and the optimal use of the quality flags in \citet{DR3-DPACP-156}, \citet{DR3-DPACP-186} and \citet{DR3-DPACP-127}.

We selected a sample of star clusters from the \citet{2020A&A...640A...1C} catalog.
\citet{DR3-DPACP-75} refined the cluster memberships using \gedr{3} astrometry. Our selection corresponds to about $230,000$ stars: the number of stars per cluster varies significantly from 40 to more than 700 with an average of $\sim 60$ stars. Open clusters contain mostly main-sequence stars with a median G=$15.6$\,mag but their populations significantly vary with the ages of the systems.
We approximated the stellar population of each cluster by an isochrone to obtain reference estimates for \teff, \logg, mass, age, and distance. Additionally, we assumed homogeneity throughout the color-magnitude diagram of \azero and \mh.
For the former, we avoid where differential extinction is more likely to be present by excluding clusters younger than $100$\,Myr from our samples. We use the PARSEC isochrones \footnote{PARSEC isochrones available from \url{http://stev.oapd.inaf.it/cgi-bin/cmd}.} for this purpose, associated with the clusters' age, distance, extinction, and metallicity from our literature catalog. Here, we summarized the statistical analysis of the accuracy of the relevant APs over the cluster members.

We compare the atmospheric and evolution APs from \gspphot, \gspspec, and \flame\ to the cluster isochrones. We emphasize that when analyzing the \gspspec\ results, we selected the stars having \linktoparampath{astrophysical_parameters}{flags_gspspec} with \texttt{f1,f2,f4,f5,f8}=0.
Table~\ref{tab:cu8par_Clusters} presents the median, and MAD of the residuals to the isochrones for \teff, \logg, \ag, \mass, and \age\ derived by \gspphot\, \gspspec\, and \flame. We note that we compared \ag\ and \age\ with the literature values independently of the isochrones.

\begin{figure*}
    \begin{center}
        \includegraphics[width=0.49\textwidth]{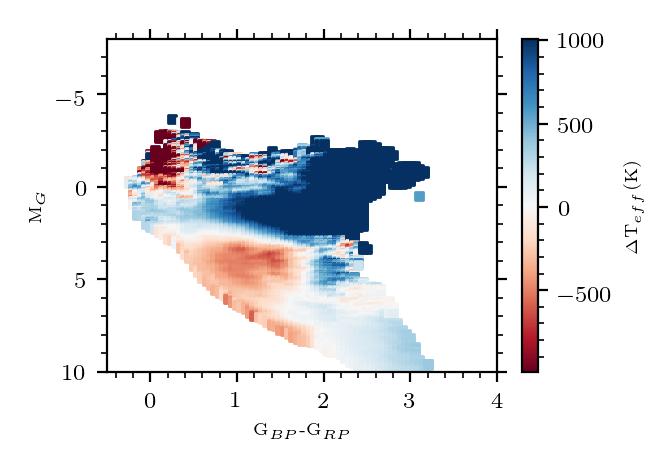}
        \includegraphics[width=0.49\textwidth]{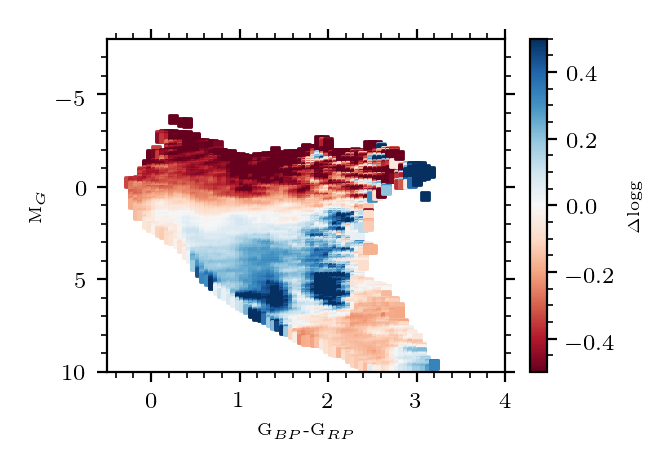}
        \caption{Residuals of \teff\ and \logg estimates from \gspphot to isochrones of star clusters for the sample described in Sect.~\ref{sec:clusters}. We show the mean residuals of the members as a function of position in the \mg\ vs. (\bpmag-\rpmag) diagram. The y-axis is corrected for extinction and distance modulus using literature values.  The color indicates $\Delta(\teff)$ and $\Delta(\logg)$ on the left and right panels, respectively.}
        \label{fig:GspphotClustersTeff}
    \end{center}
\end{figure*}

\paragraph*{\gspphot.}
We found that \teff, \logg, \ag\ from \gspphot\ are in general agreement with expectations, albeit sometimes large dispersions. It is important to note that we analyzed the ``best'' library estimates (e.g., \linktoparampath{astrophysical_parameters}{teff_gspphot}), but the results may vary with different choices of library (e.g., \linktoparampath{astrophysical_parameters_supp}{teff_marcs_gspphot}).
\gspphot\ performs better for $\gmag<16$\,mag, where the SNR of the BP/RP spectra remains high (SNR > 100).
Figure~\ref{fig:M67} illustrates our analysis with the example of Messier\,67 (aka NGC2682). In this cluster, we found 4\% of outliers defined as $\Delta{\teff}/\teff > 0.5$. But this fraction varies across the entire \gdr{3} sample.
Overall, we identified that \gspphot\ overestimated \teff\ values for giants, and underestimated them for supergiants (see Fig.\ref{fig:GspphotClustersTeff}).
In the details, we find that the distribution of the \gspphot's \logg\ values has a long tail towards overestimating values on the main sequence. Still, in contrast, \gspphot\ underestimates gravity for hot stars and giants.
We also note the issue with metallicity and the extinction estimates reported in Sect.~\ref{sec:atmosphere-primary}. Messier 67 is at $\sim 850$\,pc from us, a close distance that \gspphot's a priori assumes mostly free of extinction. This prior leads to underestimating the reddening of these stars. As a result of preserving the observed stellar SEDs, \gspphot\ underestimate \mh. \cite{DR3-DPACP-156} discuss this extinction-distance prior and related issues in detail.

\paragraph*{\gspspec.}
We also analyzed \gspspec's APs and found that \logg\ from \gspspec\ could show biases up to $-0.3$\,dex compared to isochrone predictions (similarly to Sect.~\ref{sec:atmosphere-primary}). In particular, we found significant underestimation for hot stars, and we caution the user against using \gspspec's \logg\ values for AGBs as we find them of poorer qualities. We refer to \citet{DR3-DPACP-186} for the details and especially emphasize that these comparison results depend strongly on the quality flag selection.   \citet{DR3-DPACP-186} also encouraged the user to define calibration relations for their specific use-cases.

\paragraph*{\flame} We also found that the \flame\ APs are in good agreement when we restrict our analysis to the best-measured stars, those with \linktoparampath{astrophysical_parameters}{flag_flame}=00,01.
The fact that \flame\ assumes solar metal metallicity produced poor \age\ and \mass\ estimates in low metallicity clusters, unsurprisingly. However, in the solar metallicity regime, \mass\ is in good agreement with expectations (see Table~\ref{tab:cu8par_Clusters}).
It also seems that \flame\ overestimated \age for young stars and underestimates for old stars, with the most significant discrepancies with the literature appearing for cool main-sequence stars.

\begin{figure}
    \begin{center}
      \includegraphics[width=0.49\textwidth]{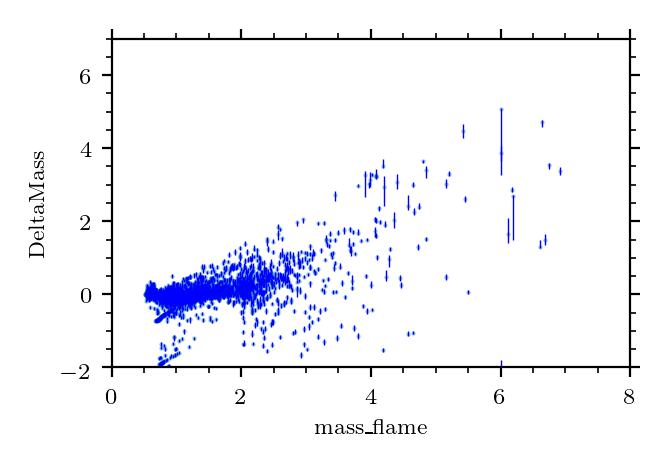}
       \caption{Residuals, Delta(Mass) in \Msun, between \linktoparam{astrophysical_parameters}{mass_flame} and the isochrone predictions for star clusters taken from \citet{2020A&A...640A...1C}. We selected estimates with  \linktoparam{astrophysical_parameters}{flag_flame}=00. Error bars indicate the \flame's uncertainties.}
       \label{fig:FlameClustersMass}
    \end{center}
\end{figure}

Using star clusters also has the advantage of assessing if the reported uncertainties are overall of the correct order.
\flame\ reported underestimated uncertainties on \mass\ and \age derived either from \gspspec\ or \gspphot\ APs. Figure~\ref{fig:FlameClustersMass} demonstrates that the \mass residuals between \gspphot\ and the isochrones disperse significantly more than the uncertainties (on average of the size of the symbols.).

\begin{figure}
    \begin{center}
        \includegraphics[width=0.49\textwidth]{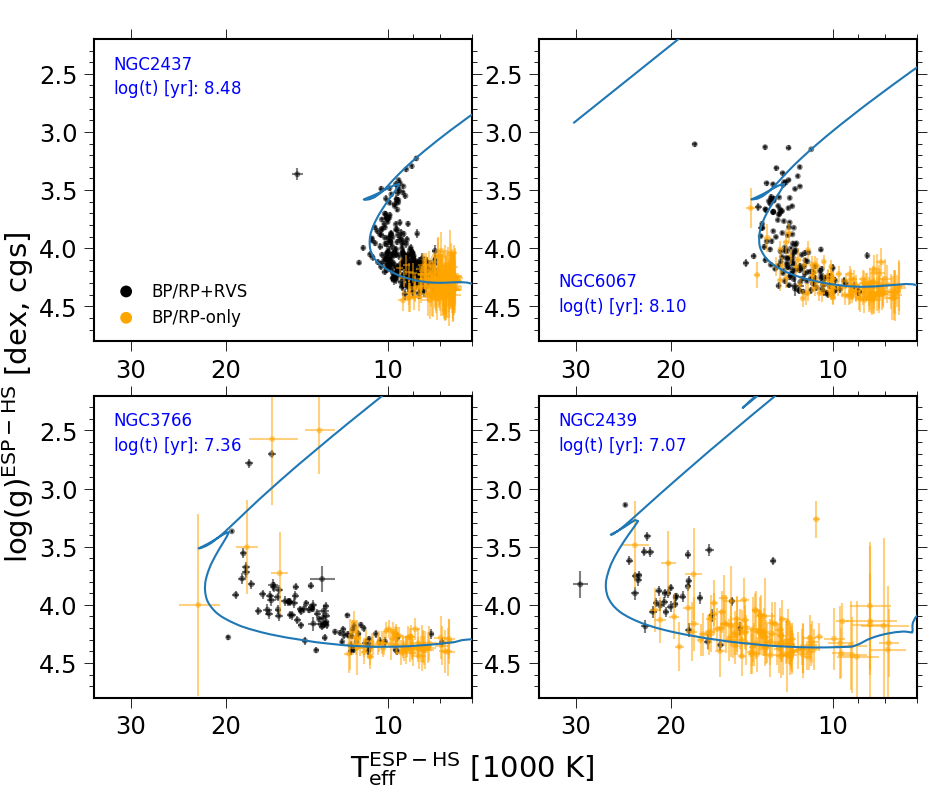}
        \caption{Cluster Kiel diagrams of ESP-HS astrophysical parameters. The PARSEC isochrones (blue) are those that correspond to the cluster age provided by \citet{2020A&A...640A...1C} assuming a Solar metallicity. Estimates obtained in both ESP-HS processing modes are shown by black and orange disks with their corresponding uncertainties.}
        \label{fig:esphsisochrones}
    \end{center}
\end{figure}

\paragraph*{\esphs.} A fraction of the OBA stellar population in open clusters went through the analysis by the \esphs\ module. Figure~\ref{fig:esphsisochrones} illustrates a selection of cluster Kiel diagrams at different ages. By comparison to the PARSEC isochrones, we found estimates commonly to the right of the isochrones in the Kiel diagrams, suggesting somewhat older cluster ages than the literature references. Such findings may relate to a systematic underestimation of \teff and \logg. Although unlikely, the literature may underestimate the clusters' ages. Still, more likely, our results may suffer from gravitational darkening due to axial rotation on the spectral energy distribution of OBA stars.

\paragraph*{\espucd.} As we detailed in \linksec{ssec:cu8par_apsis_espucd_results}{the online documentation}, \espucd detects significant overdensities at the positions of several clusters and star-forming regions. We used the BANYAN $\Sigma$ \citep{2018ApJ...856...23G} to identify UCD members of nearby young associations within $150$\,pc from the Sun. Table \ref{tab:espucd_clusters} contains the number of sources with membership probability greater than $0.5$ in each association and the effective temperature of the coolest UCD. We also include entries for associations beyond $150$\,pc derived from our clustering analysis using the OPTICS algorithm \citep{Ankerst99optics:ordering} in the space of Galactic coordinates, proper motions, and parallax. We did not use these stars to assess the performance of \espucd, but we reported our strong UCD candidates.

\subsubsection{Unresolved binaries}\label{sec:unresolved_binaries}

In \apsis, the \msc\ module aims to distinguish between the two components of binaries by analyzing their composite BP/RP spectra. It assumes these sources are blended coeval stars (same distance, extinction, and metallicity). We could not create sufficiently high-quality synthetic models of BP and RP spectra of unresolved binaries; these could not fully model these sources' instrumental (and data reduction) effects. Instead, \msc\ implements an empirical set of models constructed from observed BP and RP spectra of spectroscopic binary stars\citep[see][for details]{DR3-DPACP-157}. As a result of the limited number of unresolved binaries for reference with APs, \msc\ adopted a strong \mh\ prior centered on solar values.

\begin{figure}
    \begin{center}
        \includegraphics[width=0.49\textwidth, angle=0]{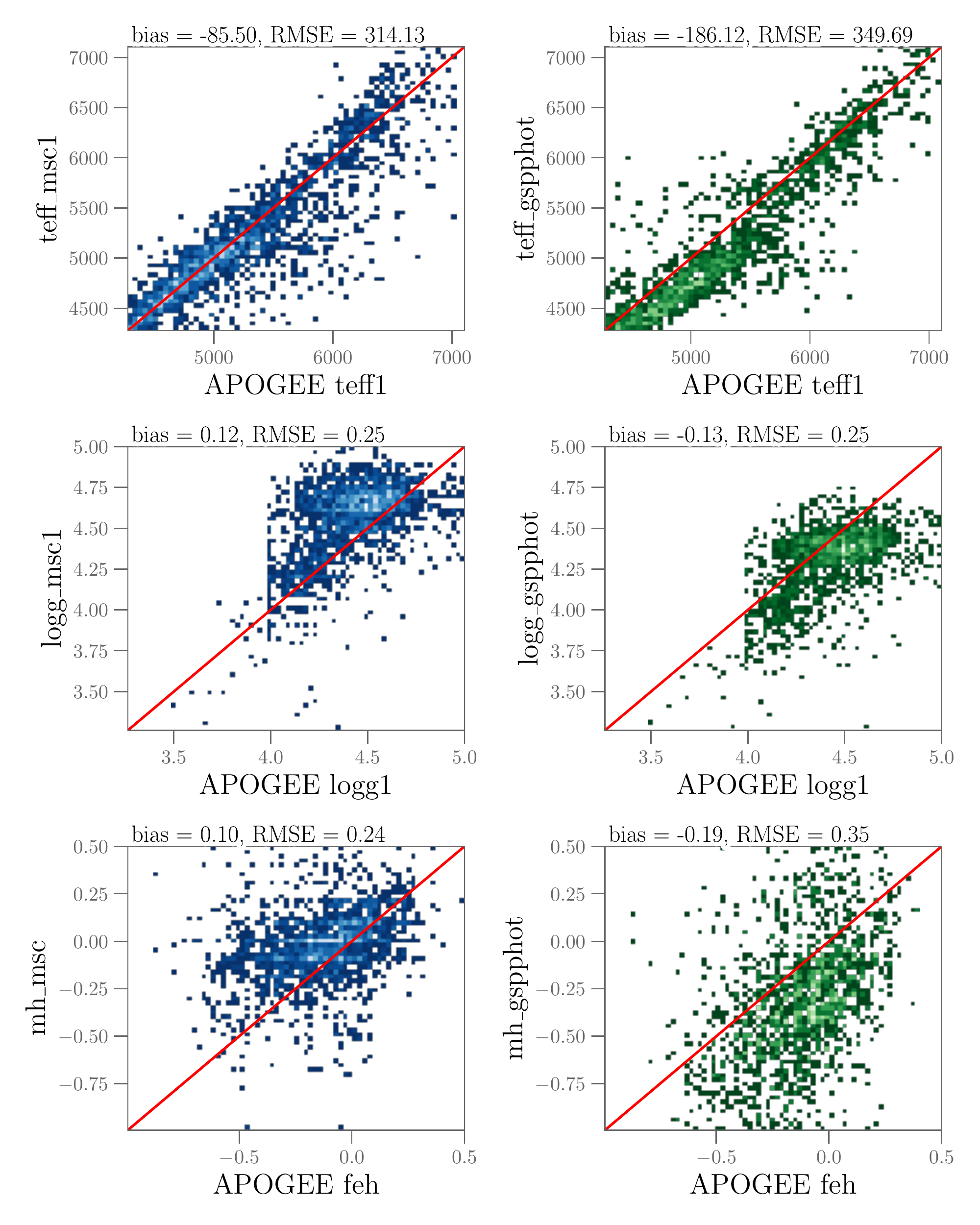}
        \caption{\msc\  and \gspphot\  inferred values on the y-axis vs. APOGEE\citep{2018MNRAS.476..528E} literature values on the x-axis for sources with common parameter range (including flux ratio smaller 5). Shown are the 3 parameters \teff, \logg, and \mh\ with their respective 1:1 line in orange. We applied the \gspphot\ postprocessing, a cut on \texttt{fidelity\_v2} $>0.5$ \citep{Rybizki2021} and a cut on \texttt{logposterior\_msc} $>-1000$.
            \label{fig:cu8par_apsis_msc_gspphot_apogee}
        }
    \end{center}
\end{figure}

\msc\ analyzes all sources with $\gmag<18.25$\,mag and therefore inherently analyzes single stars as well (assuming a binary source). Similarly, \gspphot\ takes all sources to be single stars. As internally \msc\ operates very similarly to \gspphot, we can compare their overlapping results more robustly than any other \apsis\ module.
Figure\,\ref{fig:cu8par_apsis_msc_gspphot_apogee} compares APs from \msc\ and \gspphot\  parameters with those from the binary sample of \citet{2018MNRAS.476..528E}. It is not surprising that we find a negative bias in temperature and \logg\ from \gspphot since it assumed these sources are single stars. These correspond to a luminosity-weighted average between the primary and the secondary. Commonly, this leads to a lower \teff\ and \logg to reach the observed brightness of the binary system with a single star. We find that despite its strong solar metallicity prior, the posterior of \mh from \msc\ are broad. Overall \msc\ is performing better than \gspphot\ on this particular sample of binaries.

\begin{figure}
    \begin{center}
        \includegraphics[width=0.45\textwidth]{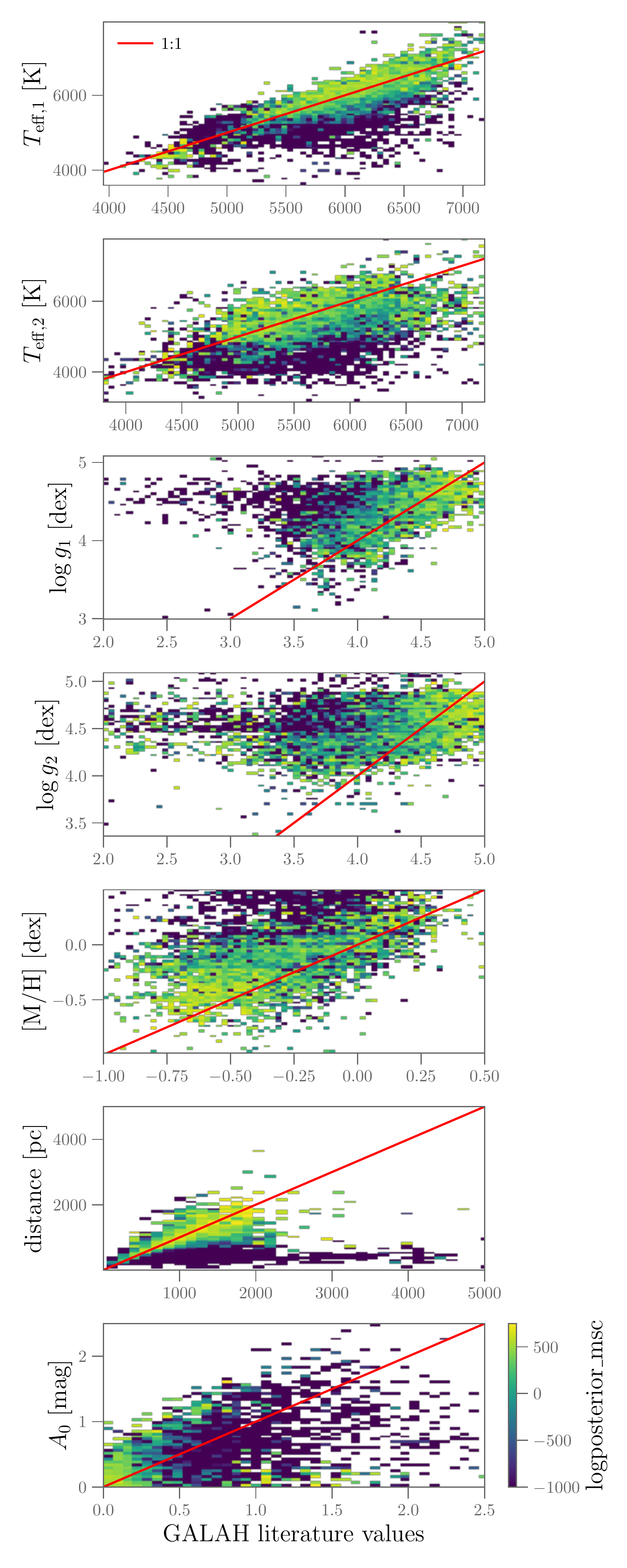}
        \caption{Comparison of \msc's APs with the GALAH catalog for $11\,263$ binary stars. From top to bottom, we compare the seven parameters inferred by \msc on the y axes with the GALAH literature values on the x axes: \teff and \logg for both components of the system, \mh, distance, and \azero. (Table~\ref{tab:product-module-atm-1} for the corresponding catalog fieldnames).
        On each panel, we indicate the 1:1 line for reference and color corresponds to the average \linktoparampath{astrophysical_parameters}{logposterior_msc} of all stars per bin. We provide associated statistics in Table~\ref{tab:cu8par_apsis_msc_Galah}.
            \label{fig:cu8par_apsis_msc_galah_binary_parameters}
        }
    \end{center}
\end{figure}

The GALAH survey \citep{2017MNRAS.465.3203M} provides another set of $11\,263$ spectroscopic binaries \citep{2020A&A...638A.145T} with a component flux ratio of less than $5$ (i.e., within the \msc\  parameter ranges).
As above, we compared \msc\ with \gspphot\ on this sample and we find their APs have comparable accuracies.
Figure\,\ref{fig:cu8par_apsis_msc_galah_binary_parameters} compares the seven APs from \msc\ with those from GALAH. We note that the plots' color-coding indicates the goodness-of-fit (using \linktoparampath{astrophysical_parameters}{logposterior_msc}) rather than a source density. Except for \azero, the goodness-of-fit is best around the identity line.
Such behavior confirms that \msc\ fits well the composite spectra of binaries when the MCMC procedure converges. The goodness-of-fit also indicates that \msc\ did not converge for many sources properly.

We can flag bad convergence as sources with low \texttt{logposterior\_msc} values.
Finding a unique threshold for all science applications is challenging. However, Table\,\ref{tab:cu8par_apsis_msc_Galah} provides the evolution of the residual statistics with the GALAH sample when changing the goodness-of-fit threshold. By construction, the residuals and the overall biases improve as the threshold increases, but we remove a significant number of sources from the sample.
Regardless of this filtering, \msc\ tend to overestimate $\logg_1$, $\logg_2$, and \mh for the GALAH sample.
We suspect that \msc's prior favoring solar metallicity leads to overestimating \mh. As a result, to match the BP and RP spectra, \msc\ compensates high \mh\ by decreasing the intrinsic luminosity, requiring higher \logg values.
However, we cannot exclude the existence of biases in the GALAH data as suggested by the fact that the GALAH catalog provides significantly lower \mh\ for the binaries than those of their single stars \citep[Sec:8.3]{2020A&A...638A.145T}. This open issue is also supported by the discrepancies with APs reported by the APOGEE binary sample \citep{2018MNRAS.476..528E} with $26$ sources in common.

\begin{table}
    \caption{\msc\  vs. GALAH sample bias and RMS comparison for different \texttt{logposterior\_msc} percentile cut-offs. \label{tab:cu8par_apsis_msc_Galah}}
    \begin{center}
        \begin{tabular}{ccccccc}
            \hline
            percentiles          & 0        & 5       & 16     & 50     & 84     & 95   \\
            count                & 11\,263  & 10\,699 & 9\,461 & 5\,637 & 1\,814 & 567  \\
            \hline
            \hline
            parameters $\downarrow$ & \multicolumn{6}{c}{sample RMS}                    \\
            \hline
            $\teff_{,1}$ [K]     & 387      & 348     & 273    & 192    & 144    & 135  \\
            $\teff_{,2}$ [K]     & 632      & 592     & 536    & 417    & 310    & 258  \\
            $\logg_1$    [dex]   & 0.40     & 0.35    & 0.33   & 0.29   & 0.25   & 0.24 \\
            $\logg_2$    [dex]   & 0.58     & 0.54    & 0.50   & 0.45   & 0.38   & 0.36 \\
            \mh [dex]            & 0.30     & 0.29    & 0.27   & 0.24   & 0.22   & 0.21 \\
            distance [pc]        & 617      & 553     & 277    & 152    & 47     & 25   \\
            \azero\ [mag]        & 0.27     & 0.24    & 0.21   & 0.19   & 0.15   & 0.13 \\
            \hline
            \hline
                                    & \multicolumn{6}{c}{sample bias}                   \\
            \hline
            $\teff_{,1}$ [K]     & -139     & -118    & -72    & -6     & 21     & 10   \\
            $\teff_{,2}$ [K]     & -418     & -392    & -350   & -245   & -144   & -60  \\
            $\logg_1$    [dex]   & 0.24     & 0.22    & 0.20   & 0.17   & 0.13   & 0.12 \\
            $\logg_2$    [dex]   & 0.35     & 0.33    & 0.30   & 0.24   & 0.17   & 0.15 \\
            \mh [dex]            & 0.21     & 0.20    & 0.19   & 0.19   & 0.19   & 0.18 \\
            distance [pc]        & -184     & -148    & -95    & -49    & -16    & -9   \\
            \azero\ [mag]        & -0.01    & 0.00    & 0.01   & 0.02   & 0.01   & 0.01 \\
            \hline
        \end{tabular}
    \end{center}
\end{table}

We also found chemically homogeneous spectroscopic parameters from \gaia\ for the components of wide binaries when compared with high-resolution data from \citet{2020MNRAS.492.1164H}.
In their sample of $25$ wide binaries, $20$ had a metallicity difference less than 0.05~dex, while the remaining five showed deviations of $\sim$0.1~dex.
From Table~3 of \citet{2020MNRAS.492.1164H}, we selected the 20 homogeneous binaries (excluding WB02, WB05, WB09, WB16, WB21) and compared the metallicities from Apsis for each of the two components\footnote{The Gaia DR2 source IDs listed in Table~3 of \citet{2020MNRAS.492.1164H} are the same as the Gaia DR3 source IDs, except for WB13B, which has DR3 source ID 3230677874682668672.}, without applying any calibrations to the data.
These are dwarf stars with \teff\ between 5000 and 6400~K and metallicities above $-0.8$~dex.
For 16 out of the 20 homogeneous binaries according to \citet{2020MNRAS.492.1164H}, the metallicities from \gspphot\ (\linktoparampath{astrophysical_parameters}{mh_gspphot}) agree within 0.15~dex. For the remaining 4 binaries they deviate by 0.2 to 0.3~dex (WB08, WB13, WB18, WB22).
18 of the 20 binaries have metallicity determinations from \gspspec\ (\linktoparampath{astrophysical_parameters}{mh_gspspec}) for both components, and all except two agree within $0.15$\,dex. The exceptions are WB14 with a difference of $0.16$\,dex, and WB15 with a difference of 0.5~dex. WB15 also has a difference in \logg\ (\linktoparampath{astrophysical_parameters}{logg_gspspec}) of 1.1~dex, whereas the two components should have equal surface gravity according to \citet{2020MNRAS.492.1164H}.
This indicates that the \gaia\ metallicities are reliable (at least in a statistical sense) in the parameter space covered by the binary sample.

We further explored the possibility of ``clean'' the \msc\  results by excluding sources with possible spurious astrometric solutions. It is not a surprise that Gaia astrometry may be affected by binarity. We applied the method from \citet{Rybizki2021} and we kept sources with \texttt{fidelity\_v2}\,$>0.5$.
After this selection, the GALAH sample shrunk from 11,263 to 9,836 sources. The RMS for the distance comparison improves from $617$ to $429$\,pc, and its bias from $-184$ to $-157$ pc (when we assume inverse parallax as the ``true'' distance). It also improves the statistics of the other parameters and overall the agreement with \gspphot's APs.

Overall, \msc's performance remains challenging to estimate. Only a few reference catalogs exist, and they rarely provide statistically significant samples (many thousands) with APs. In addition, one needs to use the astrometric measurements of binary systems with caution. We expect \gdr{4} to provide a significant improvement in the future.

\subsection{Identification and analysis of peculiar cases (outliers)}\label{sec:weird}

Galactic sources dominate the content of \gdr{3}. These with BP and RP spectra are essentially intermediate-mass stars of FGK spectral types with $\gmag < 17.65$\,mag, with the addition of a set of UCDs and extragalactic objects (see Fig.~\ref{fig:cmd_per_data}).
Outliers in this context mean objects that are not ``similarly consistent'' with the rest of the sample. The similarity in this context relates to the distance metric implemented in the clustering algorithm in the \oa\ module summarized below.

On the one hand, \apsis\ provides multiple classifications and flags that one can use to identify outliers (see Table~\ref{tab:product-module-atm-2}). For instance, one can remove stars with emission lines using \espels\ parameters, or one can generate a pure sample of solar analogs by combining APs and flags from \gspphot, \gspspec \citep[see][and other examples herein]{DR3-DPACP-123}. However, these derive from supervised classifications and comparisons against models, limiting discoveries of peculiar objects.

On the other hand, the \oa\ (outlier analysis) software is an \apsis\ module that aims at identifying groups of similar objects in the \gdr{3} sample according to their BP and RP spectra exclusively.
\oa's approach to unsupervised clustering is entirely empirical by implementing self-organizing maps \citep{Kohonen2001}.
One can further explore the resulting clusters and label them or identify new classes of objects. However, \oa\ analyzes only 10\% of the sources processed by \dsc,  those with the lowest \dsc\ combined probabilities of membership to astronomical classes.  These represent about 56 million sources in \gdr{3}. We note that the analysis scope will expand in \gdr{4}.

To compare the results from \oa to those of \dsc, we identified \oa's clusters associated with the \dsc\ classes (see \linksec{ssec:par:cu8par_apsis_oa_methods_template_matching}{Section 11.3.12.3.4 in the online documentation} for further details).
Table~\ref{tab:oadsc} presents the resulting confusion matrix between \dsc\ and \oa. We find an 83\% agreement between the two classifications for galaxies, however only 35\% agreement for quasars where \oa confused them with stars and white dwarfs. We assume that the extragalactic classification from \dsc\ is accurate as shown in \citet{DR3-DPACP-158}. We note that \dsc\ includes astrometric information in its analysis which \oa\ does not. It is thus not surprising to find significant differences. These results show that both classifications are complementary.

One way to analyze \oa's neurons (or clusters) is to compare their prototype spectra with templates. We constructed our templates from averaged spectra having reliable spectral classifications in the literature, mainly from APOGEE-DR17 and GALAH-DR3. The \linksec{ssec:cu8par_apsis_oa}{online documentation (Section 11.3.12)} details our procedure.
Based on these stellar templates, \oa\ attributed spectral labels (A, F, G, K, and M-type stars) to its relevant clusters.
We compared these labels to the \gspphot\ temperatures (\linktoparam{astrophysical_parameters}{teff_gspphot}).
We cast the \teff scale of \gspphot\ stars into: O ($\teff\geq30000$\,K), B ($10000\leq \teff < 30000$\,K), A ($7300\leq \teff < 10000$\,K),  stars), F ($5950\leq \teff < 7300$\,K), G ($5200\leq \teff < 5950$\,K), K ($3760\leq \teff < 5200$\,K) and M ($\teff < 3760$\,K),  and we constructed the confusion matrix shown in Table~\ref{tab:oagspphot}, which shows the agreement between the two modules.

\begin{figure}[h]
    \centering
    \includegraphics[width=0.98\columnwidth]{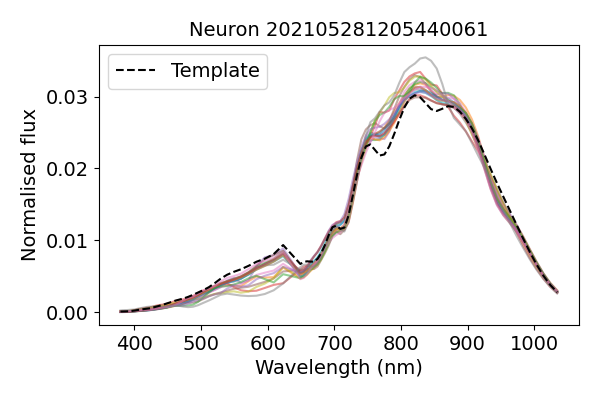}
    \caption{BP and RP spectra of 18 stars labelled as M-type by \oa, but having $\teff > 30\,000$\,K from \gspphot.
    The dashed line indicates the best stellar template for this cluster, corresponding to a M-type star.
     \label{fig:oaneuron}
    }
    \label{fig:oa_results_sources_g_mag_distribution}
\end{figure}

Overall, the agreement between both classifications is very high. However, we found $51$ O-type stars, $6$ B-type stars, and $10$ A-type stars from \gspphot\ that \oa\ classified as late-type stars.
Figure~\ref{fig:oaneuron} shows 18 BP/RP spectra from stars labeled as M-type by \oa\ but with \gspphot\ \teff > 30\,000\,K. All these objects have their SED peaked around $850$\,nm, typically expected for cool stars. As a result of visual inspection, \oa\ identified erroneous \teff\ labels from \gspphot.

On the one hand, the richness and variety of information about Milky-way stars are present in \gdr{3}. On the other hand, different interpretations and inconsistencies in the analysis we provide in the catalog warn the reader to proceed with caution.

\begin{table*}
    \caption{\dsc\ vs. \oa\ class label confusion matrix for the sample in common. \label{tab:oadsc}}
    \begin{center}
        \begin{tabular}{lrrrrrl}
            \hline
            \dsc   &                          &                       & \multicolumn{1}{c}{\oa}&                         &                        &                \\
                   & {STAR}                   & {WD}                  & {QSO}                   & {GAL}                   & {UNK}                  & {Total}        \\
            \hline\hline
            {STAR} & $21\,073\,253$ $(40$\%)  & $1\,735\,025$ $(3$\%) & $11\,834\,708$ $(22$\%) & $12\,709\,682$ $(24$\%) & $5\,942\,859$ $(11$\%) & $53\,295\,527$ \\
            {WD}   & $38\,651$ $(42$\%)       & $47\,418$ $(51$\%)    & $2\,881$ $(3$\%)        & $0$ $(0$\%)             & $3\,236$ $(4$\%)       & $92\,186$      \\
            {QSO}  & $617\,511$ $(29$\%)      & $453\,890$ $(21$\%)   & $763\,200$ $(35$\%)     & $48\,658$ $(2$\%)       & $275\,657$ $(13$\%)    & $2\,158\,916$  \\
            {GAL}  & $30\,351$ $(4$\%)        & $2\,542$ $(0$\%)      & $73\,493$ $(9$\%)       & $708\,253$ $(83$\%)     & $36\,488$ $(4$\%)      & $851\,127$     \\
            {UNK}  & $4\,110$ $(22$\%)        & $1\,320$ $(7$\%)      & $6\,481$ $(35$\%)       & $4\,183$ $(22$\%)       & $2\,510$ $(13$\%)      & $18\,604$      \\
            \hline
        \end{tabular}
    \end{center}
\end{table*}

\begin{table*}[]
\caption{\gspphot\ vs. \oa\ stellar type confusion matrix for the sample in common. \label{tab:oagspphot}}
    \begin{center}
    \begin{tabular}{@{}lrrrrrr@{}}
\hline
\gspphot   &                          &                       & \multicolumn{1}{c}{\oa}&                         &                        &                \\
                   & STAR-A       & STAR-F        & STAR-G       & STAR-K        & STAR-M        & \multicolumn{1}{c}{Total} \\ \hline\hline
STAR-O               & $146$ $(56$\%)    & $61$ $(24$\%)      & $1$ $(0$\%)       & $1$ $(0$\%)        & $50$ $(19$\%)      & $259$                       \\
STAR-B               & $4,082$ $(92$\%)  & $339$ $(8$\%)      & $6$ $(0$\%)       & $3$ $(0$\%)        & $3$ $(0$\%)        & $4,433$                     \\
STAR-A               & $23,836$ $(99$\%) & $250$ $(1$\%)      & $22$ $(0$\%)      & $10$ $(0$\%)       & $0$ $(0$\%)        & $24,118$                    \\
STAR-F               & $4,868$ $(4$\%)   & $126,786$ $(95$\%) & $1,719$ $(1$\%)   & $215$ $(0$\%)      & $34$ $(0$\%)       & $133,622$                   \\
STAR-G               & $0$ $(0$\%)       & $5,955$ $(13$\%)   & $37,699$ $(83$\%) & $1,697$ $(4$\%)    & $172$ $(0$\%)      & $45,523$                    \\
STAR-K               & $0$ $(0$\%)       & $0$ $(0$\%)        & $5,694$ $(2$\%)   & $241,823$ $(64$\%) & $131,517$ $(35$\%) & $379,034$                   \\
STAR-M               & $0$ $(0$\%)       & $0$ $(0$\%)        & $0$ $(0$\%)       & $11,613$ $(2$\%)   & $624,845$ $(98$\%) & $636,458$                   \\ \hline
        \end{tabular}
    \end{center}
\end{table*}

\section{Candidates for deeper science analyses}\label{sec:deeper}

We provide a list of six example use cases below as follows.

First is the identification of sources within some AP ranges. One should use the confidence intervals to find all sources of interest. For instance, \citet{DR3-DPACP-75} select upper main sequence stars from their apparent colors. \citet{DR3-DPACP-123} defined various ``golden'' samples of stars using our APs, stars with the most accurate and precise astrophysical parameters: for example, FGK star samples supporting many Galactic surveys, solar analogs, ultra-cool dwarfs, carbon stars, and OBA stars challenging our stellar evolution and atmosphere models.

The second is constructing the chemodynamical distribution of stars in some region of space. For instance,  \citet{DR3-DPACP-104} analyzed the chemical patterns in the positions and orbital motions of stars to reveal the flared structure of the Milky Way disk and the various orbital substructures associated with chemical patterns.

The third is constructing the three-dimensional spatial properties of the ISM.
Using published extinctions and distances, \citet{Dharmawardena2021} inferred the individual structure of the Orion, Taurus, Perseus, and Cygnus X star-forming regions and found the coherent ISM filaments that may link the Taurus and Perseus regions. One could easily replace those estimates with the ones (or a subset) we presented. Similarly, \citet{DR3-DPACP-144} explores the ISM kinematics using our DIB measurements.

A fourth is the age dating of wide binaries in the field. If an MS star has a white dwarf (WD) companion and a known distance, the age of such a binary system can then be determined precisely from the WD cooling sequence as long as the MS companion gives the chemical composition, much harder to obtain from the WD directly \citep[e.g.,][]{ 2019ApJ...870....9F, 2021ApJS..253...58Q}.

A fifth is providing the largest uniformly derived set of APs that one could use to calibrate theoretical or data-driven stellar models. For instance, \citet{2021ApJ...907...57G} developed a data-driven modeling technique to map stellar parameters  (e.g., \teff, \logg, \mh) accurately to spectrophotometric space, supporting more accurate 3D mapping of the Milky Way.

\begin{figure}
    \centering
    \includegraphics[width=\columnwidth]{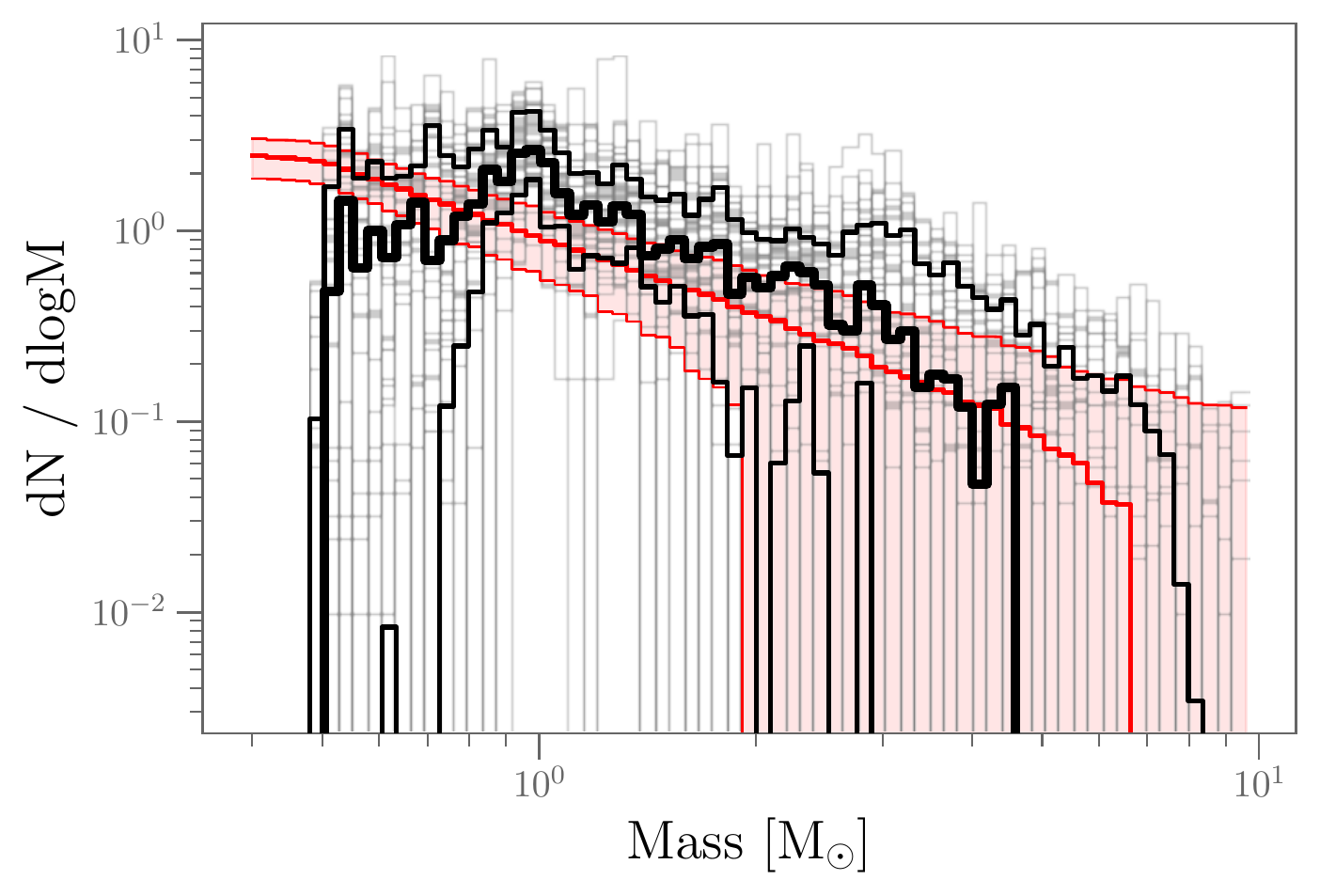}
    \caption{Mass distribution from FLAME compared with \citet{Kroupa2001} IMF. For each of the 44 open clusters from \citet{2018A&A...616A..10G},
    we plotted (grey) the recovered mass distributions from FLAME estimates. We highlighted the overall median and [16, 84th] percentile interval in black. For reference, we plot in red the expected shape of masses drawn from a Kroupa IMF accounting for the limited number of identified members. Because of the noise of low number-statistics, we expect significant scatter from cluster to cluster. The low-mass end is affected by Gaia's selection function.}
    \label{fig:flame_imf}
\end{figure}

A sixth application could be understanding the details of star formation and the dynamical evolution of star clusters. For instance, Fig.~\ref{fig:flame_imf} compares the \flame's (current) mass estimates with a simulation of stars drawn for a universal initial mass function (IMF; assumed here a \citealt{Kroupa2001}). This simulation is created by sampling the mass function (over the given mass range) for each cluster with their respective given number of Gaia identified members with mass estimates.
Although we make a comparison of current with initial stellar masses, the agreement is overall very good. The lower-mass end is affected by how many low-mass stars Gaia can extract from these clusters and thus cannot be well reproduced without a selection function. The upper-mass end agrees perfectly with our predictions from a single IMF. We note that \flame\ cannot predict masses above $10\,\msun$ with its current models.
Such analysis could support the study of cluster evaporation and mass segregation when also accounting for stellar mass loss.

Of course, this list is not exhaustive. The previous Gaia data releases led to thousands of studies ranging from solar system objects to discovering new streams and merger episodes that shaped our Galaxy.

\section{Limitations}\label{sec:limitations}

Users should keep in mind the following assumptions and limitations of our \gdr{3} catalog.

We produced APs that summarized many-dimensional posterior distributions using only quantile numbers such as mean, median, and percentile values (computed on one-dimensional marginal distributions). It is rarely possible to recover the complexity of the posterior distributions per object. One can query the MCMC chains published by \gspphot\ and \msc. These summary statistics cannot capture the full complexity of these distributions. One should not ignore the confidence intervals.

Most sources in \gdr{3} have substantial fractional parallax uncertainties. Hence, the spectro-photometric data (BP/RP) often dominate the inference of our distances and APs. However, the parallax remains generally sufficient to limit the dwarf versus giant degeneracies.

The poorer the data, the more our prior dominates our estimates. Our prior varies significantly per \apsis module. None of which included a three-dimensional extinction or Milky Way detailed model. One should expect significant differences with other AP catalogs when prior dominates. However, in reality, if the actual stellar population, extinction, or reddening distributions are very different from Galactic models, those differences may partially hint at these deviations.

To derive stellar APs, we implicitly assumed that all Gaia sources are single stars in the Galaxy (apart from \msc).  Those estimates are most likely incorrect for any non-single star (binaries, extended sources, extragalactic).

Furthermore, our stellar models also had intrinsic limitations in the range of parameters they could handle. For instance, our models did not include specific physics inherent to WDs, AGBs, and HB-stars.

Finally, by design, we infer properties for each source independently. If a set of stars is known to be in a cluster, they have a similar distance, extinction, chemical patterns, and age. It constitutes a prior that one should exploit to infer the properties of the individual stars more accurately than what we have done here.

\section{Summary}\label{sec:summary}

We have produced a catalog of distances, astrophysical, and dust extinction parameters using the Gaia BP, RP, RVS spectra, integrated G photometry, and parallaxes available with \gdr{3}.
More specifically, we provide:
\begin{itemize}
    \item 470 million distances, \teff, \logg, and \mh estimates using BP/RP,
    \item 6 million using RVS \teff, \logg, \mh, \afe estimates;
    \item 470 million radius estimates;
    \item 140 million mass, and 120 million age estimates;
    \item 5 million chemical abundance ratios;
    \item half-a-million diffuse interstellar band analysis parameters;
    \item 2 million stellar activity indices
    \item 200 million H$\alpha$ equivalent widths,
    \item and further stellar classification with 220 million spectral types  and 50 thousand emission-line stars.
\end{itemize}

We presented only a high-level overview of the validation and performance of these data products. We detail some of these tests and results in \citet{DR3-DPACP-157}, \citet{DR3-DPACP-158}, \citet{DR3-DPACP-156}, \citet{DR3-DPACP-186}, \citet{DR3-DPACP-175}, \citet{DR3-DPACP-127} and the \linksec{}{online documentation}.
Our tests comprised checking the astrophysical consistency of our data through, for example, HR or Kiel diagrams, which help to point out weaknesses in our analyses or failure in specific regions of the stellar parameter spaces.  In addition, we compared our estimates with external literature data to assess the performance of \apsis. The complexity and spread of our products often led us to restrict our tests to sub-samples and extrapolate our conclusions.

We emphasize that we did not calibrate \apsis\ APs to mimic external catalogs. Many of these external catalogs are not consistent with each other. As we do not know the true absolute scale of each AP dimension, we used external catalogs sometimes to obtain statistical relations to anchor our APs to a common ground. We recommend using these relations, but we did not apply them before the publication and instead provided the community with internally consistent APs.

First and foremost, our models have limitations in the range of parameters they can handle, and we made assumptions that we discussed in Sect.~\ref{sec:limitations}.

Our data necessarily demanded several extreme simplifications and assumptions. Therefore, one should use the data with great care. We recommend always using the flags/filters, defined in Appendix~\ref{sec:recommendations}.

Our catalog increases the availability of APs in the literature while offering results based on assumptions that differ from previous works. Such works helped to validate our results. In addition, it provides the community with values of reference to explore and understand better the content of \gdr{3}.

\gdr{3} is not an incremental improvement of the Gaia data. It multiplies the quantities of multi-messenger information of Gaia with new data products (e.g., BP, RP, RVS, APs). We increased the volume of sources with APs by a factor of $5$, but also increased the number of APs from two to $\sim 40$.
\gdr{3} represents a significant step forwards to anchor all current and future spectroscopic surveys to a common ground, and it provides us with the most comprehensive view of our Galaxy.


\begin{acknowledgements}
    We would like to thank the anonymous referee for the review of the manuscript and Carme Jordi for the very constructive feedback during the internal review of the manuscript.
    This work has made use of data from the European Space Agency (ESA)
    mission Gaia (\url{https://www.cosmos.esa.int/gaia}), processed by the Gaia Data Processing and Analysis Consortium (DPAC, \url{https://www.cosmos.esa.int/web/gaia/dpac/consortium}). Funding for the DPAC has been provided by national institutions, in particular, the institutions participating in the Gaia Multilateral Agreement.\\
    This research has used NASA's Astrophysics Data System, the VizieR catalogue access tool (CDS, Strasbourg, France).\\

    This publication made extensive use of the online authoring Overleaf platform (\url{https://www.overleaf.com/}).\\
    %
    The data processing and analysis made use of
        matplotlib \citep{Hunter:2007},
        NumPy \citep{harris2020array},
        the IPython package \citep{PER-GRA:2007},
        Vaex \citep{Breddels2018},
        TOPCAT \citep{Taylor2005},
        pyphot (\url{http://github.com/mfouesneau/pyphot}),
        QuantStack xtensor (\url{http://xtensor.readthedocs.io/}) and R \citep{RManual}.\\

    Part of the calculations have been performed with the high-performance computing facility SIGAMM, hosted by the Observatoire de la Côte d'Azur.\\

    CBJ, RA, MF, \& JR acknowledge the support from the DLR (German space agency) via grants 50QG0602, 50QG1001, 50QG1403, and 50QG2102.
    AB, ARB, AC, GC, GK, PdL, CO, PAP, MS, \& FT acknowledge financial supports from the french space agency (CNES), Agence National de la Recherche (ANR 14-CE33-014-01) and  Programmes Nationaux de Physique Stellaire \& Cosmologie et Galaxies (PNPS \& PNCG) of CNRS/INSU.\\
    AG, AK, BE, JS \& UH acknowledge financial supports from the Swedish National Space Agency (SNSA).\\

    LS acknowledges financial support by the Spanish Ministerio de Ciencia e Innovación through grant PID2020-112949GB-I00\\
    YF \& AL acknowledge the BELgian federal Science Policy Office (BELSPO) through various PROgramme de D\'eveloppement d'Exp\'eriences scientifiques (PRODEX) grants.\\
\end{acknowledgements}


\bibliographystyle{aa}
\bibliography{bibliography, 
              gaiadr3       
              }

\appendix

\section{Recommended caution and corrections}\label{sec:recommendations}

We recommend the following corrections for the \gdr{3} AP products:

- \gspphot\ provides a python tool that implements empirical calibrations models of its stellar parameters. It currently provides a metallicity, \mh, and an effective temperature, \teff, model. We trained these calibration models on literature catalogs (e.g, LAMOST DR6) based upon machine-learning algorithms. These models are not simple equations, and therefore, we provide a wrapper for the users so any update in the models will be transparently propagated. (see \citealt{DR3-DPACP-156}).

- We recommend checking for potential outliers when using \gspphot's APs by using the fractional parallax uncertainties ($\sigma_\varpi/\varpi$). It also helps identify inference priors and assumptions about the Milky-Way structure that matter.

- \gspspec\ provides an extensive flag definition detailed in \citet{DR3-DPACP-186}.

- \gspspec\ also recommend polynomial functions of \logg\ to rescale the various abundance ratios (\mh, \afe, [Mg/H], etc.). These relations are polynomial equations with coefficients given in Table~3 of \citet{DR3-DPACP-186}.

\section{Example queries}\label{sec:queries}

In this section, we list query examples that we used to produce various figures in the manuscript.

\noindent$\bullet$ Data behind Figs.~\ref{fig:cmd_per_data} and ~\ref{fig:cqd_per_module}. The following query took about $9$ hours to run. Selecting random subsets could extract statistically equivalent smaller datasets and run faster.

\begin{lstlisting}[language=sql]
select
   round((gaia.phot_g_mean_mag + 5 * log10(parallax/100)) * 10) / 10 as gmag,
   round(gaia.bp_rp * 10) / 10 as bp_rp,
   count(*) as n,
   sum(IF_THEN_ELSE(gaia.has_xp_continuous = 'true', 1, 0)) as has_xp,
   sum(IF_THEN_ELSE(gaia.has_rvs = 'true', 1, 0)) as has_rvs,
   count(aps.classprob_dsc_combmod_star) as dsc,
   count(aps.teff_gspphot) as gspphot,
   count(aps.teff_gspspec) as gspspec,
   count(aps.classlabel_espels) as espels,
   count(aps.teff_esphs) as esphs,
   count(aps.teff_espucd) as espucd,
   count(aps.activityindex_espcs) as espcs,
   count(aps.radius_flame) as flame,
   count(aps.teff_msc1) as msc,
   count(aps.neuron_oa_id) as oa
from gaiadr3.gaia_source as gaia
inner join gaiadr3.astrophysical_parameters as aps
	on aps.source_id = gaia.source_id
group by bp_rp, gmag
order by bp_rp, gmag
\end{lstlisting}

\noindent$\bullet$ The Kiel diagram from \gspphot used in Fig.\ref{fig:apsis_kiel_modules} and Fig.~\ref{fig:gspphot_kiel_libraries}.

\begin{lstlisting}[language=sql]
select
    floor(log10(teff_gspphot) / 0.05) * 0.05 as logT,
    floor(logg_gspphot / 0.05) * 0.05 as logg,
    count(*) as n
from gaiadr3.gaia_source
group by logT, logg
\end{lstlisting}

\noindent$\bullet$ Data behind Fig.~\ref{fig:GSPPhot-distances-clusters}. The following query took about $20$ minutes to run. {We note that the authors of \citet{2020A&A...640A...1C} shared their cluster catalog through the Archive}
\begin{lstlisting}[language=sql]
select
	cl_members.source_id, cl_members.cluster,
    cl_prop.agenn as logage, cl_prop.avnn as av,
	cl_prop.dmnn,  cl_prop.distpc,
	gaia.distance_gspphot, gaia.azero_gspphot
from user_tcantatg.members_2681_ocs as cl_members
inner join user_tcantatg.clusters_dr3_astrometry as cl_prop
	on cl_members.cluster = cl_prop.oc
inner join gaiadr3.gaia_source as gaia
	on gaia.source_id = cl_members.source_id
\end{lstlisting}

\noindent$\bullet$ The queries behind Fig.~\ref{fig:gspphot-gspspec-aps1}. As one cannot use the string comparison in the selection clause, we needed to run two queries.
\begin{lstlisting}[language=sql]
select round(log10(teff_gspphot) * 100) * 0.01 as logteff_gspphot,
       round(log10(teff_gspspec) * 100) * 0.01 as logteff_gspspec,
       count(*) as n
from gaiadr3.astrophysical_parameters
group by logteff_gspphot, logteff_gspspec
\end{lstlisting}
\begin{lstlisting}[language=sql]
select round(log10(teff_gspphot) * 100) * 0.01 as logteff_gspphot,
       round(log10(teff_gspspec) * 100) * 0.01 as logteff_gspspec,
       count(*) as n
from gaiadr3.astrophysical_parameters
where flags_gspspec is not NULL
and flags_gspspec like '0000000000000%'
group by logteff_gspphot, logteff_gspspec
\end{lstlisting}

\noindent$\bullet$ The queries behind Fig.~\ref{fig:apsis-FLAME-vs-GSPPhot-MG}. We generated the quantities directly during the query.
\begin{lstlisting}[language=sql]
select
	round((2 * log10(aps.radius_gspphot) + 4 * log10(aps.teff_gspphot / 5778.)) * 100.) / 100. as loglum_gspphot,
	round(log10(aps.lum_flame) * 100) / 100. as loglum_flame,
    count(*) as n
from gaiadr3.astrophysical_parameters as aps
group by loglum_gspphot, loglum_flame
order by loglum_gspphot, loglum_flame
\end{lstlisting}
\begin{lstlisting}[language=sql]
select
	round(aps.mg_gspphot * 100.) / 100. as mg_gspphot,
    round((4.74 - 2.5 * log10(aps.lum_flame) - aps.bc_flame) * 100) / 100. as mg_flame,
    count(*) as n
from gaiadr3.astrophysical_parameters as aps
group by mg_gspphot, mg_flame
order by mg_gspphot, mg_flame
\end{lstlisting}

\noindent$\bullet$ The queries behind Fig.~\ref{fig:extinction_gspphot_tge} where we used \tge's tracer selection to pick the giant stars from \gspphot.
\begin{lstlisting}[language=sql]
select
     count(*) as n, GAIA_HEALPIX_INDEX(9, aps.source_id) as hpx9,
     avg(aps.azero_gspphot) as azero_mean_gspphot, stddev(aps.azero_gspphot) as azero_std_gspphot,
     aps.libname_gspphot,
     num_tracers_used as n_tge, a0 as azero_mean_tge, a0_uncertainty as azero_std_tge
from gaiadr3.astrophysical_parameters as aps
inner join gaiadr3.total_galactic_extinction_map_opt
     on healpix_id = GAIA_HEALPIX_INDEX(9, aps.source_id)
where aps.azero_gspphot is not NULL
and teff_gspphot between 3000 and 5700 and mg_gspphot between -10 and 4
group by hpx9, libname_gspphot
\end{lstlisting}




\section{Candidate UCDs in young associations}
\label{sec:ucdsinYAs}


Table \ref{tab:espucd_clusters} lists the young associations for which we have identified candidate UCD members using BANYAN $\Sigma$ \citep{2018ApJ...856...23G}
 or the OPTICS clustering algorithm \citep{Ankerst99optics:ordering}.

\begin{table}
    \caption{Number of UCD candidates in nearby young associations according to BANYAN $\Sigma$ and our clustering analysis using the OPTICS algorithm. \label{tab:espucd_clusters}}
    \begin{center}
        \begin{tabular}{lrrl}
            \hline
            Association       & \# UCDs & Min \teff (K) & Method          \\
            \hline\hline
            CARN              & 61      & 1250          & BANYAN $\Sigma$ \\
            ARG               & 424     & 1494          & BANYAN $\Sigma$ \\
            ABDMG             & 155     & 1557          & BANYAN $\Sigma$ \\
            BPMG              & 47      & 1874          & BANYAN $\Sigma$ \\
            THA               & 42      & 1882          & BANYAN $\Sigma$ \\
            UCL               & 575     & 1991          & BANYAN $\Sigma$ \\
            COL               & 39      & 2006          & BANYAN $\Sigma$ \\
            HYA               & 52      & 2070          & BANYAN $\Sigma$ \\
            CAR               & 20      & 2105          & BANYAN $\Sigma$ \\
            OCT               & 153     & 2110          & BANYAN $\Sigma$ \\
            LCC               & 241     & 2166          & BANYAN $\Sigma$ \\
            ROPH              & 63      & 2168          & BANYAN $\Sigma$ \\
            USCO              & 508     & 2176          & BANYAN $\Sigma$ \\
            TWA               & 11      & 2189          & BANYAN $\Sigma$ \\
            TAU               & 214     & 2190          & BANYAN $\Sigma$ \\
            THOR              & 11      & 2233          & BANYAN $\Sigma$ \\
            CRA               & 7       & 2262          & BANYAN $\Sigma$ \\
            PL8               & 20      & 2279          & BANYAN $\Sigma$ \\
            IC2391            & 20      & 2321          & BANYAN $\Sigma$ \\
            PLE               & 97      & 2331          & BANYAN $\Sigma$ \\
            IC2602            & 12      & 2360          & BANYAN $\Sigma$ \\
            UCRA              & 45      & 2362          & BANYAN $\Sigma$ \\
            EPSC              & 5       & 2374          & BANYAN $\Sigma$ \\
            VCA               & 4       & 2385          & BANYAN $\Sigma$ \\
            XFOR              & 2       & 2391          & BANYAN $\Sigma$ \\
            118TAU            & 5       & 2415          & BANYAN $\Sigma$ \\
            CBER              & 7       & 2456          & BANYAN $\Sigma$ \\
            \hline
            NGC 1333 + IC 348 & 488     & 2230          & OPTICS          \\
            Serpens           & 420     & 2185          & OPTICS          \\
            Chameleon         & 69      & 2228          & OPTICS          \\
            $\gamma$2 Vel     & 266     & 2387          & OPTICS          \\
            Orion             & 1083    & 2156          & OPTICS          \\
            \hline
        \end{tabular}
    \end{center}
    BANYAN $\Sigma$ identifiers, see \cite{2018ApJ...856...23G},\\
    OPTICS clustering algorithm, see \citep{Ankerst99optics:ordering}.
\end{table}

\section{AP estimates, producers, and where to find them}\label{sec:apfields}

In this section, we compiled the various estimates of stellar parameters from \gdr{3}, which \apsis\ module producing them, and which table and field store the values in the Gaia catalog.

Table~\ref{tab:product-module-distance} lists the distance estimates discussed in Sect.~\ref{sec:distances}. Tables~\ref{tab:product-module-atm-1} and \ref{tab:product-module-atm-2} list the primary and secondary atmospheric estimates, respectively (discussed in Sect.~\ref{sec:atmosphere}). Table~\ref{tab:product-module-abundances} lists the abundances estimates (discussed in Sect.~\ref{sec:abundances}).
Table~\ref{tab:product-module-evolution-aps} lists the parameters characterizing the evolutionary state of a star. Finally, Table~\ref{tab:product-module-extinction} lists the extinction parameters and the diffuse interstellar band properties (discussed in Sect.~\ref{sec:dust}).

\begin{table*}
    \caption{Distance estimates in \gdr{3}}
    \label{tab:product-module-distance}
    \centering
    \begin{tabular}{rcl}
        \hline
        \multirow{4}{*}{distance}
         & \gspphot &
        \begin{tabular}[c]{@{}l@{}}
            \linktoparampath{gaia_source}{distance_gspphot}                      \\
            \linktoparampath{astrophysical_parameters}{distance_gspphot}             \\
            \linktoparampath{astrophysical_parameters_supp}{distance_gspphot_a}      \\
            \linktoparampath{astrophysical_parameters_supp}{distance_gspphot_marcs}  \\
            \linktoparampath{astrophysical_parameters_supp}{distance_gspphot_ob}     \\
            \linktoparampath{astrophysical_parameters_supp}{distance_gspphot_phoenix}\\
        \end{tabular}  \\
        &          &		\\
        & \msc &
       \begin{tabular}[c]{@{}l@{}}
            \linktoparampath{astrophysical_parameters}{distance_msc}		\\
       \end{tabular}		\\
   \end{tabular}
\end{table*}

\begin{table*}
    \caption{Primary atmospheric estimates in \gdr{3}: $\teff, \logg, \mh, [\alpha/\text{Fe}]$}
    \label{tab:product-module-atm-1}
    \centering
    \begin{tabular}{rcl}
        \hline
        \multirow{11}{*}{$\teff$}
         & \gspphot &
        \begin{tabular}[c]{@{}l@{}}
            \linktoparampath{gaia_source}{teff_gspphot}                          \\
            \linktoparampath{astrophysical_parameters}{teff_gspphot}             \\
            \linktoparampath{astrophysical_parameters_supp}{teff_gspphot_a}      \\
            \linktoparampath{astrophysical_parameters_supp}{teff_gspphot_marcs}  \\
            \linktoparampath{astrophysical_parameters_supp}{teff_gspphot_ob}     \\
            \linktoparampath{astrophysical_parameters_supp}{teff_gspphot_phoenix}\\
        \end{tabular}  \\
        &          &		\\
        & \gspspec &
       \begin{tabular}[c]{@{}l@{}}
            \linktoparampath{astrophysical_parameters}{teff_gspspec}		\\
            \linktoparampath{astrophysical_parameters_supp}{teff_gspspec_ann}		\\
       \end{tabular}		\\
        &          &		\\
        & \esphs   &
       \begin{tabular}[c]{@{}l@{}}
            \linktoparampath{astrophysical_parameters}{teff_esphs}		\\
       \end{tabular}		\\
        &          &		\\
        & \espucd  &
       \begin{tabular}[c]{@{}l@{}}
            \linktoparampath{astrophysical_parameters}{teff_ucd}		\\
       \end{tabular}		\\
        &          &		\\
        & \msc     &
       \begin{tabular}[c]{@{}l@{}}
            \linktoparampath{astrophysical_parameters}{teff_msc1}		\\
            \linktoparampath{astrophysical_parameters}{teff_msc2}		\\
       \end{tabular}		\\
       \hline
       \multirow{8}{*}{$\logg$}
        & \gspphot &
       \begin{tabular}[c]{@{}l@{}}
            \linktoparampath{gaia_source}{logg_gspphot}		\\
            \linktoparampath{astrophysical_parameters}{logg_gspphot}		\\
            \linktoparampath{astrophysical_parameters_supp}{logg_gspphot_a}		\\
            \linktoparampath{astrophysical_parameters_supp}{logg_gspphot_marcs}		\\
            \linktoparampath{astrophysical_parameters_supp}{logg_gspphot_ob}		\\
            \linktoparampath{astrophysical_parameters_supp}{logg_gspphot_phoenix}		\\
       \end{tabular}		\\
        &          &		\\
        & \gspspec &
       \begin{tabular}[c]{@{}l@{}}
            \linktoparampath{astrophysical_parameters}{logg_gspspec}		\\
            \linktoparampath{astrophysical_parameters_supp}{logg_gspspec_ann}		\\
       \end{tabular}		\\
        &          &		\\
        & \esphs   &
       \begin{tabular}[c]{@{}l@{}}
            \linktoparampath{astrophysical_parameters}{logg_esphs}		\\
       \end{tabular}		\\
        & \msc     &
       \begin{tabular}[c]{@{}l@{}}
            \linktoparampath{astrophysical_parameters}{logg_msc1}		\\
            \linktoparampath{astrophysical_parameters}{logg_msc2}
       \end{tabular}		\\
       \hline
       \multirow{5}{*}{[M/H]}
        & \gspphot &
       \begin{tabular}[c]{@{}l@{}}
            \linktoparampath{gaia_source}{mh_gspphot}		\\
            \linktoparampath{astrophysical_parameters}{mh_gspphot}		\\
            \linktoparampath{astrophysical_parameters_supp}{mh_gspphot_a}		\\
            \linktoparampath{astrophysical_parameters_supp}{mh_gspphot_marcs}		\\
            \linktoparampath{astrophysical_parameters_supp}{mh_gspphot_ob}		\\
            \linktoparampath{astrophysical_parameters_supp}{mh_gspphot_phoenix}		\\
       \end{tabular}		\\
        &          &		\\
        & \gspspec &
       \begin{tabular}[c]{@{}l@{}}
\linktoparampath{astrophysical_parameters}{mh_gspspec}		\\
\linktoparampath{astrophysical_parameters_supp}{mh_gspspec_ann}		\\
       \end{tabular}		\\
       \hline
       \multirow{1}{*}{$[\alpha/\text{Fe}]$}
        & \gspspec &
       \begin{tabular}[c]{@{}l@{}}
\linktoparampath{astrophysical_parameters}{alphafe_gspspec}		\\
\linktoparampath{astrophysical_parameters_supp}{alphafe_gspspec_ann}		\\
       \end{tabular}		\\
       \hline
   \end{tabular}
\end{table*}

\begin{table*}
    \caption{Secondary atmospheric estimates in \gdr{3}: classes, rotation, emission, activity}
    \label{tab:product-module-atm-2}
    \centering
    \begin{tabular}{rcl}
        \hline
        \multirow{14}{*}{classification}
         & \dsc    &
        \begin{tabular}[c]{@{}l@{}}
            \linktoparampath{astrophysical_parameters}{classprob_dsc_allosmod_star}      \\
            \linktoparampath{astrophysical_parameters}{classprob_dsc_combmod_binarystar} \\
            \linktoparampath{astrophysical_parameters}{classprob_dsc_combmod_star}       \\
            \linktoparampath{astrophysical_parameters}{classprob_dsc_combmod_whitedwarf} \\
            \linktoparampath{astrophysical_parameters}{classprob_dsc_specmod_binarystar} \\
            \linktoparampath{astrophysical_parameters}{classprob_dsc_specmod_star}       \\
            \linktoparampath{astrophysical_parameters}{classprob_dsc_specmod_whitedwarf} \\
        \end{tabular} \\
         &         &                                                                 \\
         & \esphs  &
        \begin{tabular}[c]{@{}l@{}}
            \linktoparampath{astrophysical_parameters}{spectraltype_esphs} \\
        \end{tabular}               \\
         &         &                                                                 \\
         & \espels &
        \begin{tabular}[c]{@{}l@{}}
            \linktoparampath{astrophysical_parameters}{classlabel_espels}           \\
            \linktoparampath{astrophysical_parameters}{classprob_espels_bestar}     \\
            \linktoparampath{astrophysical_parameters}{classprob_espels_dmestar}    \\
            \linktoparampath{astrophysical_parameters}{classprob_espels_herbigstar} \\
            \linktoparampath{astrophysical_parameters}{classprob_espels_pne}        \\
            \linktoparampath{astrophysical_parameters}{classprob_espels_ttauristar} \\
            \linktoparampath{astrophysical_parameters}{classprob_espels_wcstar}     \\
            \linktoparampath{astrophysical_parameters}{classprob_espels_wnstar}     \\
        \end{tabular}      \\
        \hline
        \multirow{1}{*}{rotation}
         & \esphs  &
        \begin{tabular}[c]{@{}l@{}}
            \linktoparampath{astrophysical_parameters}{vsini_esphs} \\
        \end{tabular}                      \\
        \hline
        \multirow{3}{*}{Chromospheric activity}
         & \espels &
        \begin{tabular}[c]{@{}l@{}}
            \linktoparampath{astrophysical_parameters}{ew_espels_halpha} \\
        \end{tabular}                 \\
         &         &                                                                 \\
         & \espcs  &
        \begin{tabular}[c]{@{}l@{}}
            \linktoparampath{astrophysical_parameters}{activityindex_espcs} \\
        \end{tabular}              \\
        \hline
    \end{tabular}
\end{table*}

\begin{table*}
    \caption{13 chemical abundance ratios from 12 individual elements (N, Mg, Si, S, Ca, Ti, Cr, Fe, Ni, Zr, Ce, and Nd; with the FeI and FeII species) and CN equivalent width in \gdr{3}}
    \label{tab:product-module-abundances}
    \centering
    \begin{tabular}{rcl}
        \hline
        \multirow{1}{*}{chemical abundances}
         & \gspspec &
        \begin{tabular}[c]{@{}l@{}}
            \linktoparampath{astrophysical_parameters}{fem_gspspec}   \\
            \linktoparampath{astrophysical_parameters}{feiim_gspspec} \\
            \linktoparampath{astrophysical_parameters}{cafe_gspspec}  \\
            \linktoparampath{astrophysical_parameters}{cefe_gspspec}  \\
            \linktoparampath{astrophysical_parameters}{crfe_gspspec}  \\
            \linktoparampath{astrophysical_parameters}{mgfe_gspspec}  \\
            \linktoparampath{astrophysical_parameters}{ndfe_gspspec}  \\
            \linktoparampath{astrophysical_parameters}{nfe_gspspec}   \\
            \linktoparampath{astrophysical_parameters}{nife_gspspec}  \\
            \linktoparampath{astrophysical_parameters}{sfe_gspspec}   \\
            \linktoparampath{astrophysical_parameters}{sife_gspspec}  \\
            \linktoparampath{astrophysical_parameters}{tife_gspspec}  \\
            \linktoparampath{astrophysical_parameters}{zrfe_gspspec}  \\
            \linktoparampath{astrophysical_parameters}{cn0ew_gspspec} \\
        \end{tabular} \\
        \hline
    \end{tabular}
\end{table*}

\begin{table*}
    \caption{Evolution parameter estimates in \gdr{3}.}
    \label{tab:product-module-evolution-aps}
    \centering
    \begin{tabular}{rcl}
        \hline
        \multirow{1}{*}{Luminosity $\lum$}
         & \flame   &
        \begin{tabular}[c]{@{}l@{}}
            \linktoparampath{astrophysical_parameters}{lum_flame}           \\
            \linktoparampath{astrophysical_parameters_supp}{lum_flame_spec} \\
        \end{tabular}         \\
        \hline
        \multirow{1}{*}{absolute magnitude M$_G$}
         & \gspphot &
        \begin{tabular}[c]{@{}l@{}}
            \linktoparampath{astrophysical_parameters}{mg_gspphot}              \\
            \linktoparampath{astrophysical_parameters_supp}{mg_gspphot_a}       \\
            \linktoparampath{astrophysical_parameters_supp}{mg_gspphot_marcs}   \\
            \linktoparampath{astrophysical_parameters_supp}{mg_gspphot_ob}      \\
            \linktoparampath{astrophysical_parameters_supp}{mg_gspphot_phoenix} \\
        \end{tabular}     \\
        \hline
        \multirow{3}{*}{radius $\radius$}
         & \gspphot &
        \begin{tabular}[c]{@{}l@{}}
            \linktoparampath{astrophysical_parameters}{radius_gspphot}              \\
            \linktoparampath{astrophysical_parameters_supp}{radius_gspphot_a}       \\
            \linktoparampath{astrophysical_parameters_supp}{radius_gspphot_marcs}   \\
            \linktoparampath{astrophysical_parameters_supp}{radius_gspphot_ob}      \\
            \linktoparampath{astrophysical_parameters_supp}{radius_gspphot_phoenix} \\
        \end{tabular} \\
         &          &                                                           \\
         & \flame   &
        \begin{tabular}[c]{@{}l@{}}
            \linktoparampath{astrophysical_parameters}{radius_flame}           \\
            \linktoparampath{astrophysical_parameters_supp}{radius_flame_spec} \\
        \end{tabular}      \\
        \hline
        \multirow{1}{*}{age}
         & \flame   &
        \begin{tabular}[c]{@{}l@{}}
            \linktoparampath{astrophysical_parameters}{age_flame}           \\
            \linktoparampath{astrophysical_parameters_supp}{age_flame_spec} \\
        \end{tabular}         \\
        \hline
        \multirow{1}{*}{mass $\mass$}
         & \flame   &
        \begin{tabular}[c]{@{}l@{}}
            \linktoparampath{astrophysical_parameters}{mass_flame}           \\
            \linktoparampath{astrophysical_parameters_supp}{mass_flame_spec} \\
        \end{tabular}        \\
        \hline
        \multirow{1}{*}{evolution stage \evolstage}
         & \flame   &
        \begin{tabular}[c]{@{}l@{}}
            \linktoparampath{astrophysical_parameters}{evolstage_flame}           \\
            \linktoparampath{astrophysical_parameters_supp}{evolstage_flame_spec} \\
        \end{tabular}   \\
        \hline
    \end{tabular}
\end{table*}

\begin{table*}
    \caption{Extinction and DIB parameter estimates in \gdr{3}.}
    \label{tab:product-module-extinction}
    \centering
    \begin{tabular}{rcl}
        \hline
        \multirow{5}{*}{monochromatic at $541.4$\,nm \azero}
         & \gspphot   &
        \begin{tabular}[c]{@{}l@{}}
            \linktoparampath{astrophysical_parameters}{azero_gspphot}           \\
            \linktoparampath{astrophysical_parameters_supp}{azero_gspphot_a}      \\
            \linktoparampath{astrophysical_parameters_supp}{azero_gspphot_marcs}  \\
            \linktoparampath{astrophysical_parameters_supp}{azero_gspphot_ob}     \\
            \linktoparampath{astrophysical_parameters_supp}{azero_gspphot_phoenix}\\
        \end{tabular}         \\
         &         &                                                                 \\
         & \esphs   &
        \begin{tabular}[c]{@{}l@{}}
            \linktoparampath{astrophysical_parameters}{azero_esphs}           \\
        \end{tabular}         \\
         &         &                                                                 \\
         & \msc   &
        \begin{tabular}[c]{@{}l@{}}
            \linktoparampath{astrophysical_parameters}{azero_msc}           \\
        \end{tabular}         \\
        \hline
        \multirow{4}{*}{in G band \ag}
         & \gspphot   &
        \begin{tabular}[c]{@{}l@{}}
            \linktoparampath{astrophysical_parameters}{ag_gspphot}           \\
            \linktoparampath{astrophysical_parameters_supp}{ag_gspphot_a}      \\
            \linktoparampath{astrophysical_parameters_supp}{ag_gspphot_marcs}  \\
            \linktoparampath{astrophysical_parameters_supp}{ag_gspphot_ob}     \\
            \linktoparampath{astrophysical_parameters_supp}{ag_gspphot_phoenix}\\
        \end{tabular}         \\
         &         &                                                                 \\
         & \esphs   &
        \begin{tabular}[c]{@{}l@{}}
            \linktoparampath{astrophysical_parameters}{ag_esphs}           \\
        \end{tabular}         \\
        \hline
        \multirow{1}{*}{in BP band \abp}
         & \gspphot   &
        \begin{tabular}[c]{@{}l@{}}
            \linktoparampath{astrophysical_parameters}{abp_gspphot}           \\
            \linktoparampath{astrophysical_parameters_supp}{abp_gspphot_a}      \\
            \linktoparampath{astrophysical_parameters_supp}{abp_gspphot_marcs}  \\
            \linktoparampath{astrophysical_parameters_supp}{abp_gspphot_ob}     \\
            \linktoparampath{astrophysical_parameters_supp}{abp_gspphot_phoenix}\\
        \end{tabular}         \\
        \hline
        \multirow{1}{*}{in RP band \arp}
         & \gspphot   &
        \begin{tabular}[c]{@{}l@{}}
            \linktoparampath{astrophysical_parameters}{arp_gspphot}           \\
            \linktoparampath{astrophysical_parameters_supp}{arp_gspphot_a}      \\
            \linktoparampath{astrophysical_parameters_supp}{arp_gspphot_marcs}  \\
            \linktoparampath{astrophysical_parameters_supp}{arp_gspphot_ob}     \\
            \linktoparampath{astrophysical_parameters_supp}{arp_gspphot_phoenix}\\
        \end{tabular}         \\
        \hline
        \multirow{1}{*}{in BP-RP color \ebpminrp}
         & \gspphot   &
        \begin{tabular}[c]{@{}l@{}}
            \linktoparampath{astrophysical_parameters}{ebpminrp_gspphot}           \\
            \linktoparampath{astrophysical_parameters_supp}{ebpminrp_gspphot_a}      \\
            \linktoparampath{astrophysical_parameters_supp}{ebpminrp_gspphot_marcs}  \\
            \linktoparampath{astrophysical_parameters_supp}{ebpminrp_gspphot_ob}     \\
            \linktoparampath{astrophysical_parameters_supp}{ebpminrp_gspphot_phoenix}\\
        \end{tabular}         \\
        \hline
        \multirow{1}{*}{DIB central wavelength, EW, and complexity}
         & \gspspec   &
        \begin{tabular}[c]{@{}l@{}}
            \linktoparampath{astrophysical_parameters}{dib_gspspec_lambda}      \\
            \linktoparampath{astrophysical_parameters}{dibew_gspspec}           \\
            \linktoparampath{astrophysical_parameters}{dib_p0_gspspec}  \\
            \linktoparampath{astrophysical_parameters}{dibp2_gspspec}           \\
        \end{tabular}         \\
        \hline
    \end{tabular}
\end{table*}



\end{document}